\documentclass{ws-rv9x6}
\usepackage{amsmath}
\usepackage{amssymb}
\usepackage{makeidx}

\makeindex

\allowdisplaybreaks[1]
\begin{document}
%
\setcounter{chapter}{0}
\chapter{PHASE TRANSITIONS IN\\ THE PSEUDOSPIN-ELECTRON MODEL}

\markboth{I. Stasyuk}{Phase Transitions in the Pseudospin-Electron
Model}

\author{Ihor Stasyuk}

\address{Institute for Condensed Matter Physics\\
of the
 National Academy of Science of Ukraine\\
 1 Svientsitskii Str., UA-79011 Lviv, Ukraine\\
E-mail: ista@icmp.lviv.ua}

\begin{abstract}
A review of the present  state of investigations of the
pseudospin-electron model (PEM), which is used in the theory of
strongly  correlated electron systems, is given. The model is used
to describe the systems with the locally anharmonic elements of
structure represented in the model by pseudospins. The consideration
is based on the dynamical mean  field theory approach and the
generalized random phase approximation. Electron spectrum and
thermodynamics of the model are investigated; the cases of the
simplified model, the  model with strong interaction and the
two-sublattice model are studied more in detail. The phase
transitions into other uniform or modulated states as well as
superconducting phases are described; the criteria of their
realization are established. Based on this, the description of
structural and dielectric (ferroelectric type) instabilities, phase
separation and bistability  phenomena is given. A comparison is made
with the thermodynamics of the Falicov-Kimball model (which can be
considered as a particular case of PEM). The possibility of applying
the PEM to the analysis of thermodynamics of the real HTSC systems
is discussed. Attention is paid to the unsolved problems in the
study of PEM.
\end{abstract}
\tableofcontents

\section{Introduction}\label{ista:p1}

Much attention is paid in recent years to the  investigation of
systems with  strong electron correlations, such as crystalline
compounds with the transition and rare-earth ions (transition
metals, transition metal oxides, mixed valent compounds, heavy
fermion systems, high temperature superconductors, etc.). Their
specific electronic, magnetic and conducting properties as well as
the presence of a variety of phase transitions and the phenomena
that are connected with this property  are caused to a great extent
by the splitting and reconstruction of energy spectrum due to
correlation effects. The theory of such systems is based on the
Hubbard model\index{model!Hubbard} and on its generalizations, where
the crucial idea is the one concerning the decisive role of the
strong short-range interaction of particles (electrons). At the
presence of additional (i.e., vibrational) degrees of freedom, one
can mention, among others, the pseudospin-electron model (PEM).
\index{model!pseudospin-electron (PEM)} The model appeared recently
in connection with the investigation of the high-$T_{c}$
superconductors. It was introduced to describe the contribution of
locally-anharmonic elements  of the crystal structure to their
electronic properties.

Electron system in PEM is described by the Hubbard
Hamiltonian\index{Hamiltonian!Hubbard} while the anharmonic
vibrational modes are treated using  the pseudospin
\index{pseudospin} formalism. The model Hamiltonian is as follows
\begin{equation}
H=\sum_i [U n_{i,\uparrow} n_{i,\downarrow}+(gS^{z}_i-\mu) (
n_{i,\uparrow}+ n_{i,\downarrow})-h S^{z}_{i} -\Omega S^{x}_{i}]
+\sum_{i,j,\sigma}t_{ij}a^{+}_{i,\sigma}a_{j,\sigma}.
\label{ista:eq0}
\end{equation}
Here $a_{i,\sigma}$, $a^{+}_{i,\sigma}$ are electron annihilation
 and creation operators, $n_{i,\sigma}$ is an electron occupation number;
 besides the electron correlation ($U$-term),
\index{correlation!electron $U$} the single-site part includes the
interaction with pseudospin ($g$-term) and the energy of the
tunnelling-like splitting \index{tunneling splitting} of vibrational
levels ($\Omega$-term); the field $h$ describes the asymmetry of
local potential. The electron transfer ($t$-term) is included as
well.
%

The pseudospin-electron Hamiltonian (in form (\ref{ista:eq0}) with
inclusion only of the $g$-interaction) was used by M\"uller with the
aim of describing the anharmonic vibrations in the oxygen subsystem
of the high-$T_{c}$ superconducting  crystals of the YBaCuO
type.\cite{ista:1} \index{YBaCuO-type superconducting crystals} The
YBa$_{2}$Cu$_{3}$O$_{7-\delta}$ crystal is a typical and most
studied example of  such objects. The unit cell contains, besides
two superconducting planes, the chain (at the $\delta\ll 1$
composition) elements Cu$_{1}$-O$_{1}$, connected by
Cu$_{1}$-O$_{4}$-Cu$_{2}$ bridges with Cu$_{2}$-O$_{2}$ plains
through the apical oxygen ions O$_{4}$. The vibrations of these ions
along the $c$-axis (perpendicularly to the plains) exhibit a strong
anharmonicity. Much evidence exists in support of this concept. One
can mention the EXAFS data,\cite{ista:2,ista:3} Raman scattering and
dielectric
measurements,\cite{ista:4,ista:5,ista:6,ista:7,ista:8,ista:9} local
polaron phenomena,\cite{ista:10,ista:11}  bistabilities in the
normal phase region\cite{ista:12} as well as neutron scattering
investigations\cite{ista:13} or the results of the first principle
LAPW calculations.\cite{ista:14} Despite a certain ambiguity in the
data, the conclusions were made about the existence of two different
equilibrium positions of the O$_{4}$ ion. The local double-well
picture as an approximate simplified model was supported by the
oxygen O$_{4}$ vacancy effect on the positions of apical ions
observed in  Ref.~[\refcite{ista:15}].

Moreover, a connection between positions of O$_{4}$ ions and
electron states in Cu$_{2}$-O$_{2}$ plains plays an important role
in YBa$_{2}$Cu$_{3}$O$_{7-\delta}$ crystals. The data given in
 Ref.~[\refcite{ista:16}] point  to the existence of a significant
correlation between the occupancy of electron states of the Cu$_{2}$
ion and the R$_{\rm O_{4}-Cu_{2}}$ distance as well as to the
decrease of this distance at the transition from the metallic
orthorombic phase to the semiconducting one (that takes place at
$\delta>\delta^{*}=0.55$). These and other similar facts suggest the
presence of a large electron-vibrational coupling. In the pseudospin
representation, when the pseudospin variable $S_{i}^{z}=\pm1/2$
defines the positions of O$_{4}$ ions, it is described by the
$gS_{i}^{z}n_{i\sigma}$ term.

Consideration based on the  PEM was applied to the HTSC systems
starting from
 Refs.~[\refcite{ista:1},\refcite{ista:17}--\refcite{ista:20}].
Hamiltonian similar to (\ref{ista:eq0}) was used by Hirsch and Tang
in the study of electron states in the framework of cluster
calculations. In the context of the idea concerning the effect of
anharmonicity on the superconducting transition
temperature,\cite{ista:17,ista:18,ista:19,ista:21} the possible
connection between superconducting pairing  and the lattice
anharmonicity was considered by Frick  et al.\cite{ista:22} (the
quantum Monte-Carlo calculations). In what follows, the
%
investigations
 of the PEM were  devoted to the analysis of
the electron spectrum,\cite{ista:23} the pseudospin and collective
dynamics,\cite{ista:24,ista:25}
  the charge and pseudospin pair correlations and the behaviour
of dielectric susceptibility.\cite{ista:26}

 In subsequent investigations, the main attention was  paid to the
 thermodynamics of the model in
special cases and simplifications: (i) a model with the infinitely
large correlation ($U\rightarrow\infty$) when the double occupation
of electron states on the site is excluded; (ii) simplified PEM with
$U=0$ and $\Omega=0$; (iii) simplified model ($U=0$) with the
tunnelling-like dynamics ($\Omega\neq 0$); (iv) two-sublattice PEM
for the layered
 structures of the YBaCuO-type;  (v) the cluster PEM of ferroelectric-superconductor
 heterostructures.\cite{Koerting05,Pavlenko05}

There was performed a study of
%
  phase transitions between
 the
 states with different electron concentrations
and  with
 different
orientations of pseudospins (in the regime of the fixed  chemical
potential, $\mu =\mathrm{const}$), and the phase separation effects
at a given concentration of electrons
($n=\mathrm{const}$).\cite{ista:27,ista:28} The possibility of
 the
appearance of a doubly modulated (so-called chessboard) phase or an
incommensurate phase (in the case of weak coupling) was
established;\cite{ista:29,ista:30,ista:31} the superconducting
instability in PEM was analysed.\cite{ista:32}
In the case of two-sublattice PEM the structural instabilities of
the ferroelectric type as well as the bistability phenomena  were
revealed and analysed.\cite{ista:33,ista:34,ista:35,ista:36}
The study of the PEM thermodynamics was performed mainly within the
%
  generalized random phase approximation (GRPA).\cite{ista:37} A
method of dynamical mean field theory (DMFT) was used in the case of
simplified PEM,\cite{ista:27} when the analytic formulation of the
theory is possible.
%
%

As  was shown, the  PEM also possesses  an  interesting collective
dynamics of  pseudospins. The corresponding spectrum changes its
form depending on the electron concentration and
temperature;\cite{ista:24,ista:25} the spectrum is different at high
or small values of $g$ and its shape also depends on the $h$ and
$\Omega$ parameters. In
Refs.~[\refcite{ista:a38}--\refcite{ista:a40}], the contributions
into Raman scattering intensity, connected with the mentioned
collective pseudospin excitations (that correspond to the
phonon-like vibrations of anharmonic  subsystems in the YBaCuO
structures) and electron intraband and interband transitions, were
considered.

The PEM is closely related  to the Falicov-Kimball (FK) model
\index{model!Falicov-Kimball (FK)} intensively studied in recent
years (see, for example, Ref.~[\refcite{ista:38}]), in which the
interaction between the localized and itinerant  particles
(electrons) is responsible for the similar phase transitions
(between states with different concentrations of particles and
with/without spatial modulation).
%
The simplified version of PEM corresponds  to the FK model in case
 there is no tunnelling-like splitting in the PEM  and when the
localized and the moving particles in FK model have different
chemical potentials (one can pass to FK model putting
$S_{i}^{z}=w_{1}-1/2$ and $h=\tilde{\mu}$, where $w_{i}$ and
$\tilde{\mu}$ are occupation number and chemical potential of
localized particles, respectively). The regimes of thermodynamic
averaging are usually different for both models (a fixed
concentration of the localized particles for FK model and a given
value of the field $h$ for the PEM).

It should be mentioned that for the recent few years the PEM has
found application in describing  the charge transfer in molecular
and crystalline systems with hydrogen bonds.\cite{ista:39} The model
is also very promising in investigating the thermodynamics of
processes connected with ionic intercalation in the layered
structures (see Ref.~[\refcite{ista:40}]), where different separated
positions exist in the unit cell for the intercalate ion, and the
hopping between them is possible (intercalation of Li$^{+}$ ions in
the TiO$_{2}$ matrix provides an example of such a
situation\cite{ista:41}).
A specific version of PEM was recently used\cite{ista:a45} in
modelling the electronic properties and the field effect in the
CuO$_{2}$/SrTiO$_{3}$ interfaces in HTSC/STO heterostructures.

This paper
  presents
 a review of the main results concerning the
thermodynamics and energy spectrum  of the PEM in the above
mentioned cases and approximations.
 The case of the simplified PEM ($U=0$;
$\Omega=0$ or $\Omega \neq 0$) is considered  more in  detail.
Dynamic properties of PEM are not considered  here.
 An attention is paid to the possible
  application of the PEM to the description of
inhomogeneous states, structural instabilities and bistability
phenomena  as well as transitions into the phases
 with the
charge modulation in the high $T_{c}$ superconductors  and other
systems to which the model can be applied.

\section{Thermodynamics of Simplified  PEM in Dynamical Mean Field Theory}\label{ista:p2}
The dynamical mean field theory \index{theory!dynamical mean field
(DMFT)} approach proposed by Metzner and Vollhardt\cite{ista:42} for
the Hubbard model (see also Ref.~[\refcite{ista:43}] and references
therein) is a nonperturbative scheme which is exact in the limit of
the infinite space dimension ($d\rightarrow\infty$). The method is
very successful in considering the systems with strong electron
correlations and is used with advantage in solving a variety of
problems and models.
Within the framework of DMFT, investigations of the single-particle
spectrum and the thermodynamics of the simplified PEM were performed
 for a strong coupling case ($g\gg W$, where $W$ is the half-width of the initial electron band).\cite{ista:27}
 The Fourier-transform
$G_{\sigma}(\omega_{n},\bf{k})$ of the electron Green's function
\index{function!Green's}
\begin{equation}
G_{ij}^{\sigma}(\tau-\tau')=-\langle T
a_{i\sigma}(\tau)a_{j\sigma}^{+}(\tau')\sigma(\beta)\rangle_{0}\Big/\langle\sigma(\beta)\rangle_{0},\label{ista:eq1.1}
\end{equation}
 (where $T$ denotes the $\tau$- ordering procedure)
with the scattering matrix
\begin{equation}
\sigma(\beta)=T\exp \left\{-\int_{0}^{\beta}{\rm
d}\tau\sum_{ij\sigma}t_{ij}a_{i\sigma}^{+}(\tau)a_{j\sigma}(\tau)\right\}\label{ista:eq1.2}
\end{equation}
and the averaging with the single-site part
$H_{0}=\sum\limits_{i}H_{i}$ of the Hamiltonian (\ref{ista:eq0}), is
expressed as a series in terms of the electron hopping parameter
$t_{ij}$. The Larkin's equation
\begin{equation}
G_{ij}^{\sigma}(\tau-\tau')=\Xi_{ij}^{\sigma}(\tau-\tau')+\Xi_{il}^{\sigma}(\tau-\tau'')
t_{lm}G_{mj}^{\sigma}(\tau''-\tau')\label{ista:eq1.3}
\end{equation}
separates the total irreducible (with respect to $t_{ij}$) part
$\Xi^{\sigma}$; formally
\begin{equation}
G_{\sigma}(\omega_{n},{\bf{k}})=\frac{1}{\Xi_{\sigma}^{-1}(\omega_{n},{\bf{k}})-t_{\bf
k}}\, .\label{ista:eq1.4}
\end{equation}
In the case of high dimensions $(d\rightarrow\infty)$, when the
hopping integral is scaled $(t_{ij}\rightarrow t_{ij}/\sqrt{d})$,
only single-site contributions survive in the expression for
$\Xi_{\sigma}$:\cite{ista:44}
\begin{equation}
\Xi_{ij}^{\sigma}(\tau-\tau')=\delta_{ij}\Xi_{\sigma}(\tau-\tau');\quad
\Xi_{\sigma}(\omega_{n},{\bf{k}})=\Xi_{\sigma}(\omega_{n}).\label{ista:eq1.5}
\end{equation}
Such a site--diagonal function, as it was shown by Brandt and
Mielsch,\cite{ista:45} can be calculated by mapping the
infinite--dimensional lattice problem on the atomic model
  \begin{eqnarray}\label{sss}
  \lefteqn{{\rm e}^{-\beta H}\to
  {\rm e}^{-\beta H_\mathrm{eff}}=
  {\rm e}^{-\beta H_0}
  }
  \\
  \nonumber
  &&
  \times{ T}\!\exp\bigg\{-\int_0^{\beta}\!{\rm d}\tau \int_0^{\beta}\!{\rm d}\tau'
  \sum_{\sigma} J_{\sigma}(\tau-\tau') a_{\sigma}^{+}(\tau)
  a_{\sigma}(\tau')
  \bigg\}
  \end{eqnarray}
with auxiliary Kadanoff--Baym field
$J_{\sigma}(\tau-\tau')$\cite{ista:46} which should be
selfconsistently determined from the condition that the same
function $\Xi_{\sigma}$ defines the Green's functions for lattice
(\ref{ista:eq1.4}) and atomic limit
  \begin{equation}
  G_{\sigma}^{(a)}(\omega_n)=\frac1{\Xi_{\sigma}^{-1}(\omega_n)-J_{\sigma}(\omega_n)}.
 \label{ista:eq1.7}
  \end{equation}

``Dynamical'' mean field $J_{\sigma}(\tau-\tau')$ (so-called
coherent potential)
 describes the hopping (transfer) of electron
from atom into environment at the moment $\tau$, and propagation in
environment without stray into atom until moment $\tau'$. The
connection between this ``dynamical'' mean field of atomic problem
and Green's function of the lattice can be obtained using standard
coherent potential approximation (CPA):\cite{ista:43}
\index{approximation!coherent potential (CPA)}
  \begin{equation}
  J_{\sigma}(\omega_n)=\Xi_{\sigma}^{-1}(\omega_n) -
  G_{\sigma}^{-1}(\omega_n),\label{ista:eq1.8}
  \end{equation}
where
  \begin{equation}
  G_{\sigma}^{(a)}(\omega_n)=G_{\sigma}(\omega_n)=
  \int_{-\infty}^{+\infty} \!{\rm d}t \frac{\rho(t)}{\Xi_{\sigma}^{-1}(\omega_n)-t}\label{ista:eq1.9}
  \end{equation}
is a single-site Green's function both for atomic limit and lattice.
Here summation over wave vector was changed by the integration with
the density of states (DOS) $\rho(t)$ \index{density of states
(DOS)} (the Gaussian one for the hypercubic lattice
$\rho(\varepsilon)=\frac{1}{W\sqrt{\pi}}{\rm
e}^{-\varepsilon^{2}/W^{2}}$ and semi-elliptic DOS for the Bethe
lattice $\rho(\varepsilon)=\frac{2}{\pi
W^{2}}\sqrt{W^{2}-\varepsilon^{2}}$, see Ref. [\refcite{ista:43}]).

In order to find expression for Green's function in the atomic
limit, one can use the fact that the statistical  operator of the
single-site problem (\ref{sss}) can be expressed in the
form\cite{ista:27}
\begin{equation}
  {\rm e}^{-\beta H_\mathrm{eff}} =
  P^+{\rm e}^{-\beta H_{+}} + P^-{\rm e}^{-\beta H_{-}}\label{ista:eq1.10}
  \end{equation}
because the atomic space of states splits into two independent
subspaces.

As a result, the single--electron Green's function is a sum of
Green's functions in subspaces and is equal to
  \begin{equation}
  G_{\sigma}^{(a)}(\omega_n) =
  \frac{\left\langle P^+\right\rangle}{i\omega_n+\mu-J_{\sigma}(\omega_n)-\frac g2}
  +\frac{\left\langle P^-\right\rangle}{i\omega_n+\mu-J_{\sigma}(\omega_n)+\frac g2}\, .\label{ista:eq1.11}
  \end{equation}
Here $\langle\ldots\rangle$ is the statistical averaging with the
effective Hamiltonian (\ref{sss}). Partition functions in subspaces
are
  \begin{eqnarray}
  \lefteqn{
  Z_{\pm}=\mathop{\mathrm{Sp}}{\rm e}^{-\beta H_{\pm}} = {\rm e}^{\pm\frac{\beta h}2-Q_{\pm}}
  }
  \\ \nonumber
  && ={\rm e}^{\pm\frac {\beta h}2}
  \prod_{\sigma}\left( 1+ {\rm e}^{-\beta(\mu\mp\frac g2)}\right)
  \prod_n\left(1-\frac{J_{\sigma}(\omega_n)}{i\omega_n+\mu\mp\frac g2}\right)\, .\label{ista:eq1.12}
  \end{eqnarray}
Pseudospin mean value is determined by the equation
  \begin{equation}
  \label{ista:aaa}
  \left\langle S^z\right\rangle = \frac{1}{2}\frac{Z_+-Z_-}{Z_++Z_-}
 =\frac{1}{2} \tanh\frac{1}{2}\left(\beta h-\left( Q_+[\left\langle
S^z\right\rangle]-Q_-
  [\left\langle S^z\right\rangle]\right)\right)\, .
  \end{equation}
Electron concentration mean value is determined by
  \begin{equation}
  \left\langle n\right\rangle=\frac1{\beta}\sum_{m\sigma}G_{\sigma}\left(\omega_m\right)\label{ista:eq1.14}
  \end{equation}
and the functional of the grand canonical potential can be derived
in the standard way for DMFT
  \begin{equation}
  \frac{\Phi}N = \Phi_{(a)} - \frac1{\beta}\sum_{n\sigma} \bigg\{
  \ln G_{\sigma}^{(a)}(\omega_{n}) -
  \frac{1}{N}\sum_{\bf k}\ln G_{\sigma}(\omega_{n},{\bf k})
  \bigg\},\label{ista:eq1.15}
  \end{equation}
where
  \begin{equation}
  \Phi_{(a)}=-\frac1{\beta}\ln(Z_++Z_-)\label{ista:eq1.16}
  \end{equation}
is a thermodynamic potential for atomic problem.

\begin{figure}
  \centerline{\psfig{file=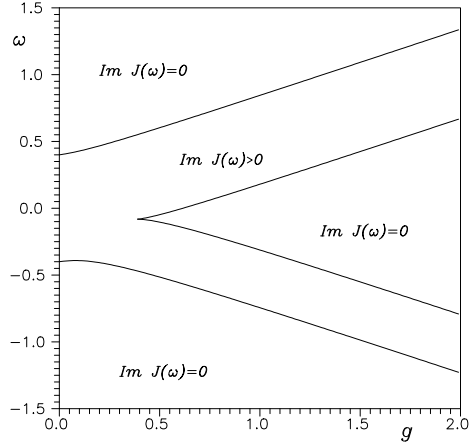,width=2.4in,angle=0}}
\vspace*{8pt}
  \caption{Electron bands boundaries (semi--elliptic DOS, $W=0.4$,
  $\left\langle S^z\right\rangle=0.2$).}
  \label{ista:f1}
  \end{figure}

The solution of the above given  set of equations and the
calculation of thermodynamic potential were performed for the case
of semi-elliptic DOS. The field $J_{\sigma}(\omega_{n})$ is
determined by the simple cubic equation
\begin{equation}
  J_{\sigma}(\omega_n)=\frac{W^2}4\left\{
  \frac{\left\langle P^+\right\rangle}{i\omega_n+\mu-J_{\sigma}(\omega_n)-\frac g2}
+\frac{\left\langle
P^-\right\rangle}{i\omega_n+\mu-J_{\sigma}(\omega_n)+\frac
g2}\right\}\, .\label{ista:eq1.17}
  \end{equation}
The solutions with $\Im {\rm m}J_{\sigma}(\omega)>0$ are considered;
the condition $\Im{\rm m}J_{\sigma}(\omega)\rightarrow0$ determines
the band boundaries. Their dependence on coupling constant at the
fixed value of $\langle S^{z}\rangle$ is shown in
Fig.~\ref{ista:f1}.
It can be seen that there exists a critical value of $g\sim 0.5 W$
when a gap in spectrum appears and we have an analogue of the Mott
transition.\index{phase!transition!Mott} In the case when the
single-electron Green's function is calculated in Hubbard-I
approximation \index{approximation!Hubbard-I}(the scattering
processes via coherent potential are not taken into account,
$J_{\sigma}(\omega)=0$ and $ \Xi_{\sigma}(\omega_{n})=
\langle P^{+}\rangle
/
(i\omega_{n}+\mu-g/2)
+
\langle P^{-}\rangle
/
(i\omega_{n}+\mu+g/2)$),
the electron subbands are always split and the gap in spectrum
exists at any values of $g$ (see below). From this point of view,
the Hubbard-I approximation is insufficient; even in the case of
strong coupling $(g\gg W)$ it  only qualitatively describes the
dependence of the subband half-widths on $\langle
S^{z}\rangle$.\cite{ista:26}

The expressions presented above allow us to investigate in the DMFT
approach the thermodynamics of the simplified PEM. It was done in
Ref.~[\refcite{ista:26}] in the $\mu=\rm const$ and $n=\rm const$
regimes.

\begin{figure}[!h]
 \centerline{\psfig{file=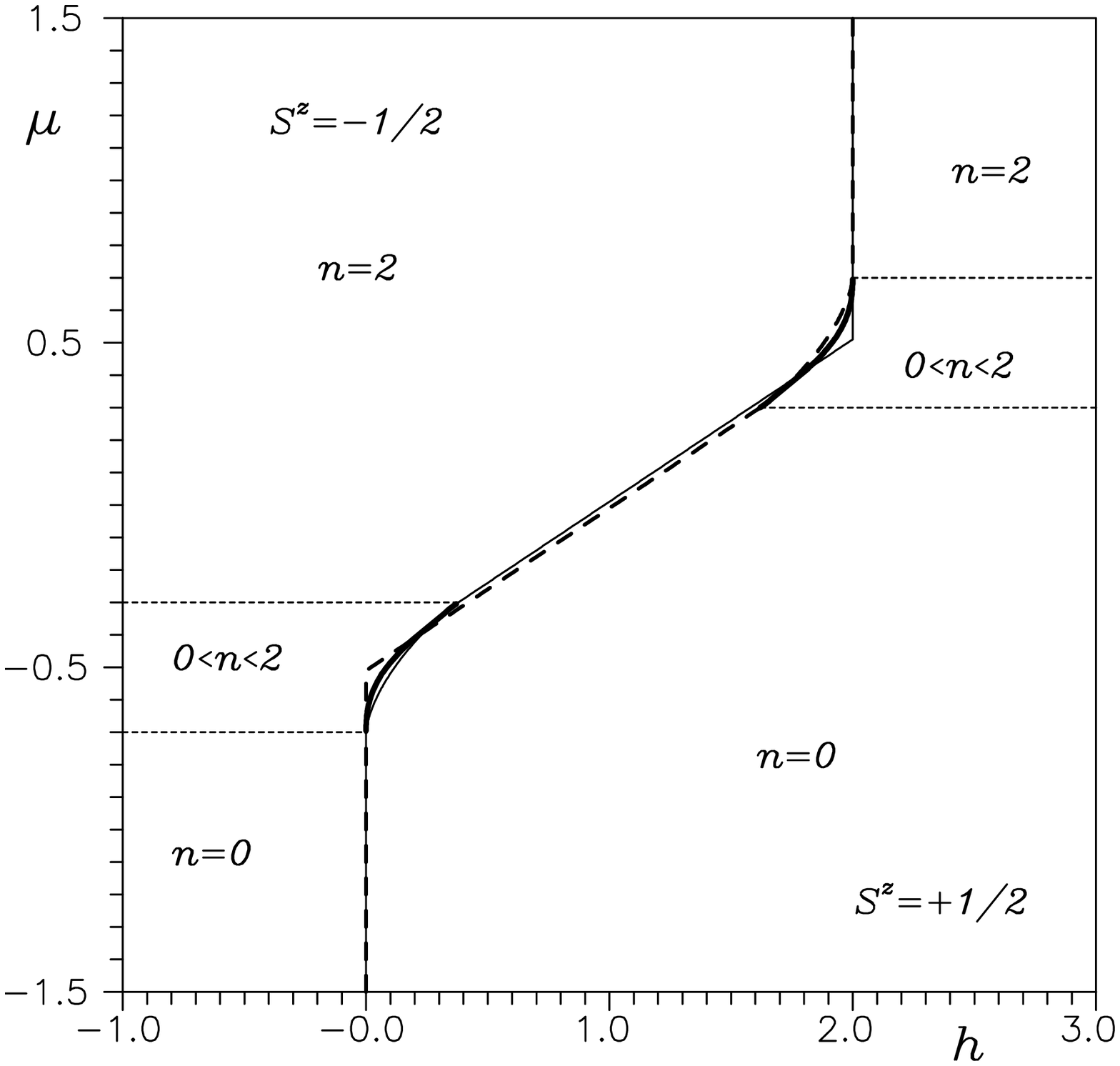,width=2.1in,angle=0} \ \
  \psfig{file=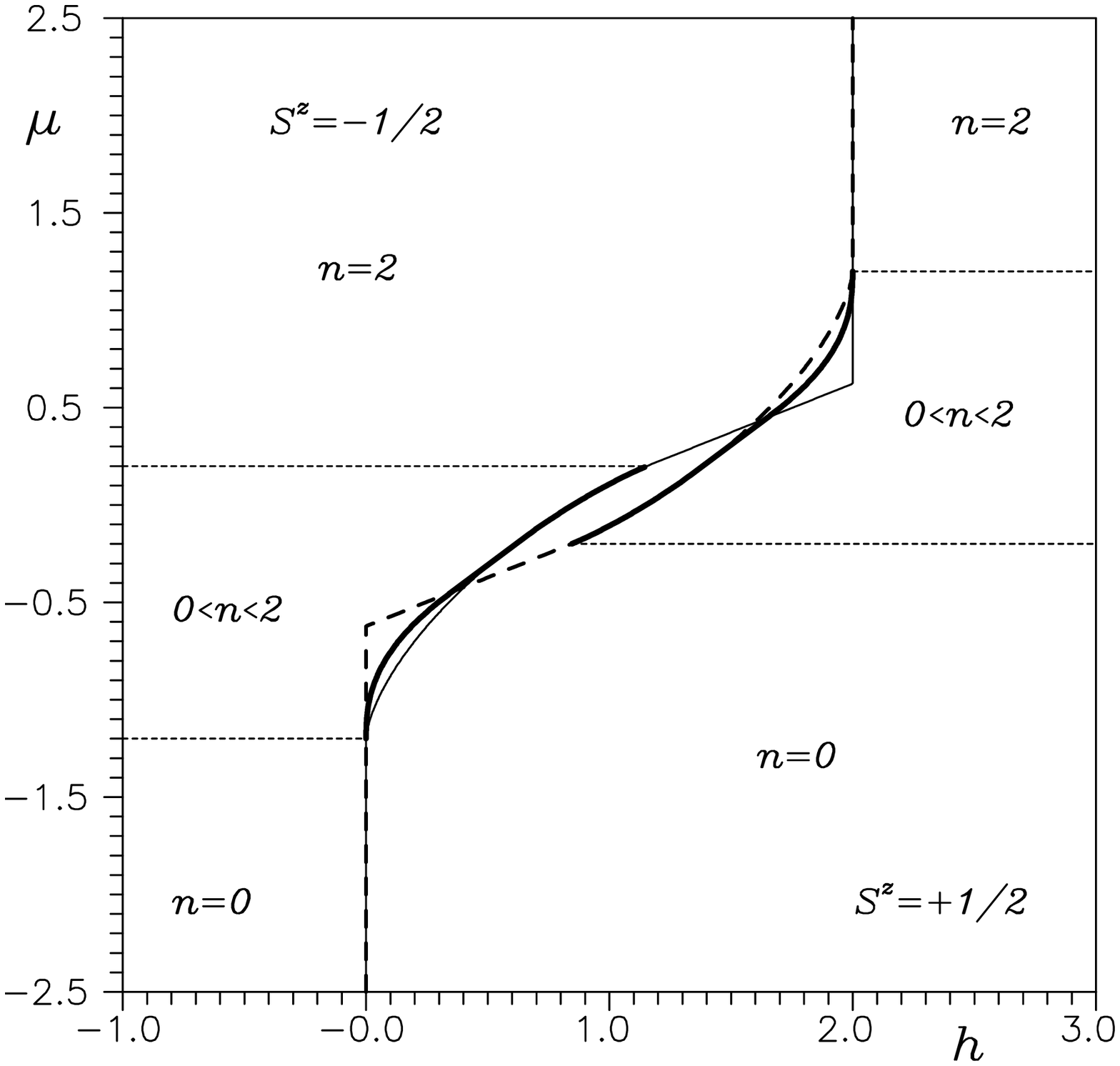,width=2.1in,angle=0}}
  \centerline{\quad (a) \hspace{5.2cm}(b)}
  \vspace*{8pt}
  \caption{Phase diagram ($\mu-h$). Dashed and thin solid lines
  surround regions with $S^z=\pm\frac12$, respectively. The
  lines of the first order phase transition are shown in bold.
  a) $g=1$, $W=0.2$; b) $g=1$, $W=0.7$.}
  \label{ista:f2}
  \end{figure}

\begin{figure}
\centerline{\psfig{file=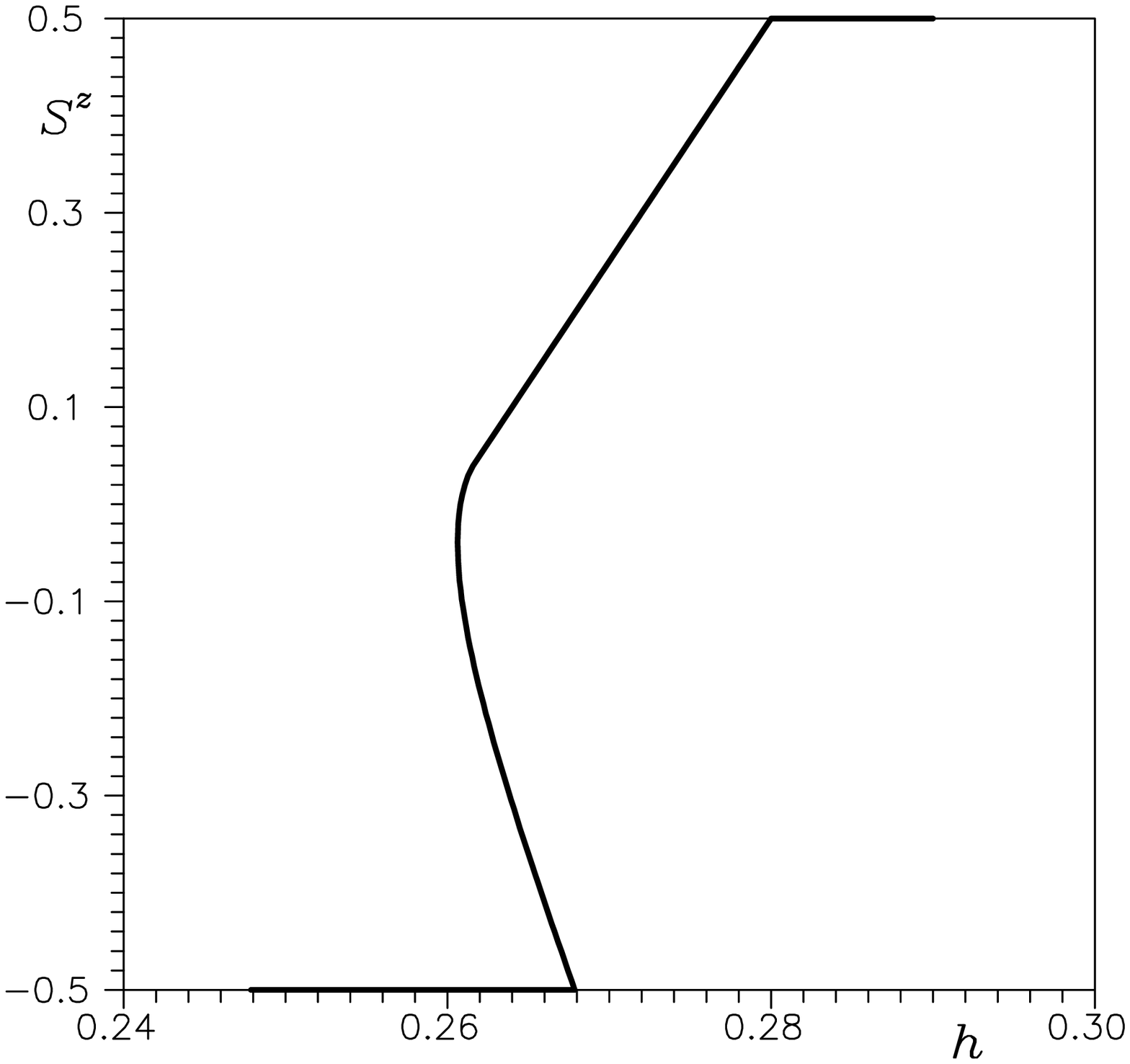,width=2.0in,angle=0} \quad
 \psfig{file=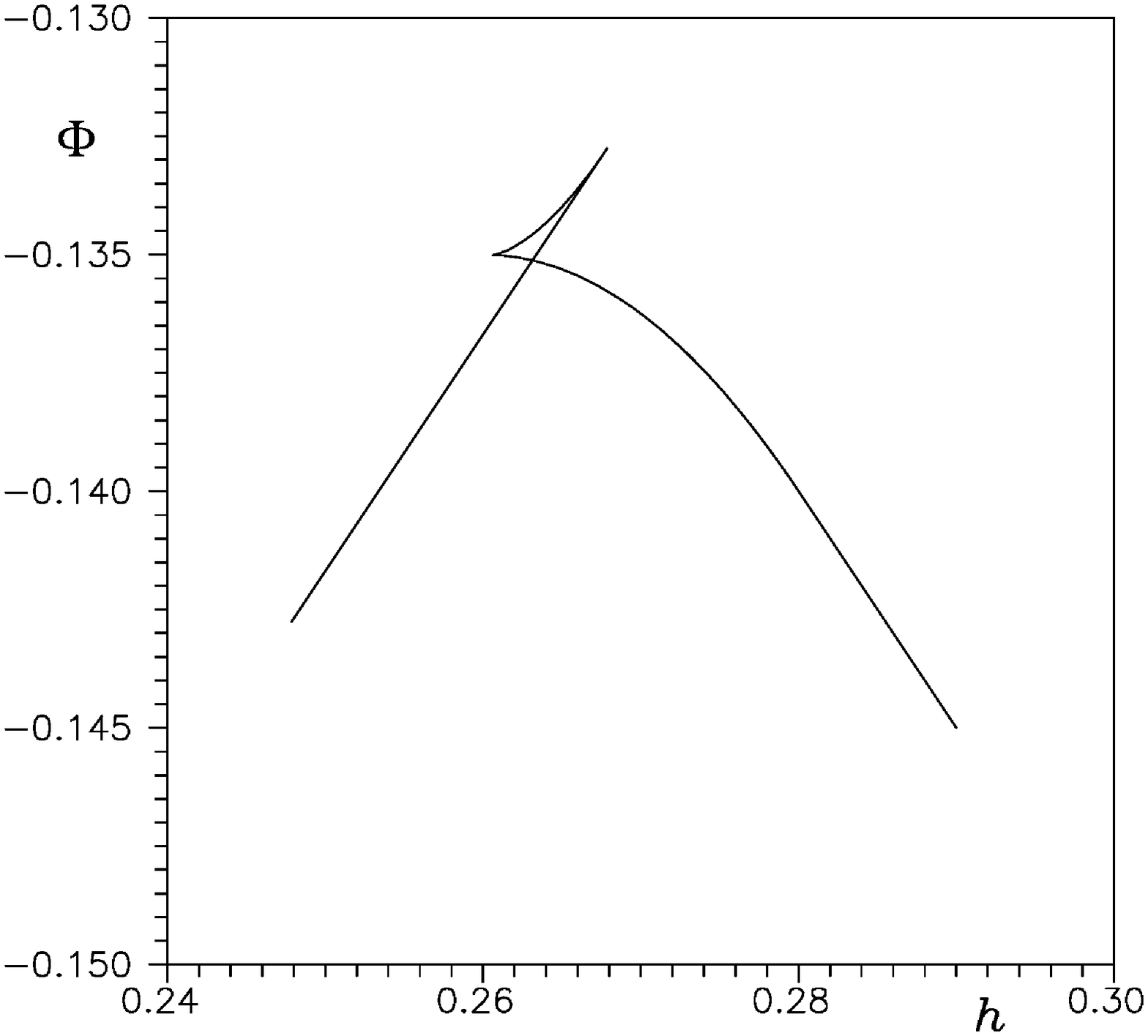,width=2.1in,angle=0}}
 \vspace*{8pt}
  \centerline{\quad (a) \hspace{5.2cm} (b)}
  \caption{Field dependencies of $\left\langle S_z\right\rangle$ (a)
  and grand canonical potential (b) for $\mu=\mathrm{const}$ regime
  when chemical potential is placed in the lower subband $\mu=-0.37$
  ($W=0.2$, $g=1$, $T=0$).}
  \label{ista:f3}
  \end{figure}

\begin{figure}
\centerline{\psfig{file=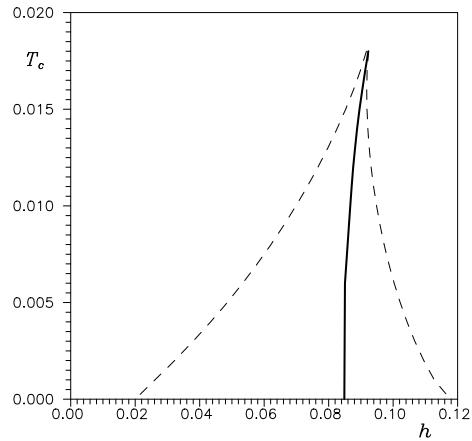,width=2.4 in,angle=0}}
\vspace*{8pt}
  \caption{Phase diagram ($T_c-h$): solid
  and dashed lines indicate the first order phase transition line
  and boundaries of the phase stability region, respectively
  ($g=1$, $W=0.2$, $\mu=-0.5$)}
  \label{ista:f4}
  \end{figure}

In the first case, the thermodynamically stable states are
determined from the minimum of the thermodynamic potential
(\ref{ista:eq1.15}). Analysis of solutions of CPA equations for
$J_{\sigma}(\omega_{n})$ together with the equation (\ref{ista:aaa})
for $\langle S^{z}\rangle$ shows that in this regime the first order
phase transitions\index{phase!transition!first-order} with the jumps
of the pseudospin mean value and electron concentration can take
place. Such transitions are realized when the $\mu$ and $h$ values
correspond to the split subbands  in an electron spectrum (see the
phase diagram\index{phase!diagram} $(\mu-h)$ at $T=0$ in
Fig.~\ref{ista:f2}).
The field dependencies of $\langle S^{z}\rangle$ and grand canonical
potential $\Phi$ in the region of the phase transition point are
shown in Fig.~\ref{ista:f3}.
Since the band structure is determined by the  pseudospin mean
value, the change of the latter is accompanied by the corresponding
reconstruction of the electron spectrum.
With the temperature increase the region of the phase coexistence
narrows. The corresponding phase diagram ($T_c-h$) is shown in
Fig.~\ref{ista:f4}. One can see that with respect to the Ising model
the phase coexistence curve is shifted in the field and deviates
from the vertical line. Hence, the possibility of the first order
phase transition with the temperature change exists in the
pseudospin--electron model for the narrow range of $h$ values.
%
%

\begin{figure}
 \centerline{\psfig{file=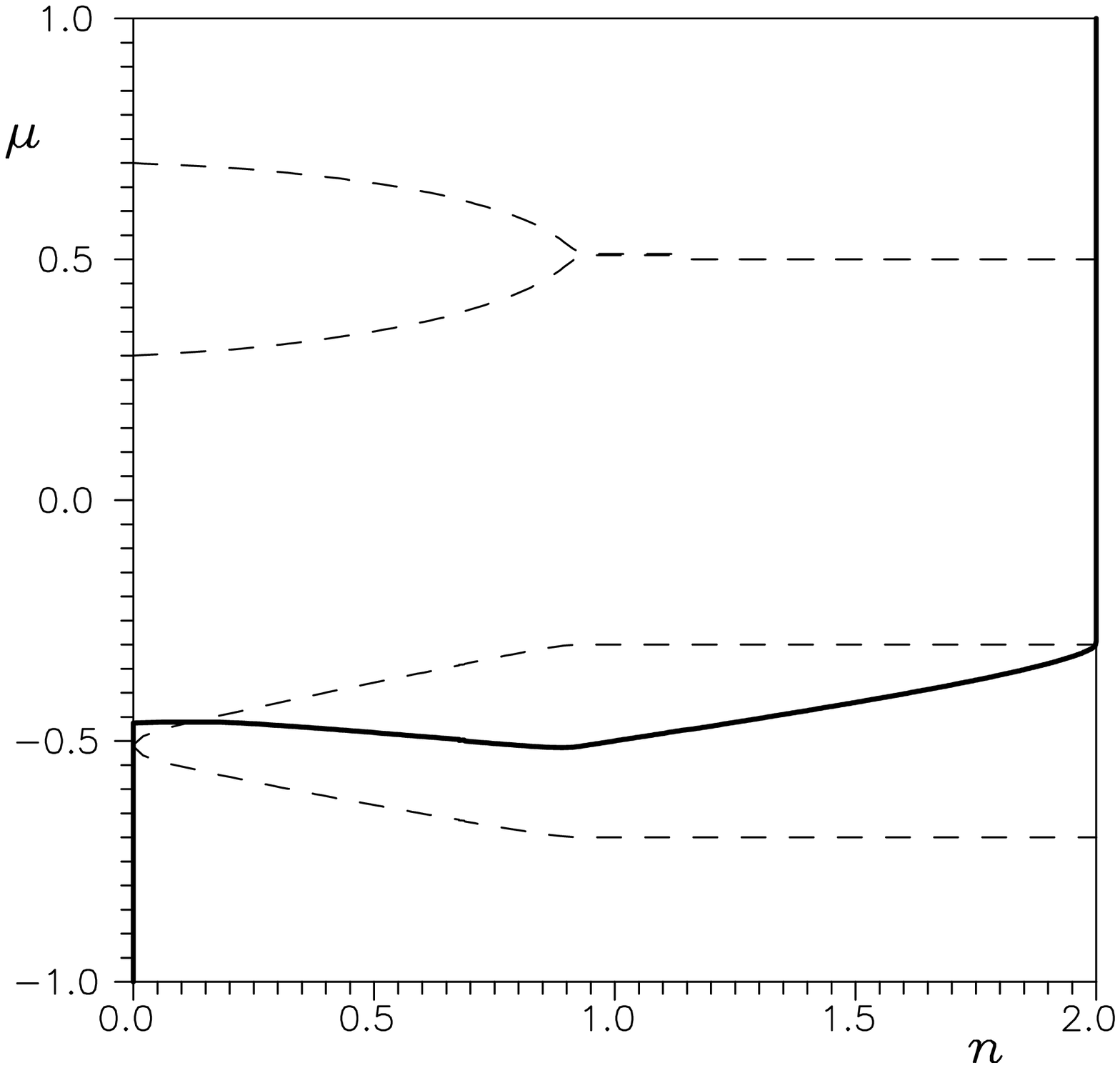,width=2.4in,angle=0}}
\vspace*{8pt}
  \caption{Dependence of the chemical potential $\mu$ and electron
  bands boundaries (dashed lines) on the
  electron concentration $n$ ($T=0.001$, $g=1$, $W=0.2$,
  $h=0.1$).}
  \label{ista:f5}
  \end{figure}

\begin{figure}
  \centerline{\psfig{file=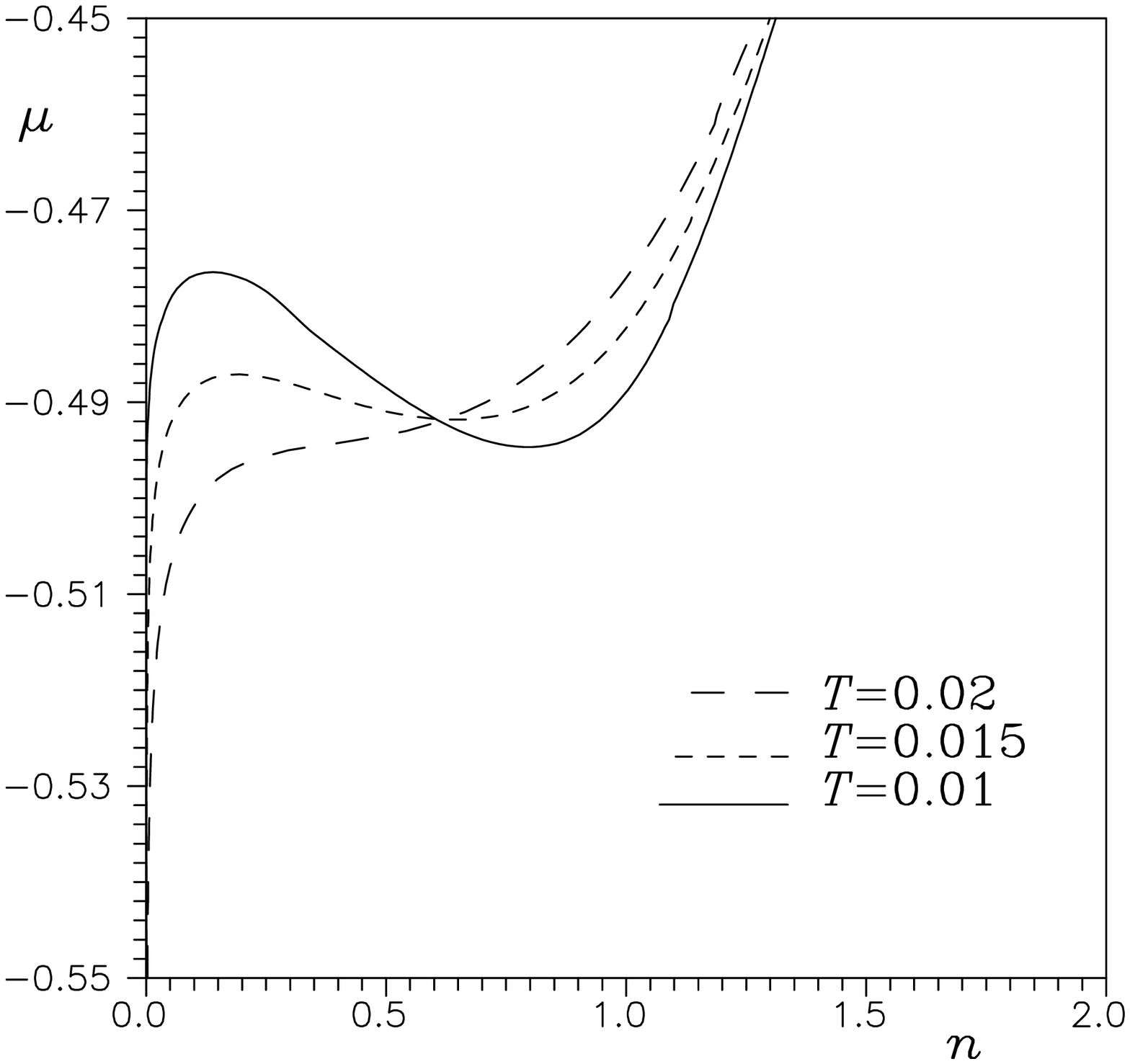,width=2.1in,angle=0}
  \quad
  \psfig{file=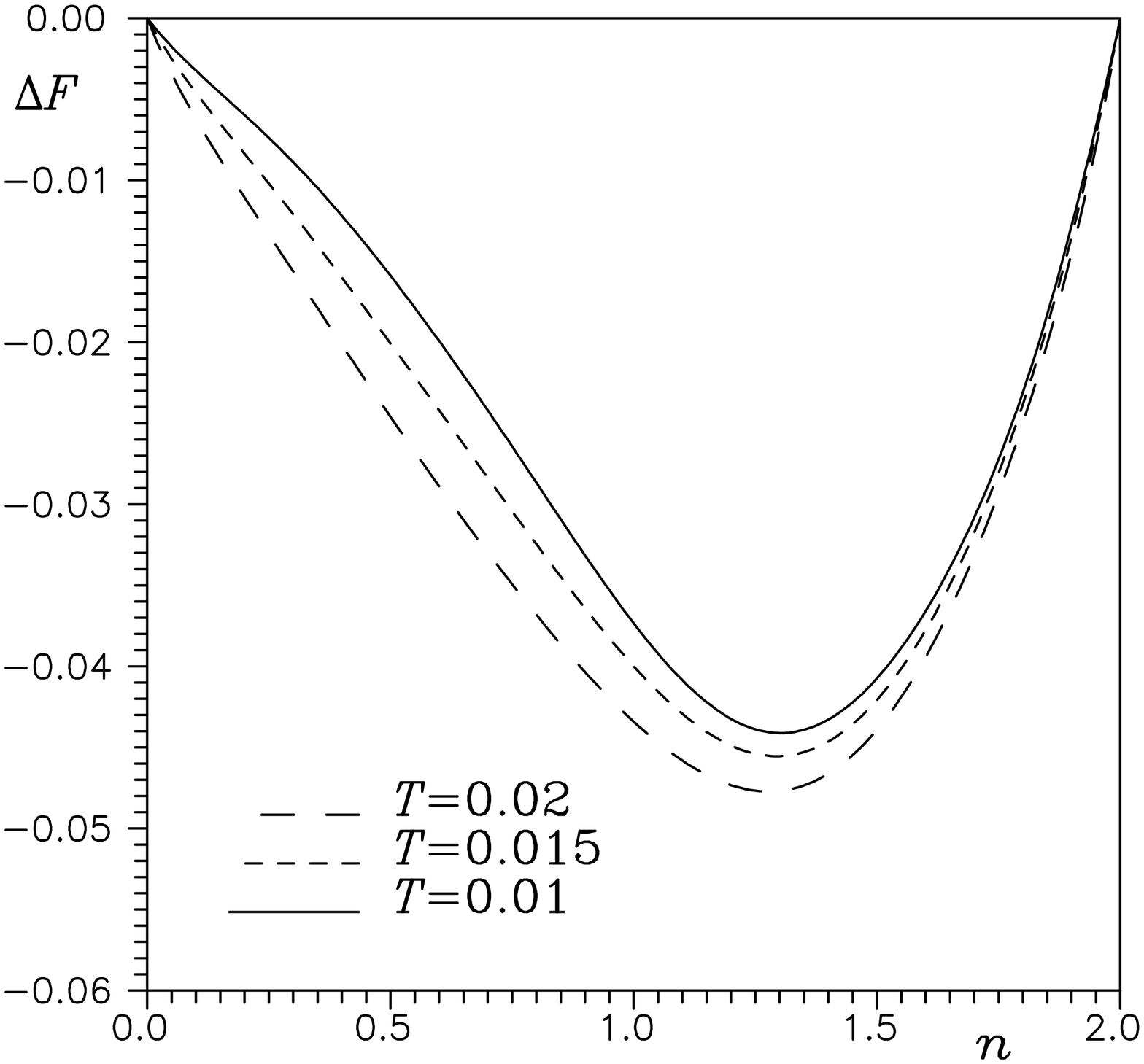,width=2.1in,angle=0}}
  \centerline{\quad (a) \hspace{5.2cm}(b)}
\vspace*{8pt}
  \caption{Dependence of the chemical potential $\mu$ (a) and the
  deviation of free energy from linear dependence
  $\Delta F=F(n)-\frac n2 F(2)-\left(1-\frac n2\right) F(0)$ (b)
  on the electron concentration $n$
  for different temperatures~$T$ ($g=1$, $W=0.2$,
  $h=0.1$).}
  \label{ista:f6}
  \end{figure}

\begin{figure}
\centerline{\psfig{file=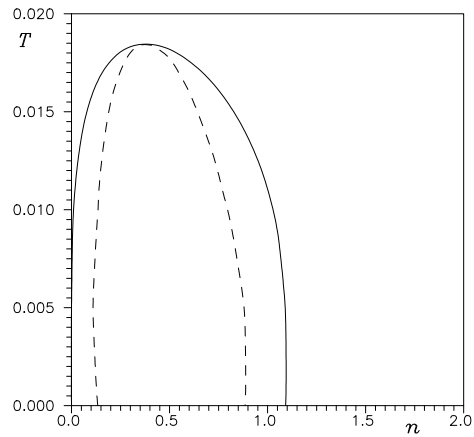,width=2.4in,angle=0}}
\vspace*{8pt}
  \noindent\null\hfill
  \hfill\null
  \caption{Phase diagram ($T-n$) for phase separated state.
  Solid line: binodal, dashed line: spinodal
  ($g=1$, $W=0.2$, $h=0.1$).}
  \label{ista:f7}
  \end{figure}

When the electron concentration is fixed (regime $n=\rm const$), the
first order phase transition transforms into the phase
separation.\index{phase!separation} The regions appear where the
derivative $\partial\mu/\partial n$ is negative, Figs.~\ref{ista:f5}
and \ref{ista:f6}.
The corresponding phase diagram $(T-n)$ is built
(Fig.~\ref{ista:f7}, see also Ref.~[\refcite{ista:26}]) with the use
of the ``Maxwell rule'' which follows in this  case from the
replacement of the original free energy in  its concavity  region by
the tangent line. The diagram describes the separation to the states
with the large and small electron concentrations (and with the
$\langle S^{z}\rangle\approx-1/2$ and $\langle
S^{z}\rangle\approx+1/2$ pseudospin averages at low temperatures,
respectively).
%
%
%

It should be noted that the problem
%
 of phase separation in strongly correlated systems is not
new (see Ref.~[\refcite{ista:47}] and references therein). It was
shown for Hubbard and $t-J$ models\cite{ista:48} that for some
values of the parameters, the system separates into hole--rich and
hole--poor regions with paramagnetic and antiferromagnetic orders,
respectively.
In our case of PEM without electron correlations, the system
separates into regions with electron spectrum that contains both
wide empty electron band and occupied localized  states (at
$n\sim0$) and partially filled wide electron band and empty
localized states (at $n\sim1$), see Fig.~\ref{ista:f5}; their
weights  are determined by the electron concentration. Localized
states of such a type
%
(polarons) result from the strong electron-pseudospin coupling
(strong interaction of electrons with the out of plane apical oxygen
vibrations)  in the case of YBaCuO--type structures, and it can be
supposed that the hopping between such polarons manifests itself in
the carrier relaxation.\cite{ista:49}

For the first time the possibility of phase separation in PEM was
mentioned in Ref.~[\refcite{ista:50}] where it was considered within
GRPA in the limit of strong  correlation $U\rightarrow\infty$ (see
below, Sec.~\ref{ista:p4} ). Here it is obtained for the opposite
case of $U=0$.
As a whole, such a picture of phase transitions into a new uniform
phase is in  agreement with the known results for the FK model in
the case of strong coupling (see Ref.~[\refcite{ista:38}]). But, as
is evident from the phase diagram obtained in
Ref.~[\refcite{ista:51}] for this model, in the region of large but
finite values of $g$ the phase with double modulation can appear. In
order to detect instabilities associated with the wave vector $\bf k
\neq 0$ one should calculate the susceptibility functions and
analyse their temperature and $\bf k$- dependencies. Such an
investigation was performed for PEM in the framework of GRPA (see
below, Secs.~\ref{ista:p3.1.2} and \ref{ista:p3.2.2}).
\section{Simplified PEM in Generalized Random Phase Approximation}\label{ista:p3}
A \index{model!pseudospin-electron (PEM)!simplified} more complete
investigation of the PEM was performed in the framework of the
generalized random phase approximation (GRPA).
\index{approximation!generalized random phase (GRPA)} Such an
approach was formulated by Izyumov and Letfulov\cite{ista:37}
 for the
calculation of the pair correlation functions and magnetic
susceptibility of the Hubbard and $t-J$ models. It is based on the
expansions in terms of electron transfer and consists in the
summation of the diagrams having a structure of sequences of the
electron loops (created by the electron Green's functions) joined by
vertices of various types appearing due to short-range
interactions.\cite{ista:26,ista:37}

Thermodynamics of PEM was studied in the GRPA for $U=0$, $\Omega=0$
(simplified model) in the cases of strong $(g\gg W)$ and weak
$(g<W)$ coupling as well as in the limit of the infinitely large
on-site electron repulsion $(U\rightarrow\infty)$. Let us first
consider the results obtained for simplified model.
\subsection{Strong Coupling Case; $U=0$, $\Omega=0$}\label{ista:p3.1}
\subsubsection{Thermodynamics of the Uniform State}\label{ista:p3.1.1}
First we consider the case of strong coupling. Here the single-site
states can be used as the basic ones and the formalism of electron
annihilation (creation) operators
$a_{i\sigma}=b_{i\sigma}P_{i}^{+}$,
$\tilde{a}_{i\sigma}=b_{i\sigma}P_{i}^{-}\quad (P_{i}^{\pm}=1/2\pm
S_{i}^{z})$ acting at a site with  certain pseudospin orientation
was introduced.\cite{ista:28} Using this representation we can write
the model Hamiltonian in the form
\begin{eqnarray}
&&H=H_{0}+H_{\rm int}\, ,\nonumber\\
&&H_{0}=\sum_{i}\left\{\varepsilon(n_{i\uparrow}+n_{i\downarrow})+\tilde{\varepsilon}
(\tilde{n}_{i\uparrow}+\tilde{n}_{i\downarrow})-h
S_{i}^{z}\right\}\, ,\nonumber\\ &&H_{\rm
int}=\sum_{ij\sigma}t_{ij}\left(a_{i\sigma}^{+}a_{j\sigma}+
a_{i\sigma}^{+}\tilde{a}_{j\sigma}+\tilde{a}_{i\sigma}^{+}a_{j\sigma}
+\tilde{a}_{i\sigma}^{+}\tilde{a}_{j\sigma}\right)\, .
\label{ista:eq1.18}
\end{eqnarray}
Here, $\varepsilon=-\mu + g/2$ and $\tilde{\varepsilon}=-\mu-g/2$
are the energies of single-site states.

 Expansion of the calculated quantities in terms of electron transfer
leads to the infinite series of terms containing the averages of the
$T$-products of the $a_{i\sigma}$, $\tilde{a}_{i\sigma}$ operators.
 The evaluation of such averages is made using the corresponding
Wick's theorem.\cite{ista:28,ista:52}
 The results are expressed in terms of the products
of nonperturbed Green's functions and averages of a certain number
of the projection operators $P^{\pm}_i$ which are calculated by
means of the semi-invariant expansion.\cite{ista:28}

Nonperturbed electron Green's function is equal to
\begin{equation}
g(\omega_{n})=\langle g_{i}(\omega_{n})\rangle;\quad
g_{i}(\omega_{n})=\frac{P_{i}^{+}}{i\omega_{n}-\varepsilon}+\frac{P_{i}^{-}}{i\omega_{n}-\tilde{\varepsilon}}
.\label{ista:1.19}
\end{equation}
In the diagrammatic representation it has the meaning of the
simplest irreducible  Larkin part in the series for the free
single-electron Green's function $G_{\bf k}(\omega_{n})$. In the
Hubbard-I type approximation (see previous section) $G_{\bf
k}(\omega_{n})$ can be written as a sum of the following chain
diagrams
%
\begin{equation}
\hspace{-.8 cm}
 \raisebox{-.7cm}{\epsfysize 1.1cm\epsfbox{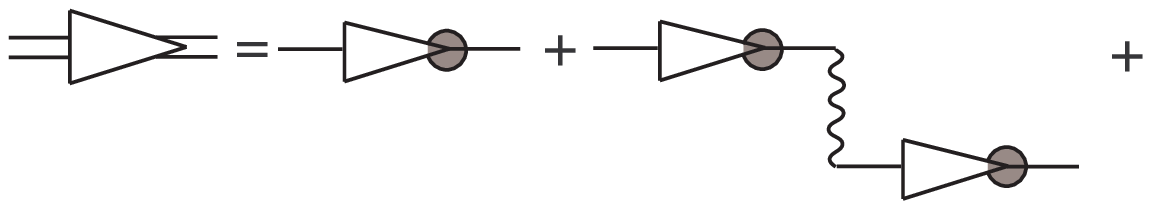}}
 \quad \dots\quad ,\label{ista:eq1.20}
\end{equation}
%
or
\begin{equation}
G_{\bf k}(\omega_{n})=\left[g^{-1}(\omega_{n})-t_{\bf
k}\right]^{-1},\label{ista:eq1.21}
\end{equation}
and its poles (after analytic continuation $i\omega_{n}\rightarrow
\omega+i\delta$) determine the electron spectrum
\begin{equation}
\varepsilon_{I,II}(t_{\bf k})=\frac{t_{\bf
k}}{2}-\mu\pm\frac12\sqrt{g^{2}+4t_{\bf k}\langle S^{z}\rangle
g+t_{\bf k}^{2}}\, .\label{ista:eq1.22}
\end{equation}
The electron subbands are  divided by a gap
 which tends to zero only at $g\rightarrow0$.

There are used here (and below) the following diagrammatic
notations:
$
 \raisebox{-.13cm}{\epsfysize .4cm\epsfbox{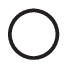}}
 {=}S^z_i
$, $
 \raisebox{-.13cm}{\epsfysize .4cm\epsfbox{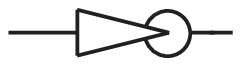}}
$ $ \displaystyle
 {=} g_i(\omega_n)
$,
wavy line is the electron intersite hopping $t_{ij}$.
Semi-invariants are represented by ovals and contain the
$\delta$-symbols on the site indices.
%

 In the adopted approximation the diagrammatic series for the pseudospin mean
value can be presented in the form
\begin{equation}
 \langle S^z\rangle= \raisebox{-.24cm}{\epsfysize 1.3cm\epsfbox{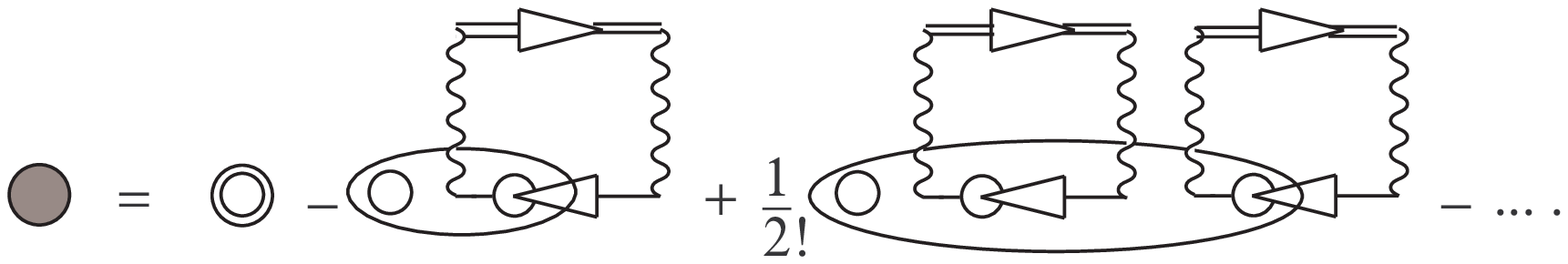}}\label{ista:eq1.23}
\end{equation}
In the spirit of the traditional mean field approach\cite{ista:53}
the
 renormalization of the basic semi-invariant by the insertion of independent
 loop fragments is taken into account in (\ref{ista:eq1.23}).

 The analytical expression for the loop is as follows:
\begin{eqnarray}
 \raisebox{-.4cm}{\epsfysize 1.1cm\epsfbox{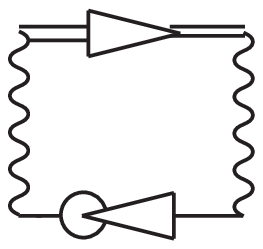}}
 &=&\frac{2}{N}\sum_{n,{\mathbf{k}}}\frac{t^{2}_{\mathbf{k}}}{g^{-1}(\omega_n)-
 t_{\mathbf{k}}}\left(\frac{P^+_i}{i\omega_n-\varepsilon}+
 \frac{P^-_i}{i\omega_n-\tilde{\varepsilon}}\right)\nonumber\\
&=&\beta(\alpha_1P^+_i+\alpha_2P^-_i).\label{ista:eq1.24}
\end{eqnarray}
%
Similarly, the  diagrammatic series for the electron concentration
mean value is as follows:
\begin{equation}
 \hspace{-1cm}
 \langle n_i\rangle {=}
\raisebox{-1.61cm}{\epsfysize
2.8cm\epsfbox{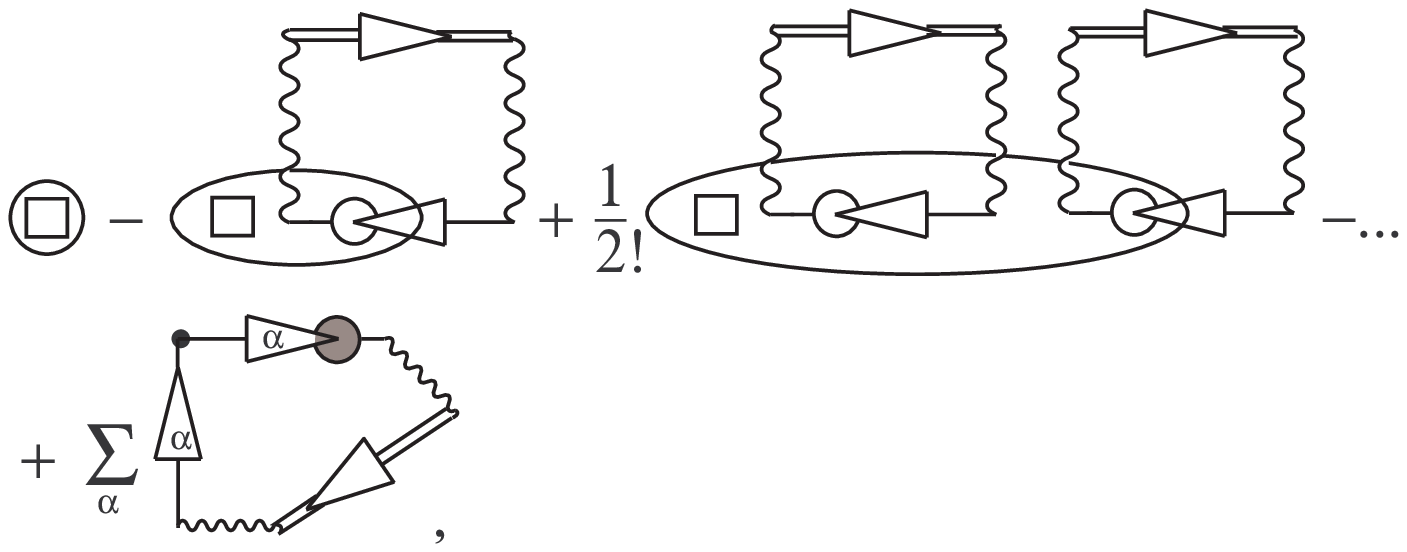}}\label{ista:eq1.25}
\end{equation}
where $
 \raisebox{-0.1cm}{\epsfysize .4cm\epsfbox{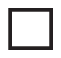}}{=}n_i,
$ $ \displaystyle
 \raisebox{-0.1cm}{\epsfysize .4cm\epsfbox{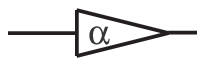}}
 {=} \frac1{{i}\omega_n{-}\varepsilon^{\alpha}},
$ $ \displaystyle
 \raisebox{-0.1cm}{\epsfysize .4cm\epsfbox{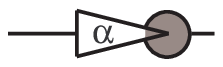}}
 {=}\frac{\langle P^{\alpha}_i\rangle}{{i}
  \omega_n-\varepsilon^{\alpha}},
$ $
 \varepsilon^{\alpha}{=}(\varepsilon,\tilde{\varepsilon}),
$ $
 P^{\alpha}_i{=}(P^+_i,P^-_i).
$

The grand canonical potential $\Phi$ and pair correlation functions
\index{function!pair correlation} ($\langle
S_{i}^{z}S_{j}^{z}\rangle$, $\langle S_{i}^{z}n_{j}\rangle$,
$\langle n_{i}n_{j}\rangle$) are calculated according to
self-consistent scheme of the GRPA: in sequences of loop diagrams in
the expressions for $\Phi$ and correlators the connections between
any two loops by more than one semi-invariant are omitted. We  have,
respectively
\begin{equation}
  \hspace{-.7cm}
  \Delta\Phi=\hspace*{-1.3cm}
  \raisebox{-2.3cm}{\epsfysize 3.5cm\epsfbox{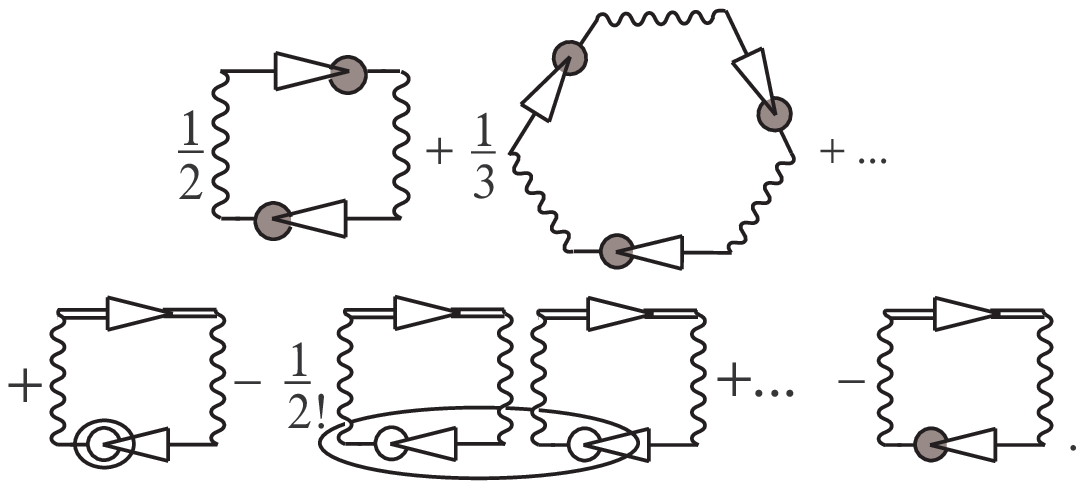}} \label{ista:eq1.26}
  \hspace{-2cm}
\end{equation}
\begin{equation}
  \hspace{-1cm}
  \langle S^z_iS^z_j\rangle=
  \raisebox{-2.15cm}{\epsfysize 2.6cm\epsfbox{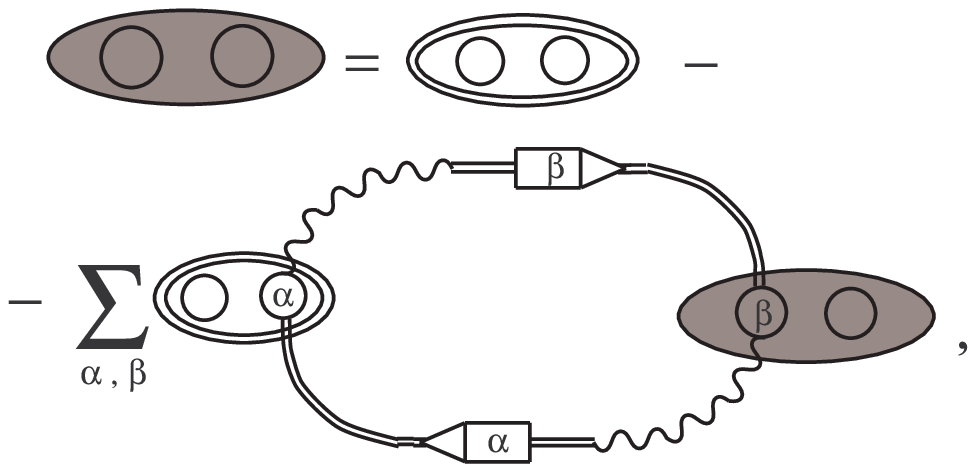}} \label{ista:eq1.27}
\end{equation}
\begin{equation}
 \hspace{-.3cm}
 \raisebox{-.12cm}{\epsfysize .4cm\epsfbox{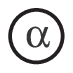}}
 =P^{\alpha}_i,\quad
 \raisebox{-1.cm}[.3cm][.9cm]{\epsfysize1.3cm\epsfbox{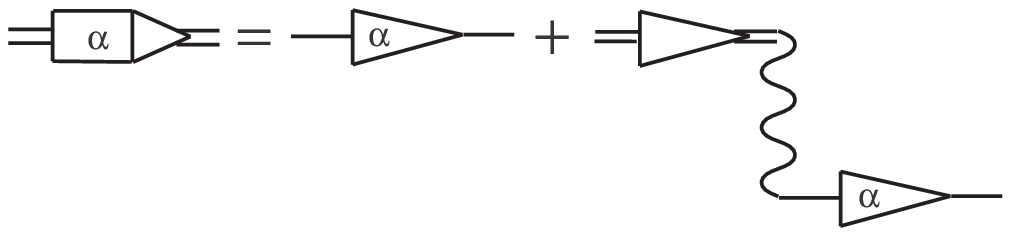}}.\label{ista:eq1.28}
\end{equation}
The first term in equation (\ref{ista:eq1.27}) takes into account a
direct action of the internal effective self-consistent field on
pseudospins:
\begin{equation}
   \hspace{-1.cm}
    \raisebox{-1.4cm}[1.2cm][1.2cm]{\epsfysize 2.8cm\epsfbox{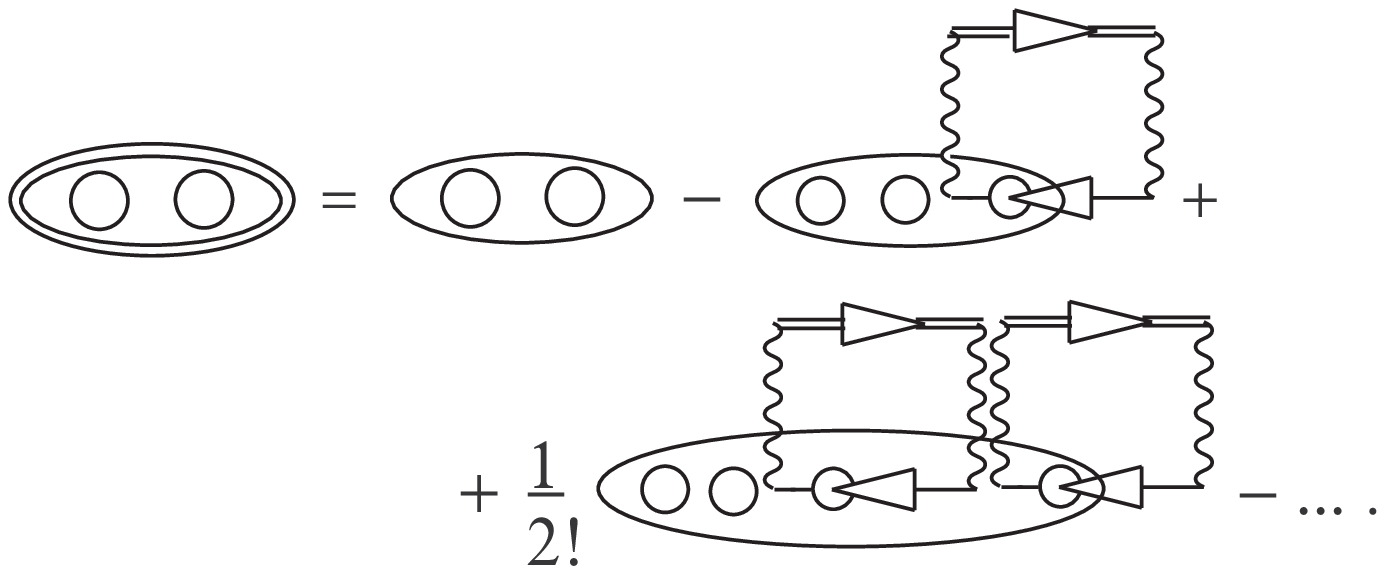}} \label{ista:eq1.29}
\end{equation}
%
leading to the renormalization  of the second-order semi-invariant
due to the inclusion of ``single-tail'' loop-like parts.
 Second term in equation $(\ref{ista:eq1.27})$ describes an interaction
between pseudospins which is mediated by electron hopping.
From (\ref{ista:eq1.23}) and (\ref{ista:eq1.24}),  the equation for
the pseudospin  mean value follows:
\begin{equation}
\langle S^{z}\rangle=\frac12{\rm
tanh}\left\{\frac{\beta}{2}(h+\alpha_{2}-\alpha_{1})+\ln
\frac{1+{\rm e}^{-\beta\varepsilon}}{1+{\rm
e}^{-\beta\tilde{\varepsilon}}}\right\}\, , \label{ista:eq1.30}
\end{equation}
where
\begin{equation}
\alpha_{2}-\alpha_{1}=\frac{2}{N}\sum_{\bf k}t_{\bf
k}\frac{\varepsilon-\tilde{\varepsilon}} {\varepsilon_{I}(t_{\bf
k})-\varepsilon_{II}(t_{\bf k})}\left[n(\varepsilon_{II}(t_{\bf k}))
-n(\varepsilon_{I}(t_{\bf k}))\right] \, .\label{ista:eq1.31}
\end{equation}
The grand canonical potential in the considered approximation has
the form:
\begin{eqnarray*}
  &&
  \Delta\Phi=\Phi-\Phi\Big|_{t=0}\!\!=
  -\frac{2}{N\beta}\sum_{\bf k}
  \ln\frac{(\cosh\frac{\beta}{2}\varepsilon_{ I}(t_{\bf k}))
  (\cosh\frac{\beta}{2}\varepsilon_{ II}(t_{\bf k}))}
  {(\cosh\frac{\beta}{2}\varepsilon)
  (\cosh\frac{\beta}{2}\tilde{\varepsilon})}\\
  &&
  +\langle S^z\rangle(\alpha_2-\alpha_1)
  -\frac{1}{\beta}\ln \cosh\left\{
  \frac{\beta}{2}(h+\alpha_2-\alpha_1)+\ln\frac
  {1+{\rm e}^{-\beta\varepsilon}}
  {1+{\rm e}^{-\beta\tilde{\varepsilon}}}
  \right\}\\
  &&+\frac{1}{\beta}\ln \cosh\left\{
  \frac{\beta}{2} h+\ln\frac
  {1+{\rm e}^{-\beta\varepsilon}}
  {1+{\rm e}^{-\beta\tilde{\varepsilon}}}
  \right\}.
  \nonumber
\end{eqnarray*}
With respect to the initial GRPA scheme,\cite{ista:26,ista:37} the
action of the internal effective self-consistent field on
pseudospins is taken into account by including the mean field type
contributions into the expressions for all thermodynamic quantities.
The electron concentration and pseudospin mean values as well as
correlation functions are calculated consistently with the
thermodynamics functions. It can be checked
explicitly\cite{ista:28,ista:29} using the relations
\begin{equation}
\frac{{\rm d}\Phi}{{\rm d}(-\mu)}=\langle n\rangle;\quad \frac{{\rm
d}\Phi}{{\rm d}(-h)}=\langle S^{z}\rangle;\quad \frac{{\rm
d}S^{z}}{{\rm d}(-\beta h)}=\langle S^{z}S^{z}\rangle_{q=0}\, .
\label{ista:eq1.32}
\end{equation}
At high temperatures, equation (\ref{ista:eq1.30}) possesses only a
uniform solution $\langle S_{i}^{z}\rangle=\langle S^{z}\rangle$.
However, there exists a possibility of phase transition between
different uniform phases with the different pseudospin mean values.
For the first time the possibility  of such a transition, which
leads to the   structural (dielectric) instability, was considered
for the PEM in the limit of the strong electron correlation
$(U\rightarrow\infty)$ in
Refs.~[\refcite{ista:25},\refcite{ista:26}] (this issue is analysed
below, see Sec.~\ref{ista:p4}). A relatively complete description of
this transition was also given in Ref.~[\refcite{ista:54}] for the
PEM with a direct interaction between pseudospins (in the $t_{ij}=0$
limit).

\begin{figure}
  \centerline{\psfig{file=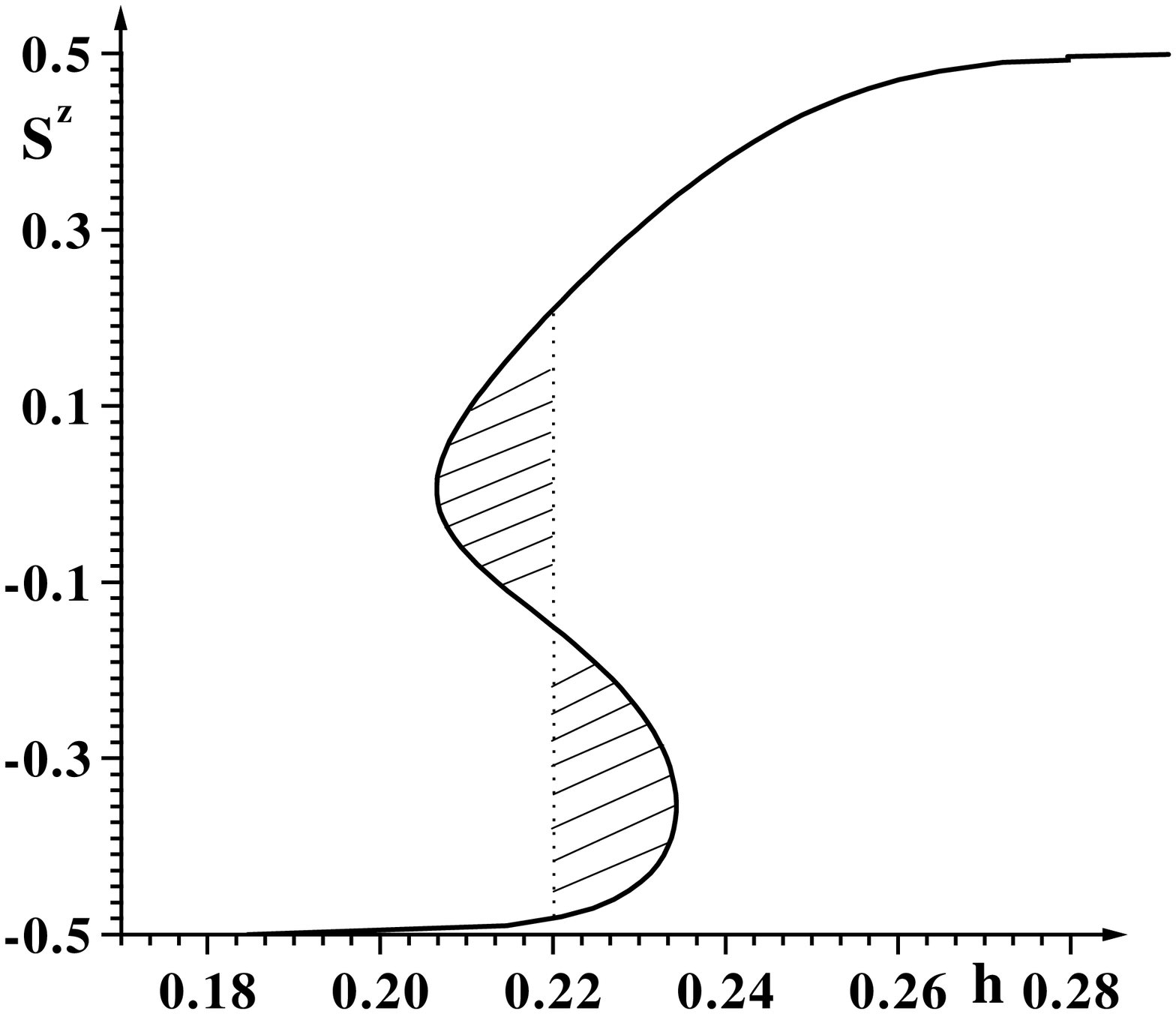,width=2.1in,angle=0}
  \quad
\psfig{file=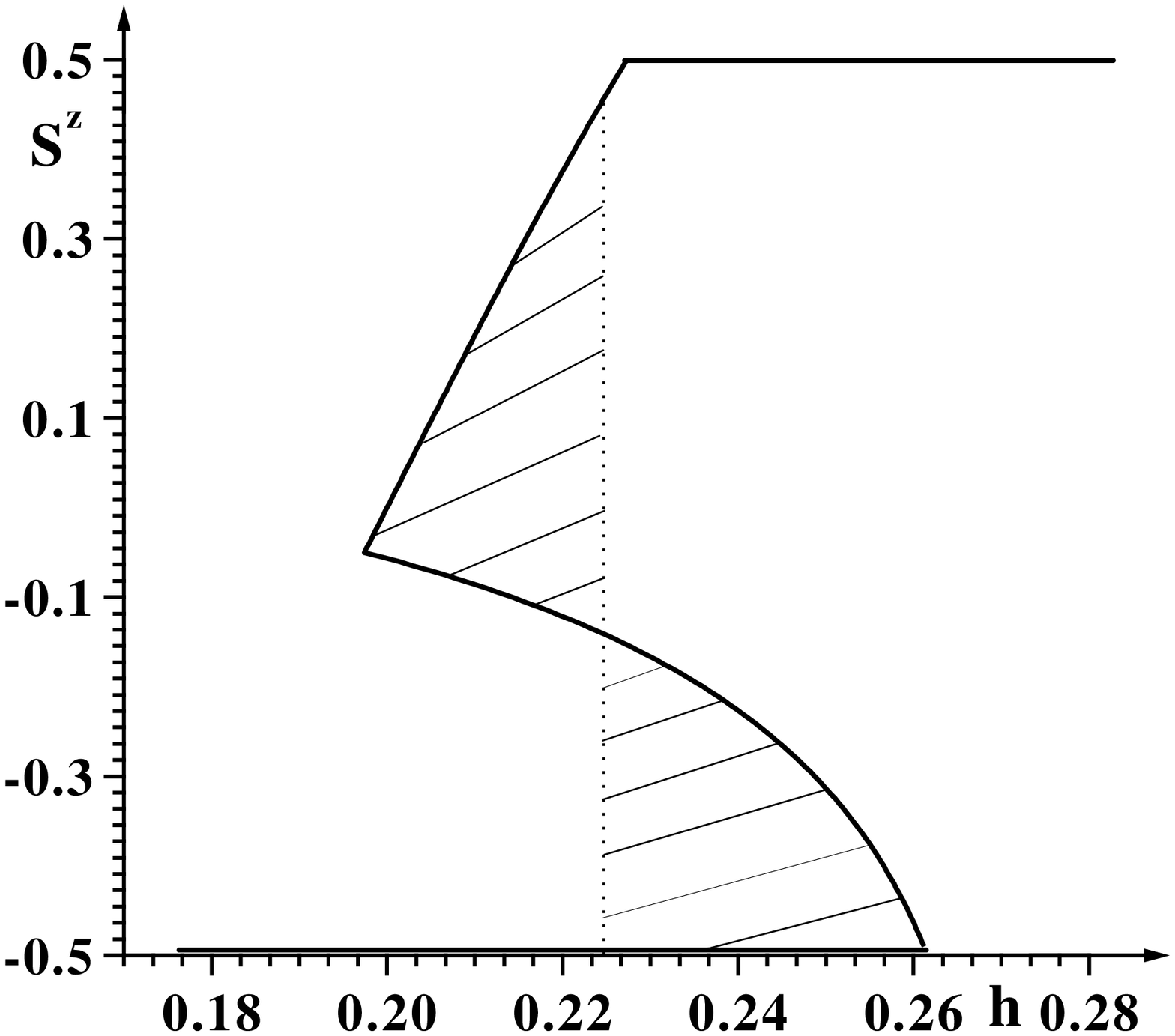,width=2.1in,angle=0}
  }
  \centerline{\quad(a) \hspace{5.2cm}(b)}
  \vspace*{8pt}
  \caption{Field dependence of $\langle S^{z}\rangle$ $(W=0.2,\,\, \mu=-0.4,\,\, g=1.0)$ for $\mu=$const regime;
  $T=0.01$ and $T=0$.}
  \label{ista:f8}
  \end{figure}

\begin{figure}
  \noindent\null\hfill
  \centerline{\psfig{file=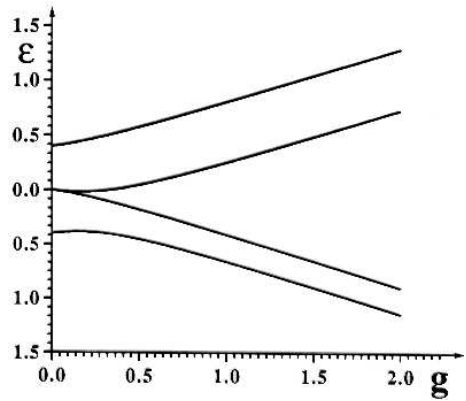,width=2.4in,angle=0}}
\vspace*{8pt}
  \hfill\null
  \caption{Electron bands boundaries ($t_{0}=0.4$, $\langle S^{z}\rangle=0.2$).}
  \label{ista:f9}
  \end{figure}

\begin{figure}
\centerline{\psfig{file=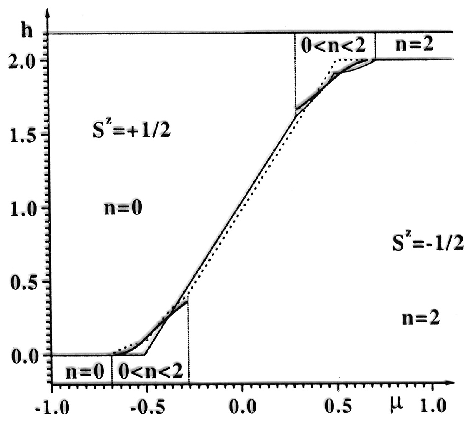,width=2.1in,angle=0}
  \quad
\psfig{file=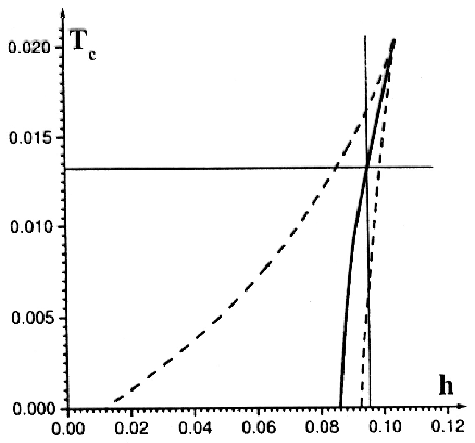,width=2.1in,angle=0}
  }
  \centerline{\quad(a) \hspace{5.2cm}(b)}
  \vspace*{8pt}
  \caption{(a): ($h-\mu$) phase diagram ($T=0$, $t_{0}=0.2$, $g=1$).
(b): ($T_{c}-h$) phase diagram ($g=1$, $t_{0}=0.2$, $\mu=-0.5$).
  }
  \label{ista:f10}
  \end{figure}

\begin{figure}
\centerline{\psfig{file=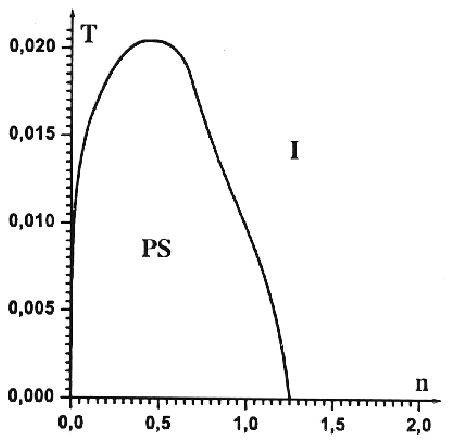,width=2.4in,angle=0}}
\vspace*{8pt}
  \noindent\null\hfill
  \hfill\null
  \caption{($T-n$) phase diagram ($h=0.1$, $t_{k=0}=0.2$, $g=1$).}
  \label{ista:f12}
  \end{figure}

Within the GRPA scheme presented here, the uniform-uniform phase
transition in the simplified PEM was analysed in
Ref.~[\refcite{ista:28}]. The solutions of the set of equations
(\ref{ista:eq1.32}) which correspond to the absolute minimum values
of $\Delta\Phi$ were determined and analysed. The calculations were
numerically performed
 for the square lattice with the nearest-neighbour
hopping  ($d=2$ DOS with the bandwidth 2W). The obtained picture of
the phase transition in the $\mu={\rm const}$ and $n={\rm const}$
regimes is very similar to the case of DMFT approach. When the
chemical potential is fixed, there exist jumps of the pseudospin
mean value and electron concentration on the phase transition line
(Fig. \ref{ista:f8}). The $(h-\mu)$ and $(T_{c}-h)$ phase diagrams
 are nearly the same by their shape as in the
DMFT case despite the fact that electron energy spectrum in GRPA is
split at any values of $g$, see Fig.~\ref{ista:f9}, \ref{ista:f10}.
The same conclusion can be made when we compare the ($T-n$) phase
diagrams obtained in the $n={\rm const}$ regime; the corresponding
phase separation areas are shown in Fig. \ref{ista:f12}. The
difference in positions of critical points in the ($h$, $T$) plane
calculated in the DMFT and GRPA approaches is also small (see
Figs.~\ref{ista:f4} and \ref{ista:f10}b).
  %
%
%
%

In figures discussed above, the case is presented when the chemical
potential is placed in the lower energy subband. If $\mu$ is placed
in the upper subband, the results are transformed according to the
internal symmetry of the Hamiltonian and the following replacements
should be made
\begin{equation}
\mu\rightarrow -\mu,\quad h\rightarrow2g-h,\quad
n\rightarrow2-n,\quad S^{z}\rightarrow-S^{z}. \label{ista:eq1.33}
\end{equation}

Summing up, we  can conclude that the generalization  of GRPA
scheme,\cite{ista:28,ista:29} which  takes  into account the
mean-field loop-like contributions to the semi-invariant averages,
makes it possible to calculate the thermodynamic functions and to
investigate the first order phase transitions  between different
uniform phases. At the transitions, there always remains a gap (in
the strong coupling case, $g\gg W$) in the electron spectrum. With
the change of  the mean value of the pseudospin a reconstruction of
the  electron spectrum takes place, at which the widths of the
electron subbands change which results in  the jump-like  change of
electron concentration (at the given chemical potential $\mu$, it
corresponds to a charge transfer from /to the electron reservoir).

The phase  coexistence  curve in the  ($h$, $T$) plane is tilted
from the vertical line; therefore, there is a possibility of the
first  order phase transition with the temperature change (in the
narrow interval of the field $h$ values). It should be noted that
the existence of the shifted and tilted coexistence curve (as the
result of the local  pseudospin-electron interaction) was obtained
for the first time in Ref.~[\refcite{ista:54}] for a PEM with the
direct interaction between pseudospins. In the case  considered
here, such a direct interaction is not included, but  due to
electron transfer there appears an  indirect  one (the latter is
formed by the loop-like contributions \fbox{$\Pi$}$_{\bf q}$). The
relative role of the direct and indirect interactions between
pseudospins is analysed more in detail in  Sec.~\ref{ista:p5}, on
the example of the two-sublattice PEM.

Herein below, considering the $\langle S^{z} S^{z} \rangle$ pair
correlation function, we shall look at the possibility of the
spatially modulated charge (charge density wave, CDW) \index{charge
density wave (CDW)} and pseudospin orderings in the PEM with $U=0$,
$\Omega=0$ in the strong coupling case.

\subsubsection{The Chess-Board Phase}
\label{ista:p3.1.2}
The analysis of the $\langle S^{z}S^{z}\rangle_{\bf q}$ correlator
temperature behaviour shows that for certain values of model
parameters, the high temperature phase may be unstable with respect
to fluctuations with $\bf{q}\neq0$.\cite{ista:29} Solution of Eq.
(\ref{ista:eq1.27}) for pseudospin correlator has the form
\begin{equation}
 \hspace{-.7cm}
 \langle S^z S^z\rangle_{\bf q}
   =\frac{1/4-\langle S^z\rangle^2}
   {1+\mbox{\fbox{$\Pi$}}_{\bf q}(\frac{1}{4}-\langle S^z\rangle^2)},\label{ista:eq1.34}
\end{equation}
where \fbox{$\Pi$}$_{\bf q}$ characterizes an interaction between
pseudospins via electron subsystem:
\begin{equation}
 \mbox{\fbox{$\Pi$}}_{\bf q}=\sum\limits_{\alpha,\beta}(-1)^{\alpha+\beta}
 \raisebox{-.75cm}[.5cm][.5cm]{\epsfysize 1.5cm\epsfbox{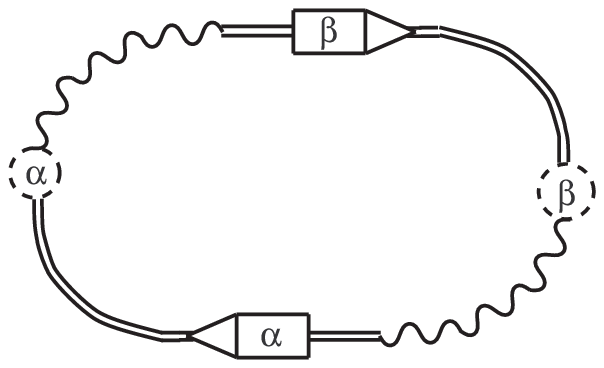}}\;\;,\label{ista:eq1.35}
\end{equation}
and its singularities
\[
 1+\mbox{\fbox{$\Pi$}}_{\bf q}(\frac{1}{4}-\langle
 S^z\rangle^2)=0
\]
 give the instability points of uniform
phase.

It follows from  calculations\cite{ista:29} that for some model
parameter values, the uniform phase becomes unstable with respect to
fluctuations with ${\bf q}=(\pi,\pi)$ (the chess-board phase).
\index{phase!chess-board}
%
%
%

To consider  the thermodynamics of the chess-board phase
analytically, we take into account the modulation of the pseudospin
and electron distribution, introducing two kinds of sites: $\langle
S_{1}^{z}\rangle$ and $n_{1}$ correspond to one sublattice and
$\langle S_{2}^{z}\rangle$ and $n_{2}$ to the other one. In this
case the single-electron Green's function for the $l$ sublattice is
equal to:
\begin{equation}
G_{l}({\bf k},\omega_{n})=\frac{g_{l}(\omega_{n})}{1-t_{\bf
k}^{2}g_{1}(\omega_{n})g_{2}(\omega_{n})},\label{ista:eq1.36}
\end{equation}
where $g_{l}(\omega_{n})$ ($l=1,2$) is the nonperturbated Green's
function for sublattice $l$. The single-electron spectrum is
determined from the equation
\begin{equation}
x^4-(g^2/2+t_{\bf k }^2)x^2- gt_{\bf k }^2(\langle
 S^z_1\rangle+\langle S^z_2\rangle)x
 +g^4/16-g^2t_{\bf k }^2\langle
 S^z_1\rangle\langle S^z_2\rangle=0\, .\label{ista:eq1.37}
\end{equation}
The roots $
 \varepsilon_{1}(t_{\mathbf k })\geq
 \varepsilon_{2}(t_{\mathbf k })\geq
 \varepsilon_{3}(t_{\mathbf k })\geq
 \varepsilon_{4}(t_{\mathbf k })
$ of the equation (\ref{ista:eq1.37}) form four subbands.
 The widths of subbands depend on the mean values of pseudospins.

 The branches $\varepsilon_{1}(t_{\bf k})$, $\varepsilon_{2}(t_{\bf k})$,
 on the one side , and $\varepsilon_{3}(t_{\bf k})$,
 $\varepsilon_{4}(t_{\bf k})$ on the other side, form two pairs of
 bands which are always separated by a gap.
 The equation for pseudospin mean values (\ref{ista:eq1.30}) can
 be now written  in the form:
\begin{equation}
 \langle S^z_l\rangle=
 \frac{1}{2}\tanh\left\{\frac{\beta}{2}(h{+}\alpha^l_2{-}\alpha^l_1)+
 \ln{\frac{1{+}{\rm e}^{-\beta\varepsilon}}
 {1{+}{\rm e}^{-\beta\tilde{\varepsilon}}}} \right\},\, l=1,2\, \label{ista:eq1.38}
\end{equation}
where expressions for the effective self-con\-sis\-tent fields are
\begin{equation}
  \alpha^l_2{-}\alpha^l_1{=}
 \frac 2N\sum_{\bf k}t^2_{\bf k}(\varepsilon-\tilde{\varepsilon})
 \sum^4_{i=1}
  A^l_in[\varepsilon_i(t_{\bf k}){-}\mu],\label{ista:eq1.39}
\end{equation}
\[
 A^l_i{=}\frac{\varepsilon_i(t_{\bf k})+g\langle S^z_{l'}\rangle}
{(\varepsilon_i(t_{\bf k}){-}\varepsilon_j(t_{\bf k}))
(\varepsilon_i(t_{\bf k}){-}\varepsilon_p(t_{\bf k}))
(\varepsilon_i(t_{\bf k}){-}\varepsilon_m(t_{\bf k}))},\,
 i\not=j,p,m,\quad l\not=l'.
\]
 Expression for the electron mean number  follows from (\ref{ista:eq1.32}):
\begin{eqnarray}
 \label{ista:bbb}
 \hspace*{-1.5cm}
 \lefteqn{\langle n_1{+}n_2\rangle{=}\frac{2}{N} \sum_{\bf k}
 \sum^4_{i=1}n[\varepsilon_i(t_{\bf k}){-}\mu]}
 \\
 \nonumber
&& -2\Big[
  \big(\langle P_1^+\rangle{+}\langle P_2^+\rangle\big) n(\tilde{\varepsilon})
 +\big(\langle P_1^-\rangle{+}\langle P_2^-\rangle\big)
 n(\varepsilon)\Big],
\end{eqnarray}
and the grand canonical potential (\ref{ista:eq1.26}) can be written
for the two-sublattice case in the following analytic form:
\begin{eqnarray}
 \label{ista:eq1.41}
  \Delta\Phi &=&-\frac{2}{N\beta}\sum_{\bf k}
 \ln\frac{
\prod\limits^4_{i=1} \cosh\big[\frac{\beta}{2}(\varepsilon_i(t_{\bf
k}){-}\mu)\big]}
 {(\cosh\frac{\beta}{2}\varepsilon)^2
 (\cosh\frac{\beta}{2}\tilde{\varepsilon})^2}+
 \sum_{l=1,2}\langle S_l^z\rangle(\alpha^l_2-\alpha^l_1)\nonumber \\
 &{+}&\sum_{l=1,2}\left[-\frac{1}{\beta}\ln\cosh
  \left\{\frac{\beta}{2}(h{+}\alpha^l_2{-}\alpha^l_1)
   +\ln\frac{1{+}{\rm e}^{-\beta\varepsilon}}
           {1{+}{\rm e}^{-\beta\tilde{\varepsilon}}}
  \right\}\right.\nonumber\\
  &{+}&\left.\frac{1}{\beta}\ln \cosh
 \left\{ \frac{\beta}{2}h
  +\ln\frac{1{+}{\rm e}^{-\beta\varepsilon}}
           {1{+}{\rm e}^{-\beta\tilde{\varepsilon}}}
 \right\}\right].
\end{eqnarray}
As previously in the investigation of equilibrium conditions, we
consider  two different thermodynamic regimes.

\noindent$a)$ The $\mu={\rm const}$ regime.

\begin{figure}[!h]
\centerline{\psfig{file=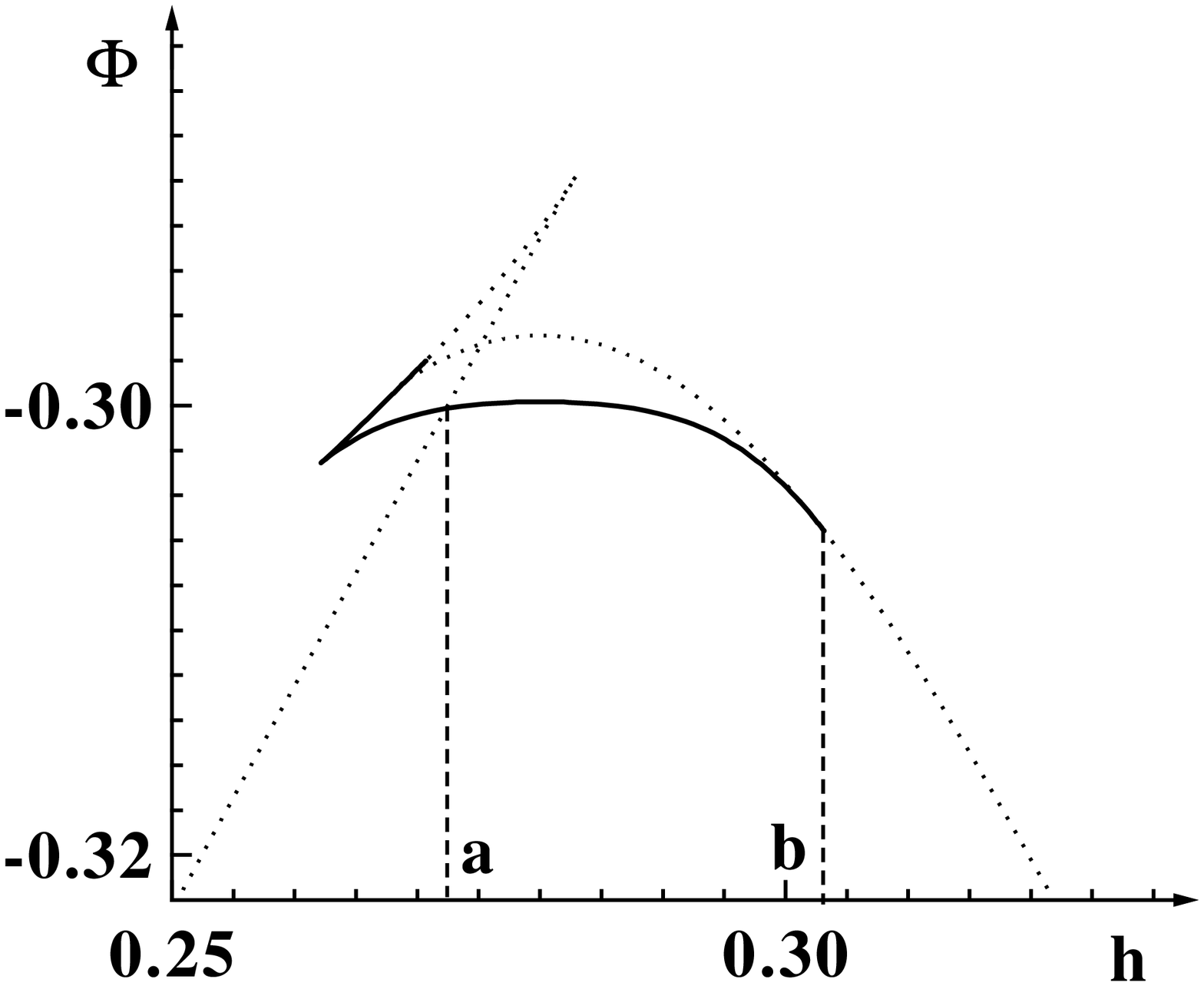,width=2.1in,angle=0} \ \ \ \
\psfig{file=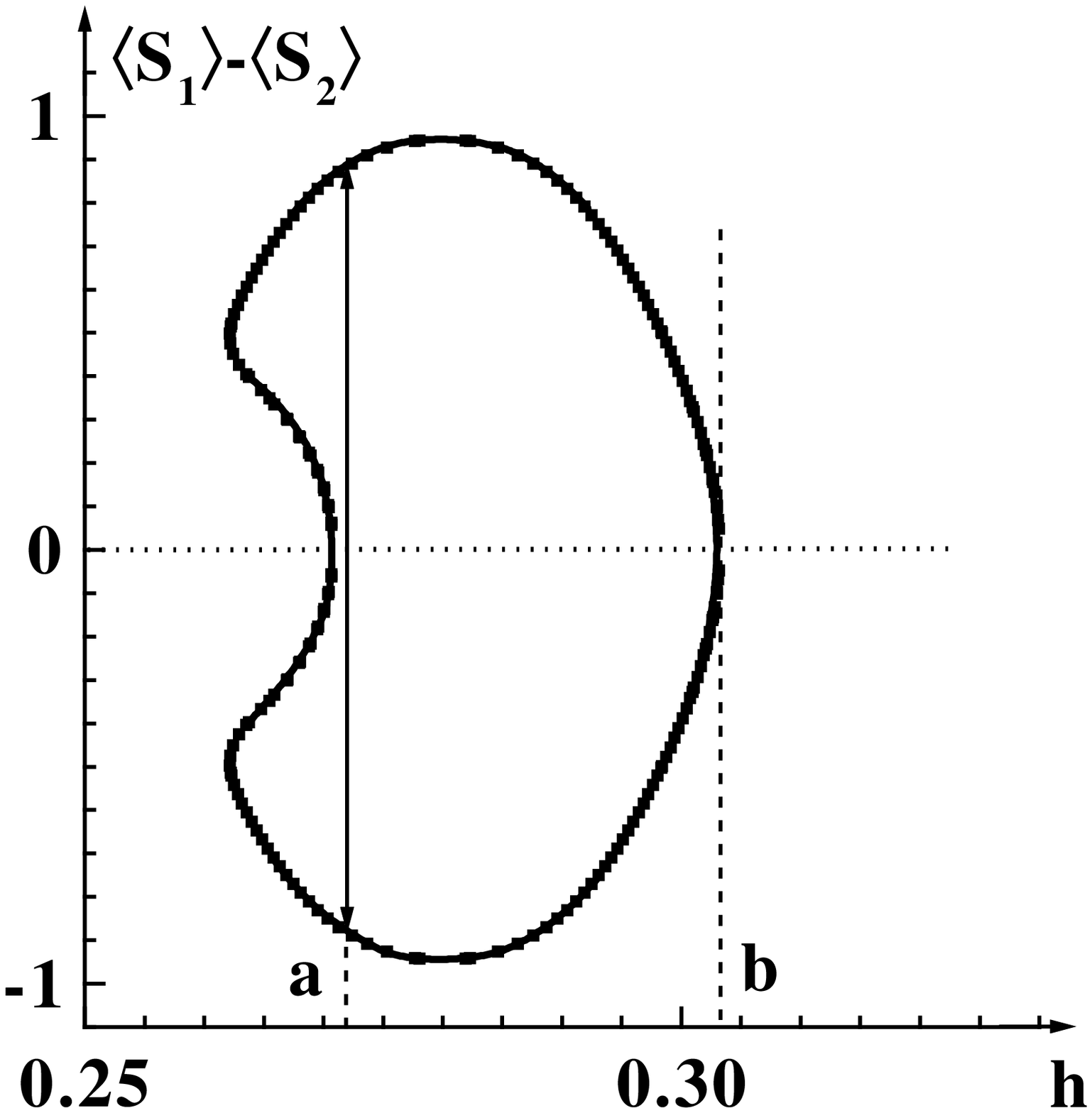,width=1.8in,angle=0}}
 \vspace*{8pt}
\caption{Field dependence of the grand canonical potential and order
parameter ($T=0.005$, $\mu =-0.36$, $W=0.2$, $g=1$).
 Dotted and solid lines correspond to the uniform and the chess-board
phases, respectively. See the text for a whole description.}
\label{ista:f14}
\end{figure}

\noindent The equilibrium is defined by the minimum condition of
$\Phi$ (\ref{ista:eq1.41}). Numerical analysis of solutions of
equations (\ref{ista:eq1.38})-(\ref{ista:bbb}) which satisfy this
criterion, was performed in Ref.~[\refcite{ista:29}]. The examples
of the calculated field dependencies of $\langle
S_{1}^{z}{-}S_{2}^{z}\rangle$ (the order parameter for the
chess-board phase) and grand canonical potential are presented in
Fig.~\ref{ista:f14} for low temperatures (the case $g\gg W$ is
considered). It is seen from comparison of the $\Phi$ values for the
uniform and the chess-board phases that the modulated phase is
thermodynamically stable at intermediate values of the $h$ field in
the region between points $a$ and $b$.
These points correspond to the first and second order phase
transitions, respectively. In the first case, the jump-like change
of order parameter  is accompanied by characteristically similar
changes of the subband widths and, as a result, by changes of the
electron concentration.

\begin{figure}[!h]
\centerline{\psfig{file=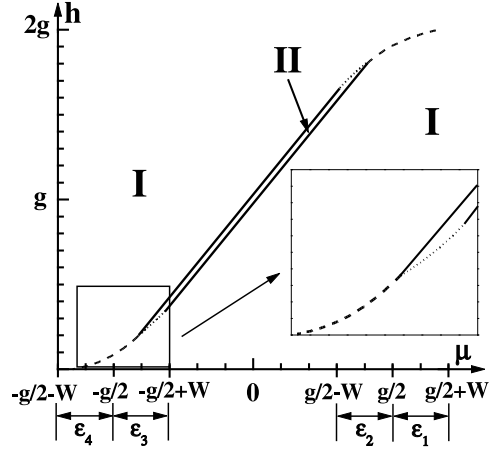,width=2.5in,angle=0}}
\vspace*{8pt}
  \noindent\null\hfill
  \hfill\null
  \caption{($h-\mu$) phase diagram
($T=0.005$, $W=0.2$, $g=1$). I: uniform phase, II: chess-board
phase.
 Dashed lines: first order phase transitions between the
uniform phases with different pseudospin mean values.
 Dotted lines: first order phase transitions between the
uniform and the chess-board phases.
 Solid lines: second order phase transitions.}
\label{ista:f15}
  \end{figure}

The resulting phase diagram ($h-\mu$)  at low temperatures is shown
in Fig.~\ref{ista:f15}.  The chess-board phase exists as an
intermediate phase between the uniform phases with different
$\langle S^{z}\rangle$ and $n$ values. Transitions between the
uniform and modulated phases are of the first or second order and
can be realized in the case when $\mu$ is placed in
$\varepsilon_{2}$ or $\varepsilon_{3}$ subbands  or between them.
Transitions between different uniform phases (described in the
previous section), which are of the first order, take place when the
chemical potential is placed within the $\varepsilon_{1}$,
$\varepsilon_{4}$ and partially  within $\varepsilon_{2}$,
$\varepsilon_{3}$ subbands.

\begin{figure}
\centerline{\psfig{file=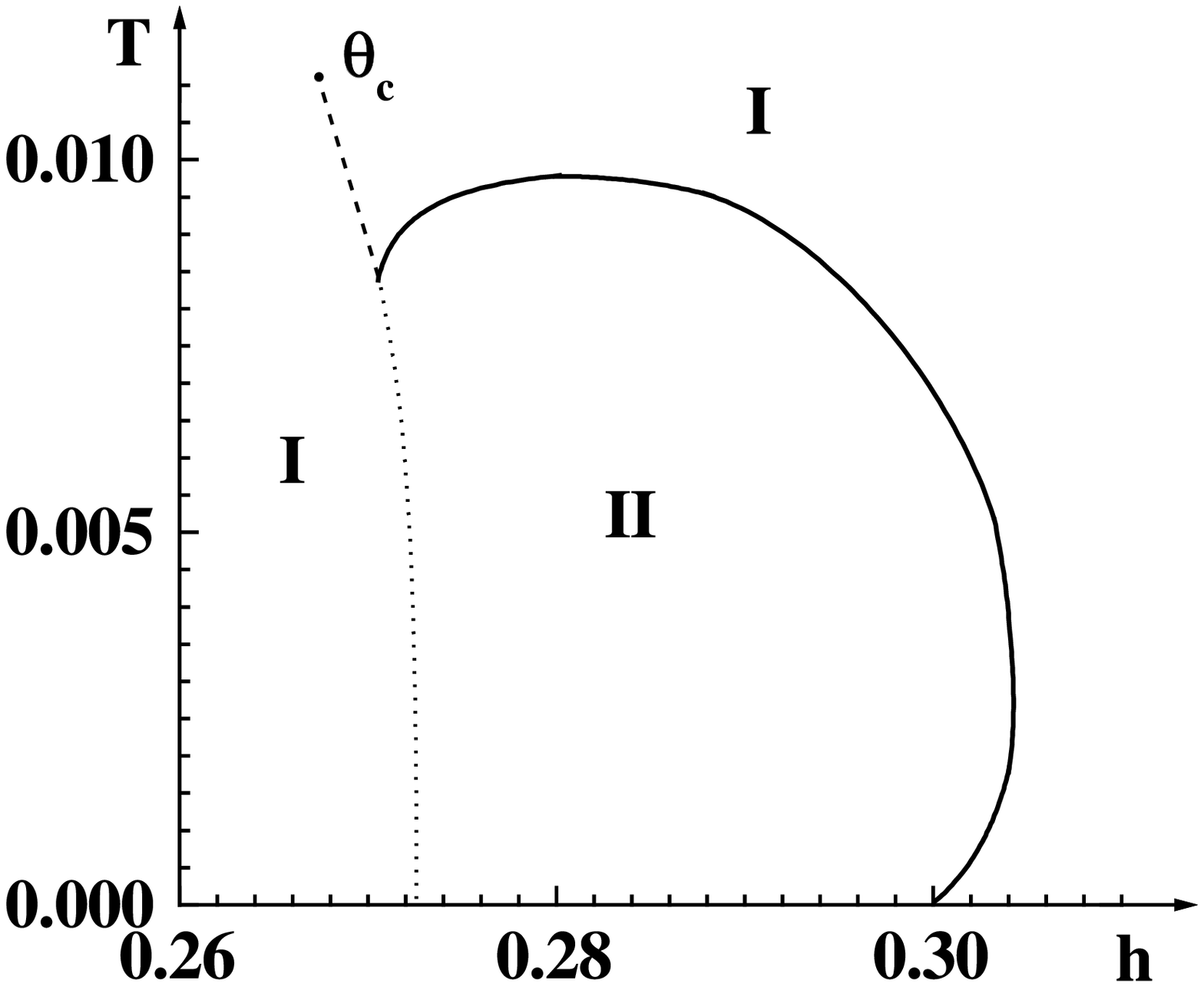,width=2.1in,angle=0}
  \quad
\psfig{file=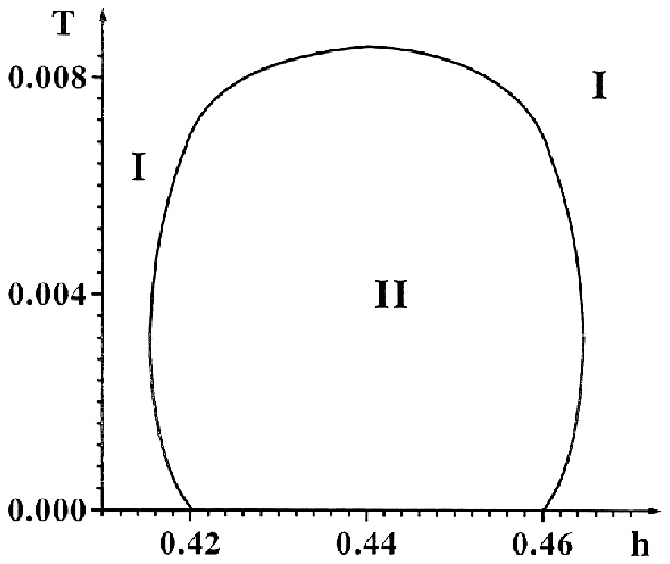,width=2.1in,angle=0}
  }
  \centerline{\quad(a) \hspace{5.2cm}(b)}
  \vspace*{8pt}
  \caption{(a): Phase diagram ($T-h$) ($\mu =-0.36$, $W=0.2$, $g=1$).
 I: uniform phase, II: chess-board phase.
 Dashed lines: first order phase transitions between
different uniform phases (bistability).
 Dotted line: first order phase transitions between the
uniform and the chess-board phases.
 Solid lines: second order phase transitions.
(b): Phase diagram ($T-h$) ($\mu =-0.28$, $t_{{\bf k}=0}=0.2$,
$g=1$).
 }
 \label{ista:f16}
  \end{figure}

The shape of the ($T-h$) phase diagram strongly depends   on the
$\mu$ value. In the case when $\mu$ is placed in the
$\varepsilon_{3}$ subband, such a diagram is shown in
Fig.~\ref{ista:f16}a.
With the temperature increase, the first order phase transition
between the uniform and the chess-board phases transforms into the
first order phase transition between uniform phases and, finally,
disappears in the critical point {$\bf \theta_c$}.
The diagram  shows the possibility of the first order phase
transitions between uniform phases and either first or second order
ones between the uniform and the chess-board phases at the change of
temperature.

In the case when chemical potential lies between the
$\varepsilon_{1}$, $\varepsilon_{2}$ and $\varepsilon_{3}$,
$\varepsilon_{4}$ subbands, the transitions between phases I
(uniform) and II (chess-board) are of the second order; the
corresponding ($T-h$) diagram is shown in Fig.~\ref{ista:f16}b.

\noindent$b)$ The $n={\rm const}$ regime.

\begin{figure}
\centerline{\psfig{file=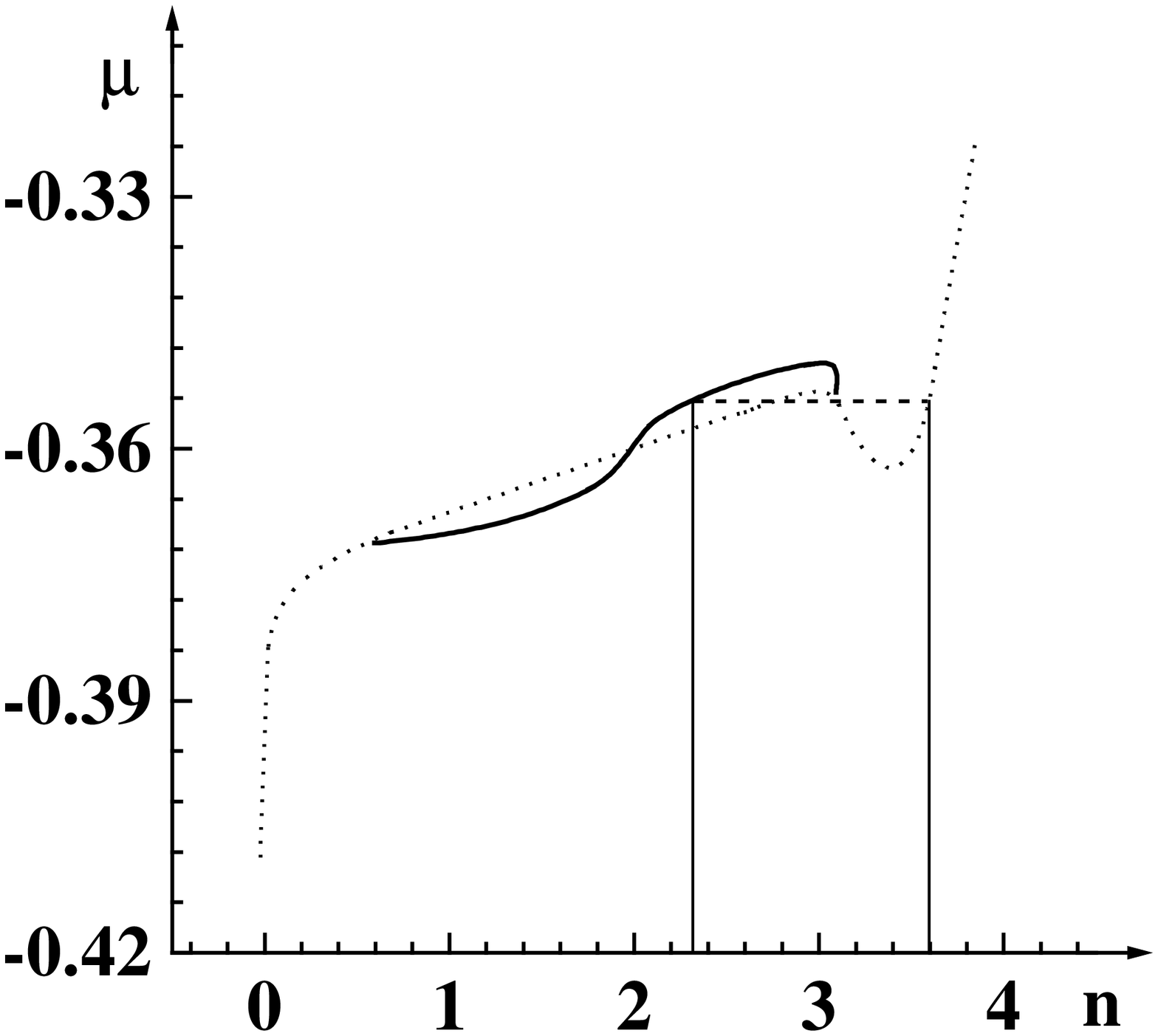,width=2.0in,angle=0}\quad
\psfig{file=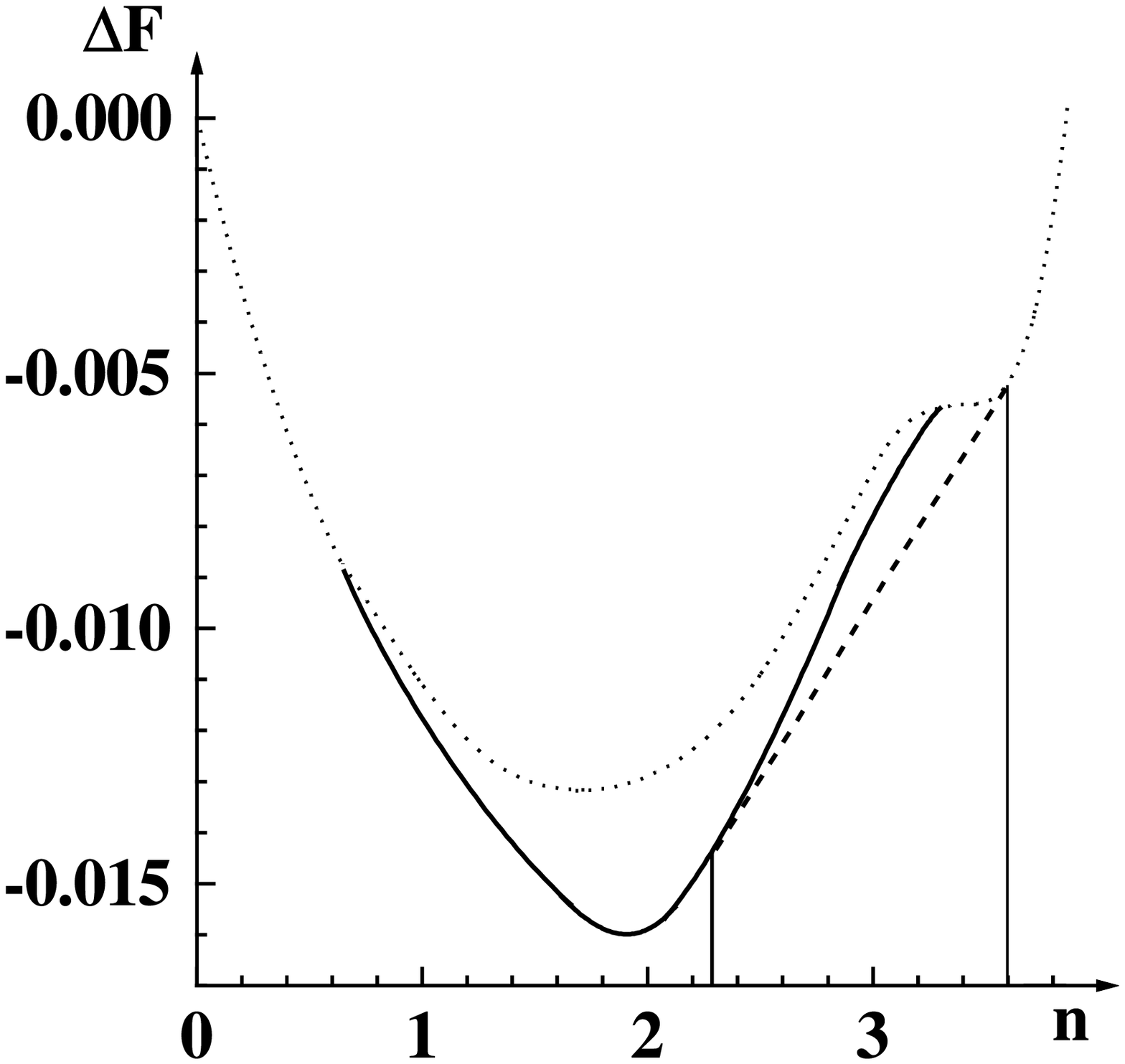,width=2.0in,angle=0}} \vspace*{8pt}
\caption{Dependence of the chemical potential $\mu $ on the electron
concentration $n$ and deviation of the free energy from linear
dependence ($T=0.005$, $h=0.28$, $W=0.2$, $g=1$).
 Dashed line: phase separation area.
 Dotted line: uniform phase.
 Solid line: chess-board phase.}
 \label{ista:f18}
\end{figure}

\noindent In this regime, the equilibrium is defined by a minimum of
the free energy $F=\Phi+\mu N$.
This condition forms a set of equations (\ref{ista:eq1.38}) and
(\ref{ista:bbb}) for the pseudospin mean values and chemical
potential.
 The obtained dependencies of $F$ and $\mu$ on the electron concentration
are presented in Fig.~\ref{ista:f18}.
%

%
\begin{figure}
\centerline{\psfig{file=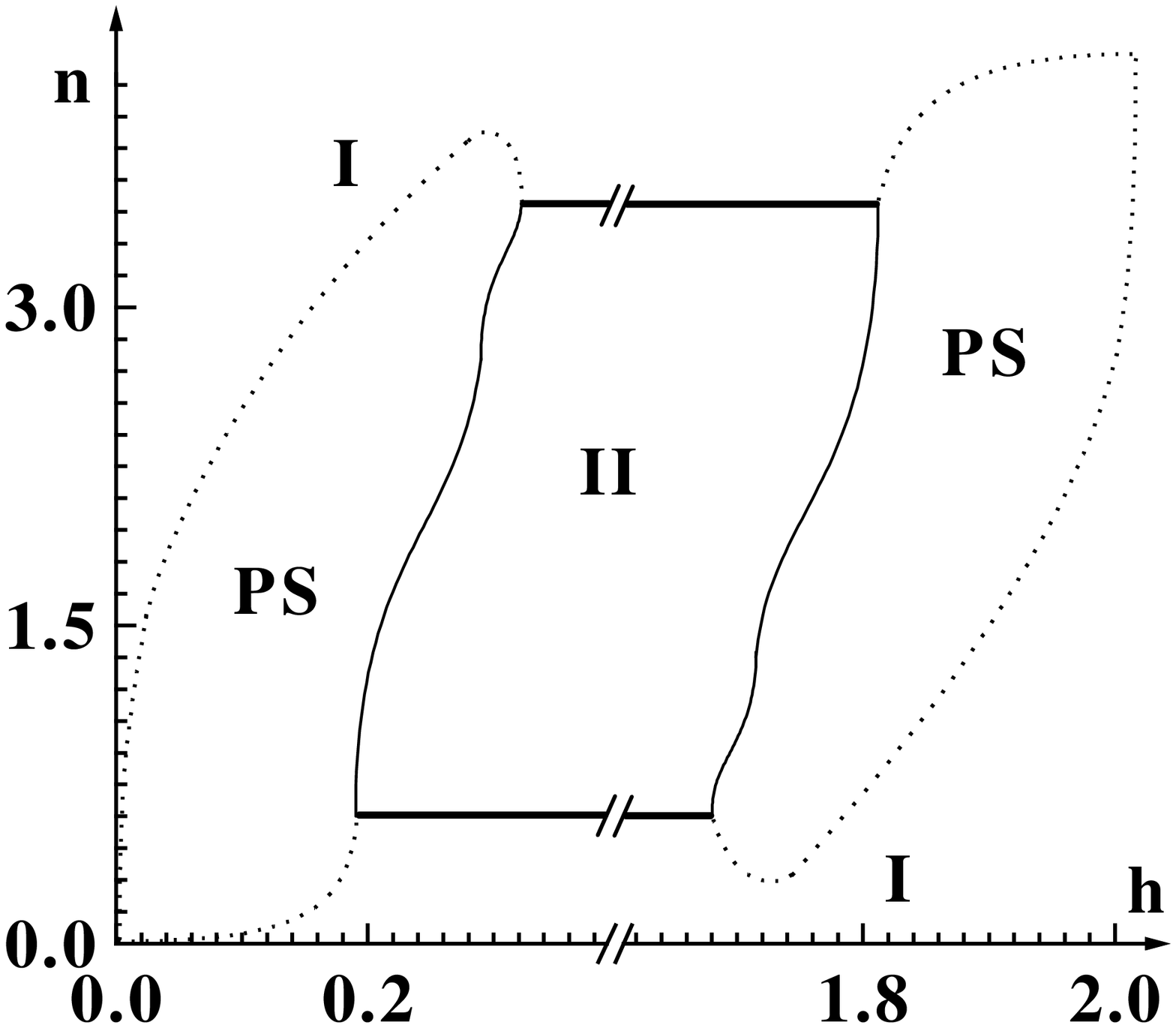,width=2.4in,angle=0}}
\vspace*{8pt} \caption{Phase diagram ($n-h$) ($T=0.005$, $W=0.2$,
$g=1$).
 I:  uniform phase,
 II: chess-board phase,
 PS: phase separation area.}
 \label{ista:f19}
\end{figure}

 One can see the regions with ${\rm d}\mu /{\rm d}n \leq 0$
where
 the phase separation into
regions with different phases (the uniform and the chess-board
phases in this case) and with different electron concentrations and
pseudospin mean values take place.

Based on the  obtained results, the phase diagram ($n-h$) was
constructed (Fig.~\ref{ista:f19}). The phase separation
 into regions with the uniform and the
chess-board phases takes place when the chemical potential is placed
within the subbands $\varepsilon_{2}$, $\varepsilon_{3}$. That
agrees with the results obtained in the $\mu={\rm const}$ case when
within this area we had the first order phase transition between the
corresponding phases.

 The phase separated and the chess-board phase regions narrow with the
tempera\-ture increase, but thick solid lines in Fig. \ref{ista:f19}
approach faster one  another  and, for high enough temperatures, we
have only the phase separation into the regions with uniform phases.

The fact of the existence of doubly-modulated (chess-board) phase in
the PEM as intermediate one between  segregated  phases (at low
temperatures) is similar  to the results obtained in DMFT for the
Falicov-Kimball model  by Freericks and Lemansky.\cite{ista:55}
Investigation of temperature $T_{c}$ of instability of the
high-temperature phase of the FK model as a function of the ordering
wave vector (determined by the divergence of the relevant
susceptibility) showed that near half-filling ($\rho_{e}=0.5$) the
chess-board phase is stable, while the segregated phase exists at
the occupations near electron band edges. The  region of modulated
phase narrows at the increase of $g$ and disappears at
$g\rightarrow\infty$.\cite{ista:51} In this limit the
thermodynamically stable states were  analyzed\cite{ista:56} based
on the calculations of the grand canonical potential  and the phase
separation diagrams (where besides  the spinodal lines the lines of
the first  order transition temperatures were shown) were built; the
cases of different electron concentrations $\rho_{e}$ were
considered.

Nevertheless, one cannot claim the one-to-one correspondence. The
calculations of Freericks and Lemansky were performed  in another
thermody\-namic regime: the concentration of heavy particles
$\rho_{i}$ was fixed (that  would correspond to the given value of
$\langle S^{z}\rangle$ for PEM). The  closest to our case are the
results obtained by  Brandt  and Mielsch.\cite{ista:57} They
constructed   the $(T-\rho_{i})$ diagrams  at the finite $g$ values
for the FK model. The obtained sequence of phases
(segregated-modulated-segregated) when $\rho_{i}$ increases (from
$\rho_{i}=0$ up to $\rho_{i}=1$) is the same  as for PEM at the
increase of the field $h$.

\subsection{Weak Coupling Case; $U=0$}\label{ista:3.2}

Now let us consider the case of weak coupling $(g<W)$. We can base
on the perturbation theory approach taking the mean field
Hamiltonian as the zero-order one.\cite{ista:30} We use the
approximation
\begin{equation}
gn_{i}S_{i}^{z}\rightarrow gn_{i}\langle S_{i}^{z}\rangle+ g\langle
n_{i}\rangle S_{i}^{z}-g\langle n_{i}\rangle \langle
S_{i}^{z}\rangle \label{ista:eq1.42}
\end{equation}
in this case, based on the arguments that for the simplified PEM
with $U=0$ the interaction constant $g$ plays the role which is
similar to that of $U$ in the Hubbard model, and at the decrease of
$g$ below the critical value $(g\sim W)$ the system should pass to
the mean field regime of the Hartree-Fock type.

Having in mind the possibility that the system can be in the uniform
or modulated state, we shall consider, as above, two different
cases. The first one corresponds to the homogeneous pseudospin
ordering and spatially uniform mean distribution of electrons. The
second one is the case of modulation with doubling of the lattice
period (chess-board phase).

\subsubsection{Uniform Phase}\label{ista:p3.2.1}

The Hamiltonian of PEM in the mean-field approximation (MFA)
\index{approximation!mean-field} reads
\begin{eqnarray}
&& H=H_{el}+H_s+U,\\\label{ista:eq1.43} &&
H_{el}=\sum_{i,\sigma}(g\eta-\mu)n_{i,\sigma}+
\sum_{i,j,\sigma}t_{i,j}a^{+}_{i,\sigma}a_{j,\sigma}\, ,\nonumber\\
&& H_s=\sum_i [(gn-h)S^{z}_i+\Omega S^{x}_i] \, ,
\\ \label{ista:eq1.44} && U=-g \sum_{i}n\eta=-Ngn\eta.\nonumber
       \end{eqnarray}
The parameters $\langle n_{i}\rangle=n$ and $\langle
S_{i}^{z}\rangle=\eta$ are determined from the set of equations
\begin{eqnarray}
&&n=\frac{1}{N}\sum_{{\bf k} \sigma} ({\rm e}^{\beta(g\eta+t_{\bf
k}-\mu)}+1)^{-1}\equiv\frac{1}{N}
\sum_{{\bf k}\sigma}f(g\eta+t_{{\bf k}}),\label{ista:eq1.45}\\
&&   \eta=\frac{h-gn}{2\lambda} \tanh(\frac{\beta\lambda}{2})  \ ;
\quad
  \lambda=\sqrt{(g n-h)^2+\Omega^2}. \nonumber
\end{eqnarray}
The grand canonical potential in the MFA is given by the expression
\begin{eqnarray}
&& \frac{\Phi}{N}=-\frac{T}{N}\sum_{{\bf k},\sigma} \ln(1+{\rm
e}^{\frac{\mu-t_{\bf k}-g\eta}{T}})-T\ln(2\cosh\frac{\beta
\lambda}{2})-gn\eta \label{ista:eq1.46}
\end{eqnarray}
where $\lambda=\sqrt{(gn-h)^{2}+\Omega^{2}}$.

Similarly to the strong coupling case, we can distinguish the
regimes of the constant electron chemical potential $\mu={\rm
const}$ (where the stable states can be found from the minimum
$\Phi$ condition) and the given electron concentration $n={\rm
const}$ (when one should find the minimum of the free energy $F=\mu
n+\Phi$).

\noindent$a)$ {Thermodynamics in the $\mu={\rm const}$ regime.}

We consider at first the simplest case $\mu=0$, $h=g$. Here the
solution $\eta=0$, $n=1$ of the set of equations (\ref{ista:eq1.45})
exists at any temperature and describes a disordered phase.
Furthermore, at low temperatures there appears a non-zero solution
$\eta\neq0$, $n\neq1$. A critical temperature $T_{c}$ is determined
from the equation
\begin{eqnarray}
   &&1+\frac{g^2}{2\Omega}\tanh\frac{\beta\Omega}{2}\Pi_0=0,\label{ista:eq1.47}
     \end{eqnarray}
which in the case $\Omega\rightarrow0$ reduces to the form
 \begin{eqnarray}
&&    1+\frac{\beta}{4}g^2\Pi_0=0,\label{ista:eq1.48}
     \end{eqnarray}
           Here
          \begin{eqnarray}
 && \Pi_0=\frac{2}{N}\sum_{\bf k} f'(t_{\bf k})=2\int_{-W}^{W}{\rm d}t\rho(t)f'(t)\, .\label{ista:eq1.49}
     \end{eqnarray}

In the low temperature limit we have $T_{c}=\frac{g^{2}}{2}\rho(0)$
(when the DOS at the Fermi level $\rho(\mu)\big|_{\mu=0}$ is
finite). In the case $\Omega\neq0$ there exists such a critical
value $\Omega_{cr}=g^{2}\rho(0)$ ($\Omega_{cr}=\frac{g^{2}}{2W}$ for
the rectangular DOS), above
which
        (at $\Omega>\Omega_{cr}$) the phase transition to ordered phase
        disappears. This is equivalent to the existence of a critical
 value of $g$: at given $\Omega$ the phase transition is possible when
 $g>g_{cr}=\sqrt{\frac{\Omega}{\rho(0)}}$.
 In the case of the DOS
      with  logarithmic singularity,  the critical
       temperature exists at any values of the tunneling splitting
       parameter \index{tunneling splitting} $\Omega $  and at
       $\frac{\Omega}{g}  \gg  \frac{g}{W}$ we have an asymptotic
       expresssion:
         \begin{eqnarray}
    &&      T_c\approx {2{\rm e}W}
    \exp(\frac{-\Omega W\pi^2}{2g^2}).\label{ista:eq1.50}
           \end{eqnarray}
   The physical nature of the  phase transition considered here
  at $h=g, \mu=0$ (the fixed $\mu$ regime) is  as follows:
  the appearance of an ordered phase is connected with its stabilization
  due to the shift of the electron band down to  the  low energy
   values
   under the effect of the internal field; this  ensures the
  corresponding gain in the electron energy (it should be mentioned that the electron band
  spectrum in this case remains unsplit in the uniform phase; only the shift of the band as a whole
  can take place).

   This mechanism
   remains  the  main reason of the phase transition when the
 initial   electron band is not half-filled. In this case
  (when $\mu\neq 0$)  we performed
 the investigation  using the numerical calculations when
  the set of equations (\ref{ista:eq1.45}) is solved and
  using the expression (\ref{ista:eq1.46}) for the grand canonical potential $\Phi$. The
  selection of  solutions was
 carried out using the condition of the absolute minimum of $\Phi$.

 As can be seen,\cite{ista:30} below $T_{c}$ the system undergoes  the first
 order phase transition
 with
 jumps of the
       mean values of the electron concentration and pseudospin at the
       change
        of the field $h$; the phase transition point is
        determined using the
        Maxwell rule. The similar transition takes place at the change of
        the chemical potential at fixed $h$.
          The presence
           of the tunneling-like splitting decreases the  temperature of the
           phase transition at the fixed values of $\mu$ and $h$.

\begin{figure}
\centerline{\psfig{file=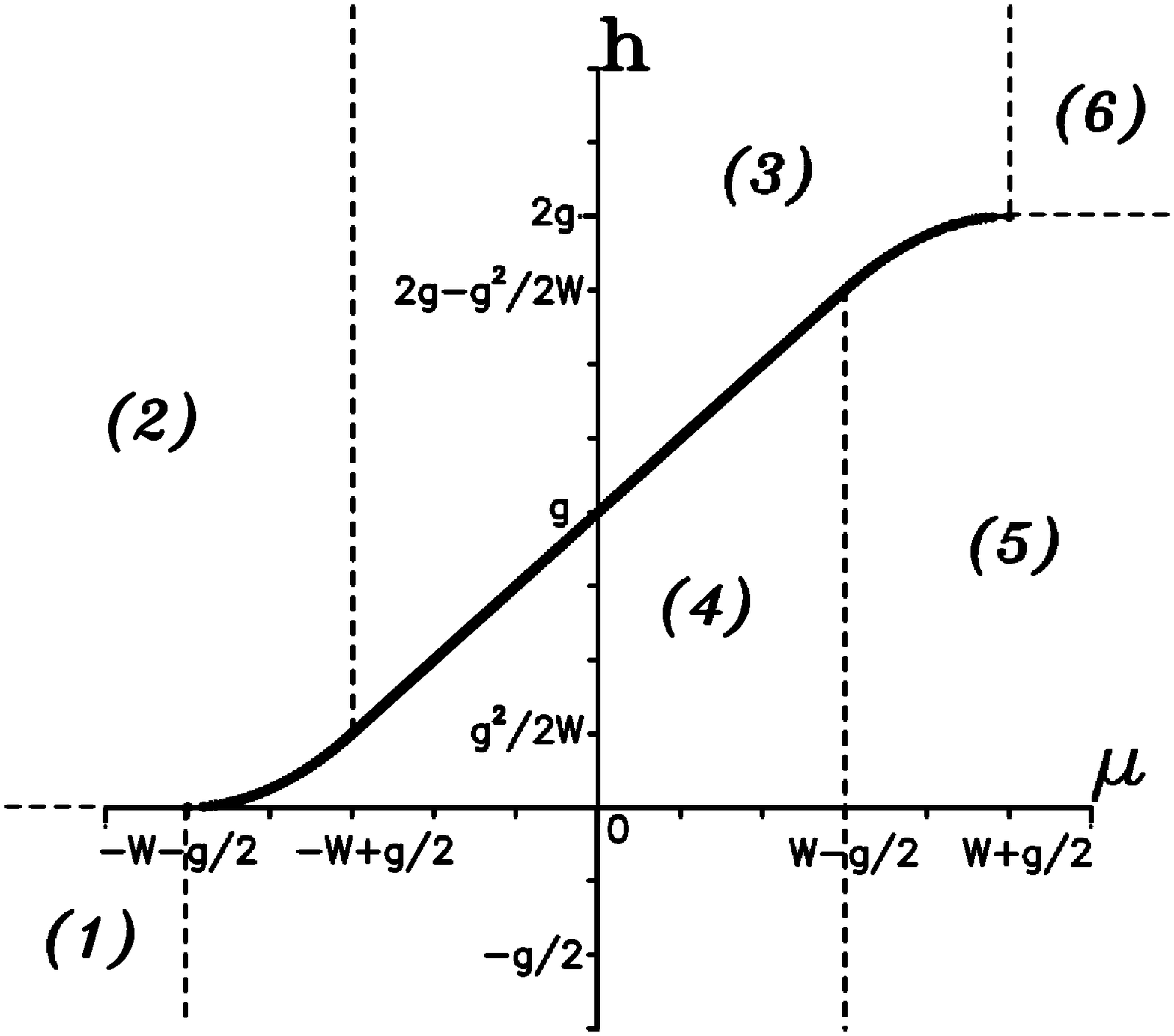,width=2.1in,angle=0}
  \quad
\psfig{file=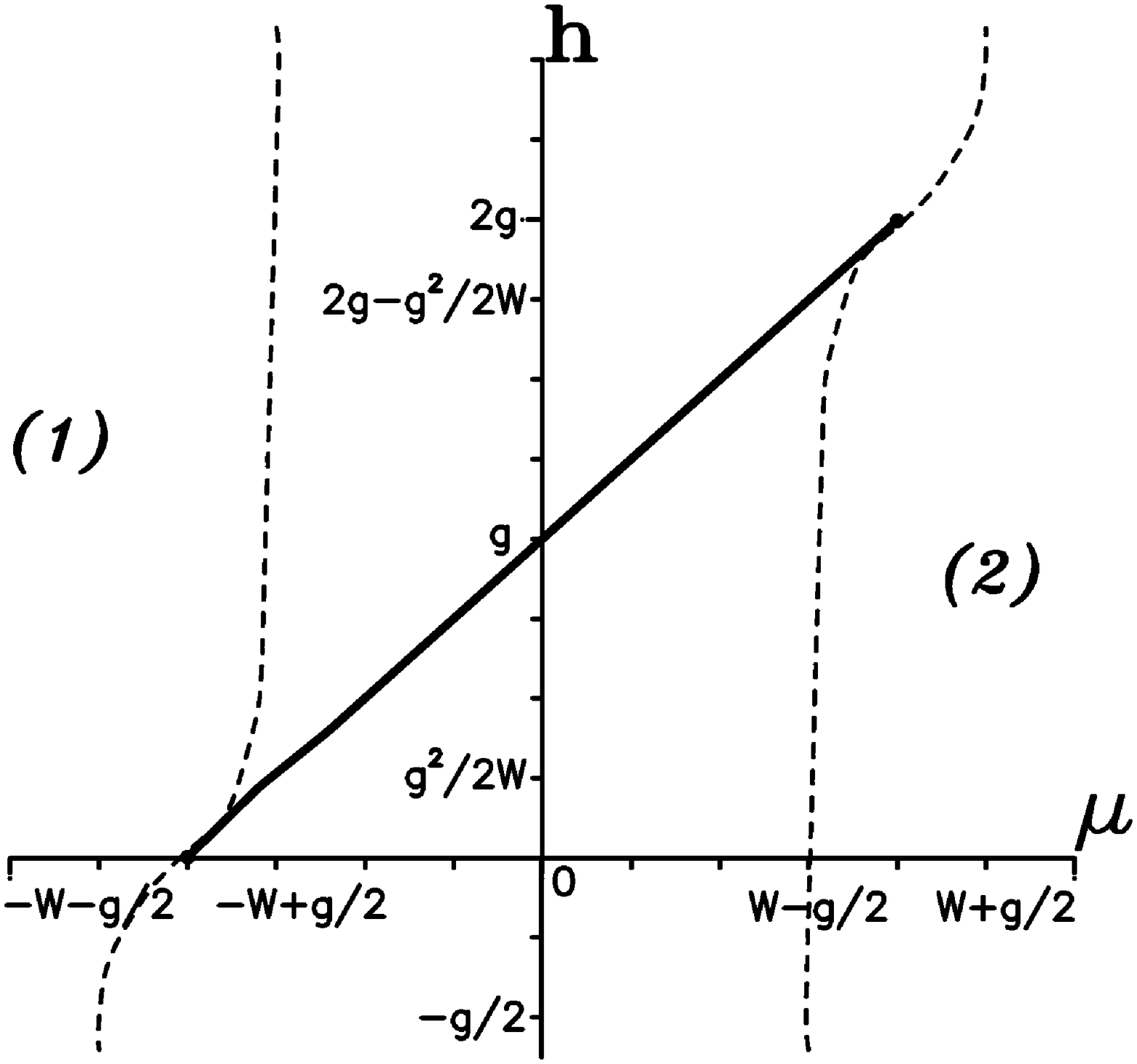,width=2.1in,angle=0}
  }
  \centerline{\quad(a) \hspace{5.2cm}(b)}
 \vspace*{8pt} \caption{(a): The ground state diagram ($T=0,
\Omega=0$). Regions with different $n,\eta $ values
 are separated by dashed lines and a solid line (the phase transition
 line);
 (1) $ n=0,\eta =-1/2$; (2) $n=0,\eta =1/2$; (3) $\eta
 =1/2, n=1+\frac{\mu}{W}
  -\frac{g}{2W}$;
(4) $\eta =-1/2, n=1+\frac{\mu}{W}+\frac{g}{2W}$;
 (5) $n=2,\eta =-1/2$; (6) $n=2,\eta =1/2$.
(b): The ($h-\mu$)
 phase diagram ($T=0.004, \Omega=0.12$). Regions with
different $n,\eta $ values
 are separated by dashed lines and a solid line (the phase transition
 line):
 (1) $n\approx 0,\eta \approx \frac{1}{2}\frac{h}{\sqrt{h^2+{\Omega}^2}}$;
  (2) $n\approx 2,\eta \approx
 \frac{1}{2}\frac{h-2g}{\sqrt{(h-2g)^2+{\Omega}^2}}$.
  \label{ista:f20} }
\end{figure}

  The regions of coexistence of  phases \index{phase!coexistence}
  with different values of the electron concentration and
 pseudospin are shown in the plane $(\mu,h$) at
   $\Omega=0$ and $\Omega\neq 0$
    in Figs.~\ref{ista:f20}.

\begin{figure}
\centerline{\psfig{file=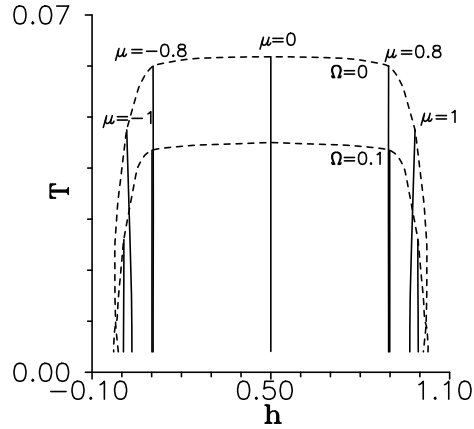,width=2.4in,angle=0}}
\vspace*{8pt} \caption{The critical temperature lines (dashed lines)
and the phase transition
 lines (solid lines) for $\Omega=0$ and $\Omega=0.1$ (the case of the
 rectangular DOS).}
\label{ista:f22}
\end{figure}

\begin{figure}[!h]
\centerline{\psfig{file=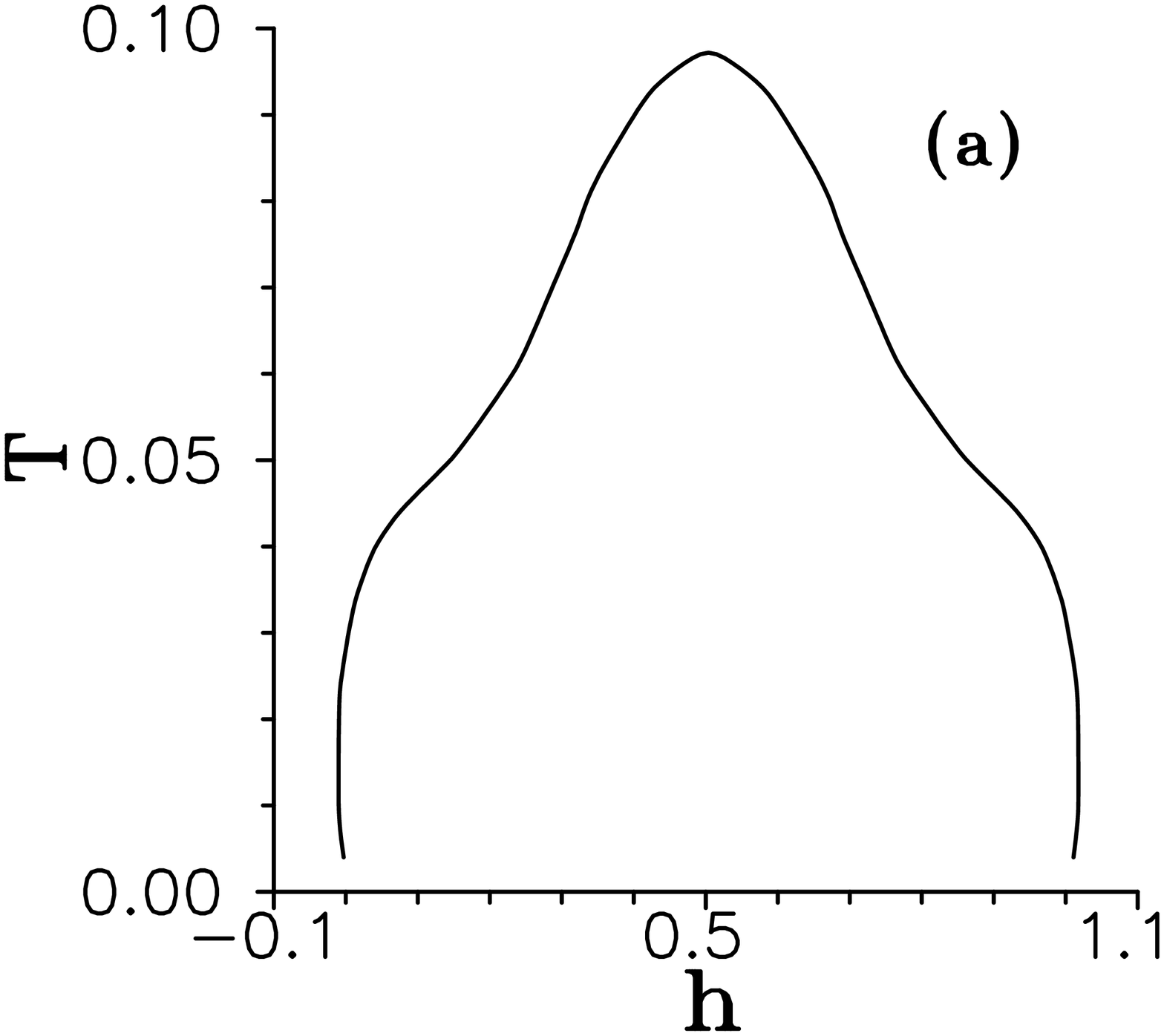,width=2.0in,angle=0} \quad
\psfig{file=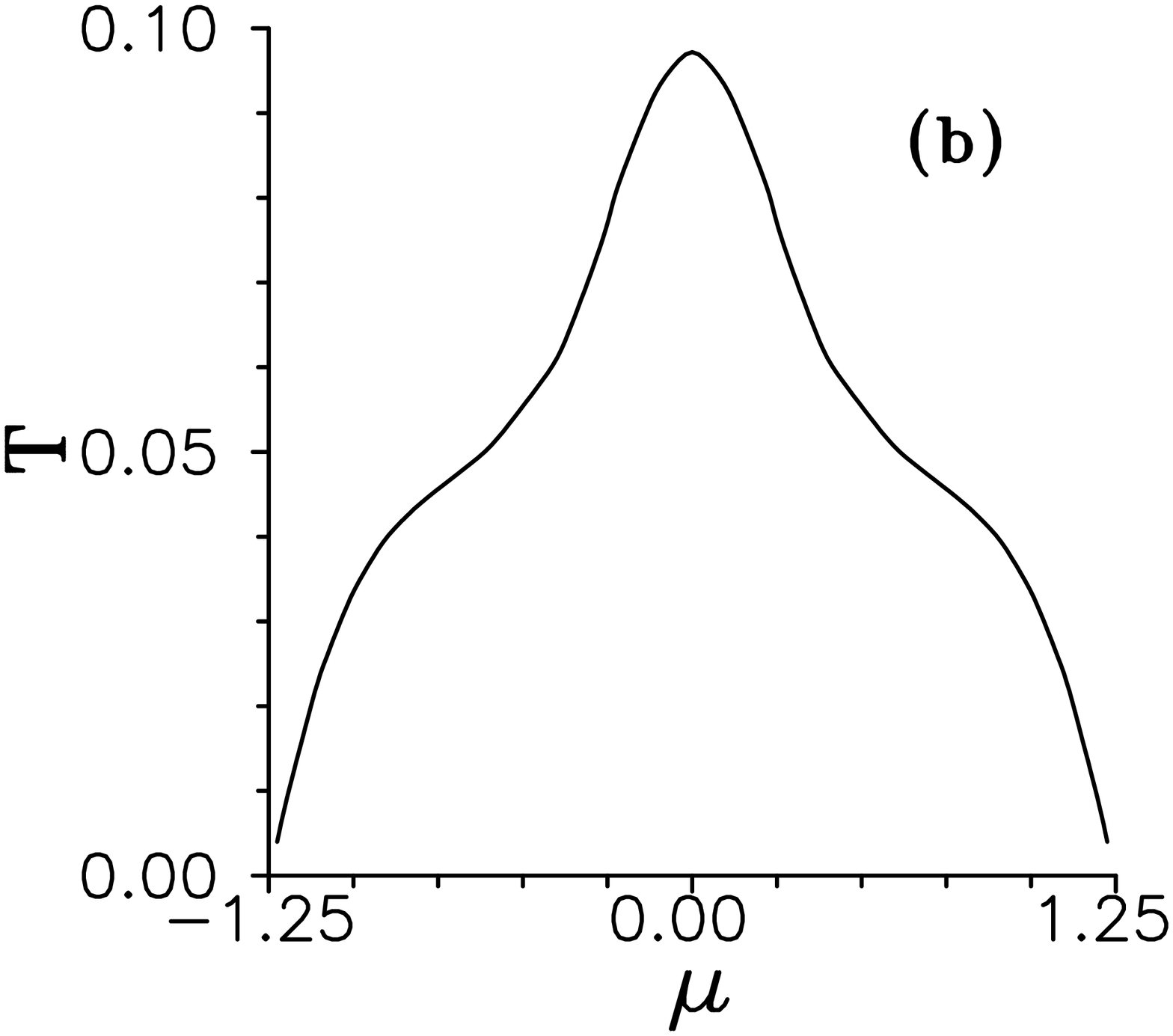,width=2.0in,angle=0}}
\vspace*{8pt} \caption{The critical temperature lines on the a)
$(h,\, T)$ and
  b) $(\mu, \,T)$ planes for
  $\Omega=0$ (direct momentum summation is used).}
\label{ista:f23}
\end{figure}

Phase transition lines in the plane $(T,h)$ at different values of
$\mu$ are shown in Fig.~\ref{ista:f22}. Such a line is vertical for
the
case $\mu=0$  only;
  for the case $\mu\neq 0$, the line is bent. This makes the first order phase transition possible at the change of
  temperature (with the jumps of the parameters $\eta, n$). The slopes  of
  the  phase equilibrium curves are opposite for $\mu>0$ and $\mu<0$.
    The lines of
 the critical points
    are shown  for the cases $\Omega=0, \Omega\neq 0$;
 the calculations are carried out using rectangular density of states.
 Similar phase diagrams are obtained using direct momentum
 summation for square lattice (Fig.~\ref{ista:f23}). As in the case of rectangular
 DOS, the maximum $T_{c}$ is achieved at $\mu=0$ but the critical
 temperature line has a more pronounced peak.
%

\noindent$b)$ {Thermodynamics in the $n=\,\,$const regime.}

 The values $n_{1},n_{2}$ and $\eta_{1},\eta_{2}$ between which
 the jumps of the electron concentration and pseudospin mean value
 take place at the phase transitions in the $\mu={\rm const}$ case,
 correspond to the phases which coexist in the phase transition
 points. In the regime $n={\rm const}$, there is a phase
 separation on the phases with the above mentioned values of $n$
 and $\mu$. For example, at the parameter value $\Omega=0$,
 $h=0.7$, $T=0.008$, the system is unstable with respect to the
 phase separation in the region $n_{1}=1.149<n<n_{2}=1.645$.\cite{ista:30}

\begin{figure}[b]
\centerline{\psfig{file=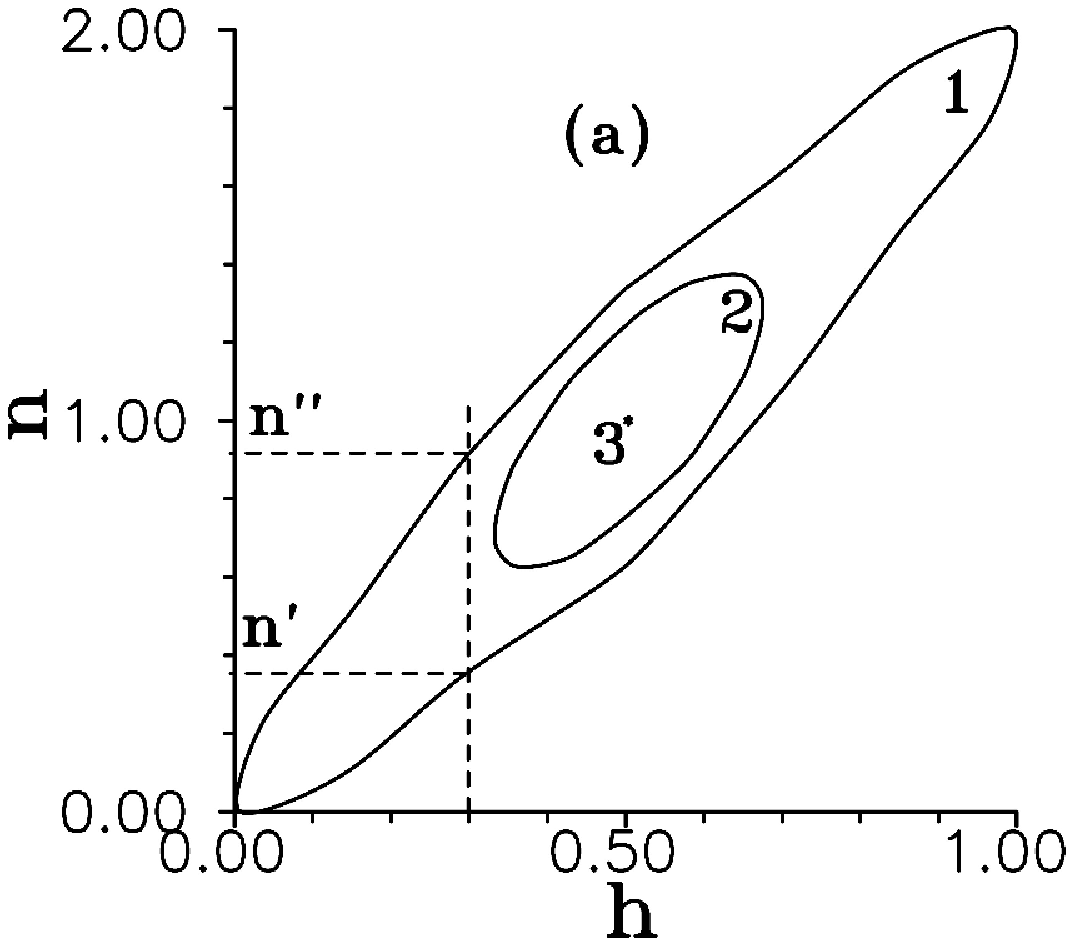,width=2.0in,angle=0} \quad
\psfig{file=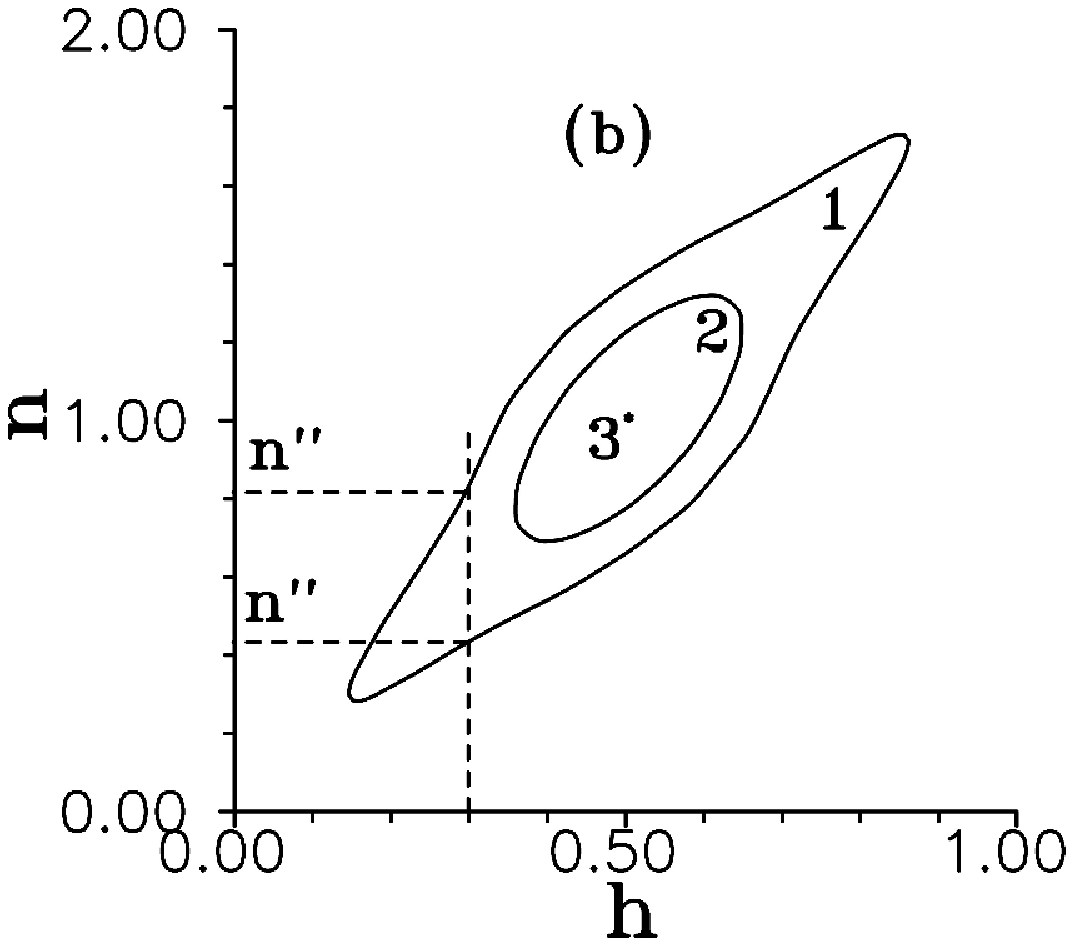,width=2.0in,angle=0}}
\vspace*{8pt} \caption{$(n-h)$--phase diagram
 in the cases a) $\Omega=0$; b)
  $\Omega=0.1$. Phase separation regions for
different temperatures are shown. a) 1: $T=0.008$, 2: $T=0.08$, 3:
$T=0.0972$;
 b) 1: $T=0.008$, 2: $T=0.075$, 3: $T=0.0903$.}
\label{ista:f24}
\end{figure}

 Phase separation  regions are shown in Fig. \ref{ista:f24} at different
 temperatures (the calculations were carried out for square lattice using direct momentum
     summation when solving the set of equations (\ref{ista:eq1.45})). At the
     increase of
       temperature the separation region narrows and at
      $T>T_{c}$ it disappears.  The presence of the tunneling-like
     splitting leads to the decrease of  the area
      of phase separation region and to the lowering of $T_c$.
      %
%

      It can be noted that at weak coupling, when only uniform states are
      considered,
      we have one phase separation
 region in the $(h,n)$ plane while in the strong coupling case ($g\gg W$)
 there exist two such regions at the distance of order $g$ along
 $h$ axis; the modulated phase is placed between them\cite{ista:29} (see also Sec.~\ref{ista:p3.1.2}).

 \subsubsection{Phase with Double Modulation}\label{ista:p3.2.2}

 Let us consider now the thermodynamics of the simplified PEM in
 the case of doubly modulated phase. The possibility  for such a phase
 was shown at a strong coupling $(g\gg
 W)$; the necessary condition was the location of the chemical
 potential between the split electron subbands. Now (at $g<W$) the
 band is unsplit and in such a situation there must be another
 mechanism of stabilizing the lattice  modulation.

\begin{figure}[h]
\centerline{\psfig{file=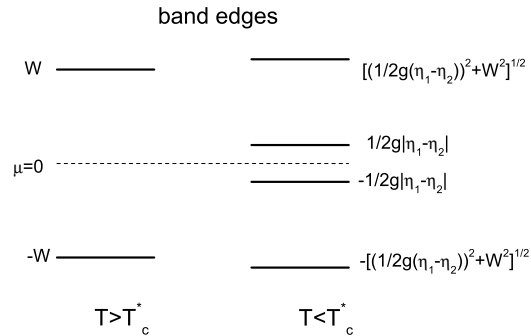,width=3.0in,angle=0}}
\vspace*{8pt}
\caption{The band edges for the cases $T>T^{*}_{c}$ (homogeneous
phase) and $T<T^{*}_{c}$ (the double modulation case); $\mu=0,
h=g$.} \label{ista:f25}
\end{figure}

 At the double modulation the crystal can be divided into two
 sublattices $(\alpha=1,2)$, and the parameters $\eta_{\alpha}=\langle
 S_{i\alpha}^{z}\rangle$, $n_{\alpha}=\sum\limits_{\sigma}\langle
 n_{i\alpha\sigma}\rangle$ can be introduced ($i$ is  an unit cell
 index). Similarly to the homogeneous phase, the mean
 field  approximation is used. The modulation leads to the
 splitting in the electron spectrum due to difference between the
 internal field acting in sublattices
 \begin{equation}
 \lambda_{{\bf k}\alpha}=g\frac{\eta _1+\eta _2}{2}+(-1)^{\alpha}
      \sqrt{(g\frac{\eta _1-\eta _2}{2})^2+t^{2}_{\bf k}}\label{ista:eq1.51}
 \end{equation}
(the similar effect takes place in the FK model\cite{ista:58}). The
initial
band is divided into two subbands separated by the gap
      $\Delta=g|\eta_1-\eta_2|$ (see Fig. \ref{ista:f25}).

      Contributions from both sublattices are present in the equation for
      the electron concentration in sublattices
      \begin{equation}
      n_{\alpha}=\frac{1}{N}\sum_{{\bf k} \sigma}(\frac{1+\cos 2 \phi}{2}
 ({\rm e}^{\beta(\lambda_{{\bf k} \alpha}-\mu)}+1)^{-1}
 +\frac{1-\cos 2 \phi}{2}
 ({\rm e}^{\beta(\lambda_{{\bf k} \beta}-\mu)}+1)^{-1}),\label{ista:eq1.52}
      \end{equation}
       where
       \[
      \cos2\phi=\frac{-g\frac{\eta _1-\eta _2}{2}}
   {\sqrt{(g\frac{\eta _1-\eta _2}{2})^2+t^{2}_{\bf k}}}\]
      can be obtained diagonalizing the mean-field two-sublattice Hamiltonian of the model.\cite{ista:30}
      Another equation which appears
      as a result of the averaging of the operator
      $S_{i\alpha}^z$ has the form
           \begin{eqnarray}
       && \eta _{\alpha}=\frac{h-gn_{\alpha}}{2
       \tilde{\lambda}_{\alpha}}
 \tanh(\frac{\beta \tilde{\lambda}_{\alpha}}{2}).\label{ista:eq1.53}
      \end{eqnarray}
Here
\begin{equation}
\tilde{\lambda}_{\alpha}=\sqrt{(gn_{\alpha}-h)^{2}+\Omega^{2}}\, .
\label{ista:eq1.54}
\end{equation}

In the mean field approximation, the grand canonical potential for
the double modulation case has the form
\begin{eqnarray}
   &&\frac{ 2\Phi}{{N}}=-\frac{T}{N}\sum_{{\bf k},\sigma}
      \ln((1+{\rm e}^{-\frac{\lambda_{{\bf k}1}-\mu}{T}})
      (1+{\rm e}^{-\frac{\lambda_{{\bf k}2}-\mu}{T}}))\nonumber\\
     && -T\ln(4\cosh\frac{\beta\tilde{\lambda_1}}{2}
      \cosh\frac{\beta\tilde{\lambda_2}}{2})-g(n_1\eta _1+n_2\eta _2). \label{ista:eq1.55}
              \end{eqnarray}

\begin{figure}[h]
\centerline{\psfig{file=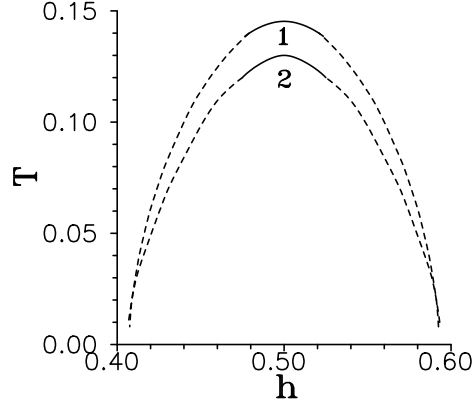,width=2.4in,angle=0}}
\vspace*{8pt}
\caption{The phase transition lines (solid
 and dashed lines are the lines of the second and of the first order phase
transitions, respectively) from the uniform
 phase to the phase with double modulation ($1:$ $\Omega =0$; $2:$
 $\Omega=0.2$).}
\label{ista:f26}
\end{figure}

The solution of the set of equations for the $n_{\alpha}$ and
$\eta_{\alpha}$ parameters and the investigations of
thermodynamically stable states were carried out
numerically.\cite{ista:30} It was established that phase transitions
from the uniform to the low temperature modulated phase are of the
second or  first order. This is illustrated in Fig.~\ref{ista:f26},
where the phase transition lines at $\mu=0$ are shown for $\Omega=0$
and $\Omega\neq0$ (here a direct momentum summation is used in
calculations);   the tricritical points
 \index{point!tricritical} are also present here.


The difference $\delta n=n_{1}-n_{2}$ (as well as the difference
$\delta\eta=\eta_{1}-\eta_{2}$) can play a role of the order
parameter  for the modulated phase. Coming from the equations for
$\delta n$ and $\delta\eta$ we obtain the following condition of the
appearance of nonzero solutions
\begin{eqnarray}
 && 1=\frac{g}{N}\sum_{{\bf k}\sigma}\frac{1}{t_{\bf k}}\left({\rm e}^{\beta(g\eta-t_{\bf k}-\mu)}
 +1\right)^{-1}\nonumber\\
  &&{}\times\left[\beta g\frac{(h-g n)^2}{\lambda^2}
\left(\frac{1}{4}-{\langle\sigma^z\rangle}^2\right)+g
\langle\sigma^z\rangle\frac{\Omega^2}{\lambda^3}\right]\, .
\label{ista:eq1.56}
\end{eqnarray}
Proceeding from this equation, we can find a critical temperature
 $T^{*}_c$ as the maximum temperature (among the set of temperatures
 which
 are obtained for
  different $h$
  values) which fulfills this equation
 at a fixed value of the chemical potential. This temperature is the
 point of the second order phase transition to modulated phase at the
 corresponding
  value of the field~$h$.

 In the symmetric case, when $\mu=0$, $h=g$ and in the
 high-temperature phase $n=1$, $\eta=0$, the equation (\ref{ista:eq1.56}) reduces
 to the form
 \begin{eqnarray}
 && 1=-\frac{g^2}{\Omega}\tanh\frac{\beta\Omega}{2}\frac{1}{N}
 \sum_{\bf k } \frac{1}{t_{\bf k}}f(t_{\bf k}).\label{ista:eq1.57}
 \end{eqnarray}
 The critical temperatures $T_{c}^{*}$, obtained in the cases of
 rectangular DOS and DOS with logarithmic singularity, are,
 respectively,
\begin{equation}
T_{c}^{*}\approx\frac{ {\rm e}W}{2}\exp\left(-\frac{2\Omega
W}{g^{2}}\right)\label{ista:eq1.58}
\end{equation}
and
\begin{equation}
T_{c}^{*}\approx 2{\rm e}W\exp\left(-\frac{\pi\sqrt{\Omega
W}}{g}\right)\, .\label{ista:eq1.59}
\end{equation}
In  both cases, they are higher than the corresponding temperatures
$T_{c}$ for transitions between uniform phases and remain finite at
high $\Omega$ values.

\begin{figure}[h]
\centerline{\psfig{file=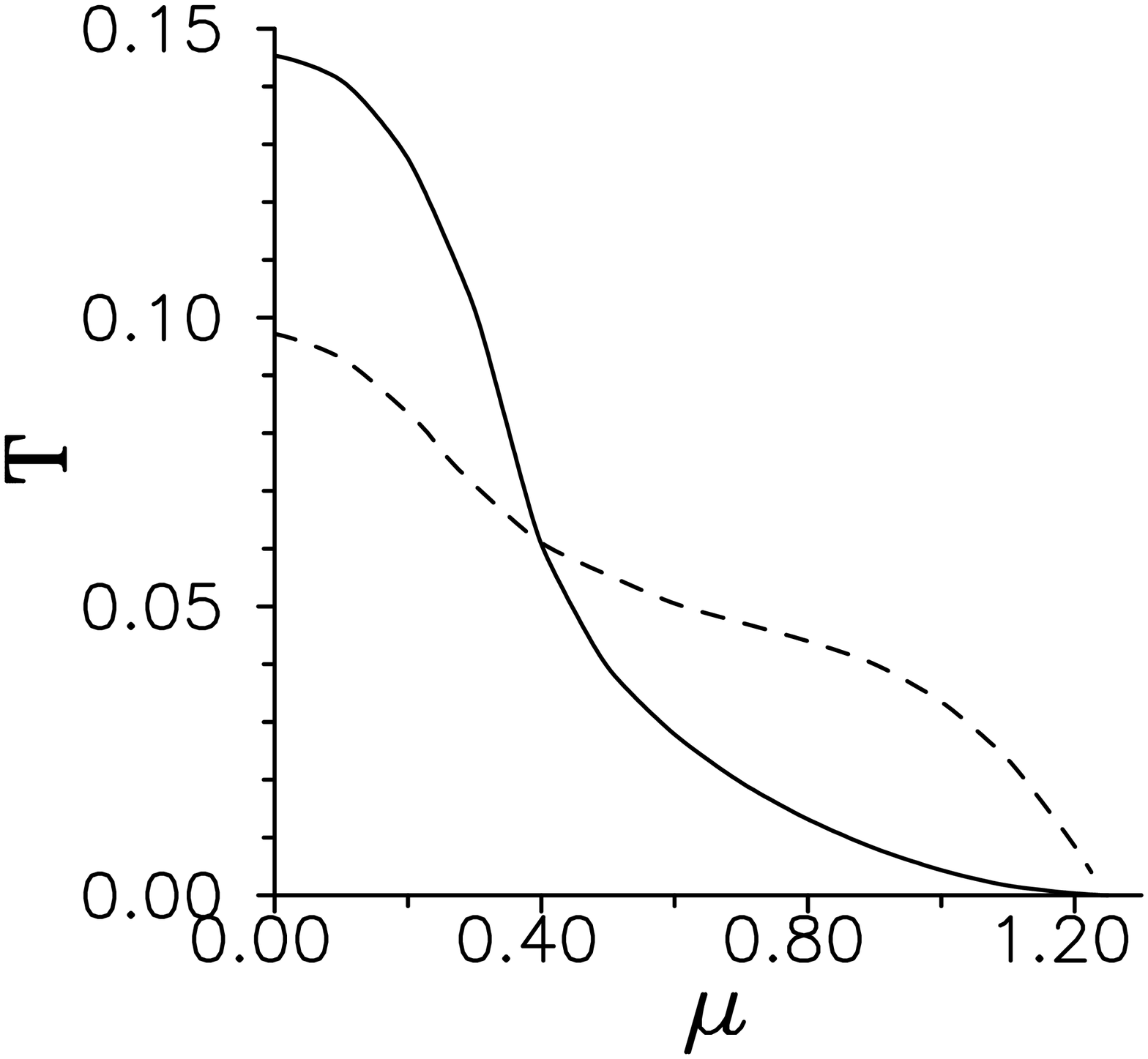,width=2.0in,angle=0} \quad
\psfig{file=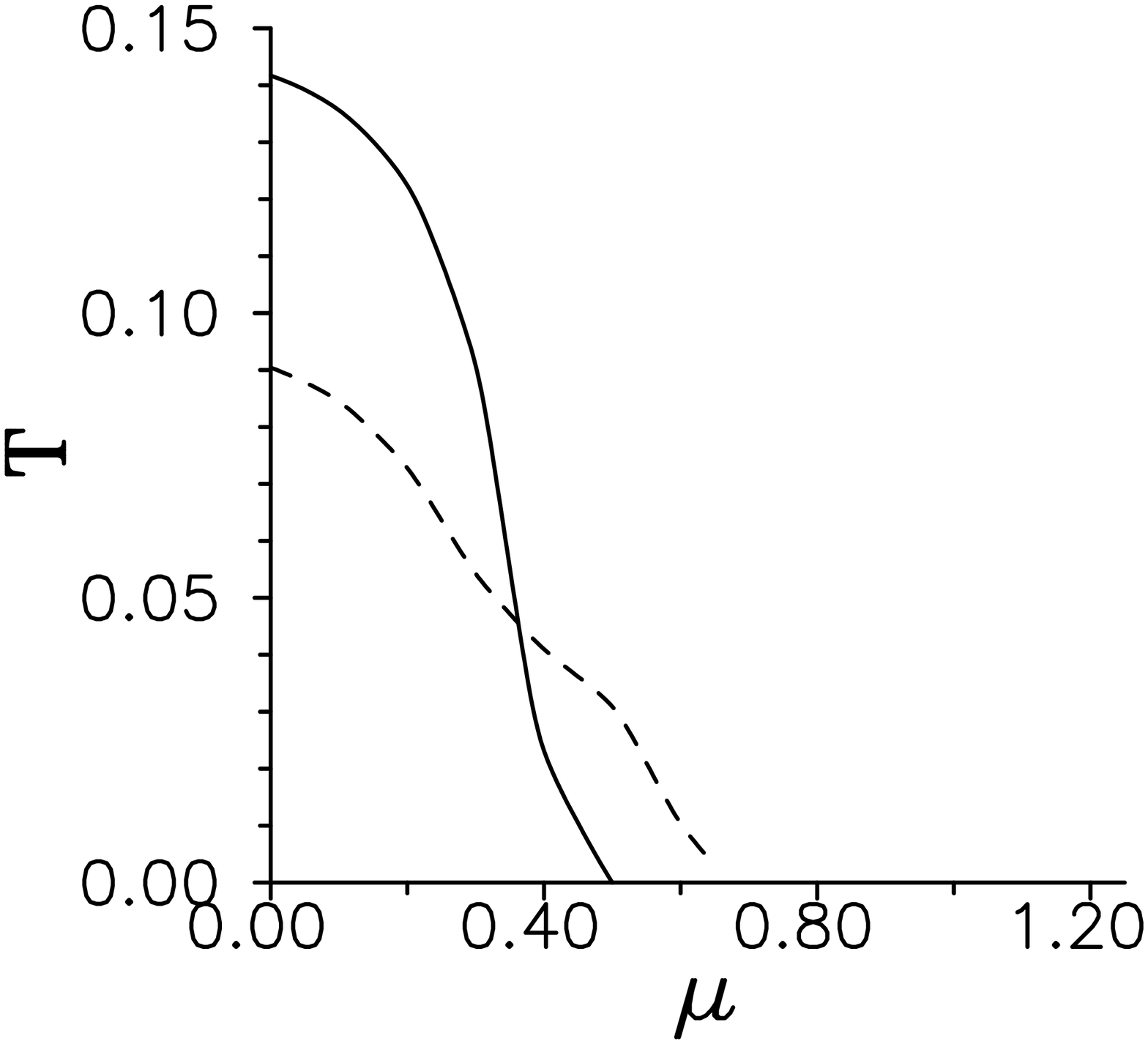,width=2.0in,angle=0} }
\centerline{\quad (a) \hspace{5.1cm}(b)}
 \vspace*{8pt}
\caption{The dependence of the critical temperature on the chemical
potential; a) $\Omega=0$ b) $\Omega=0.1$. Solid line refers to the
case of
 the phase with double modulation, dashed line refers to the
  transition into the
 homogeneous phase.}
 \label{ista:f27}
\end{figure}

It can be seen that $T_{c}^{*}>T_{c}$ also at $\mu\neq0$, but for
the $\mu$ values, which are less than the certain ($|\mu|<\mu_{0}$)
value. The typical dependencies of $T_{c}^{*}$ and $T_{c}$ on $\mu$
are shown in Fig. \ref{ista:f27} in the cases $\Omega=0$ and
$\Omega\neq0$. One can conclude that in the case of the electron
band occupation close to the half-filling, the transition to
modulated phase should be realized. The transition between two
different uniform phases is possible only when $\mu$ is placed near
the band edges.


\begin{figure}[!h]
\centerline{\psfig{file=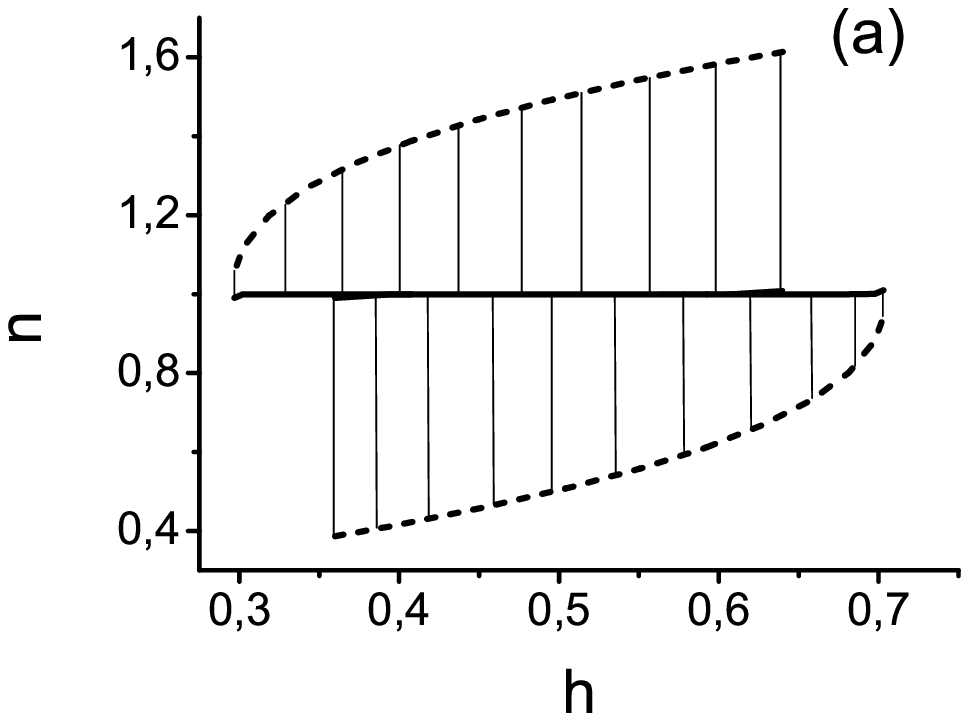,width=2.1in,angle=0} \quad
\psfig{file=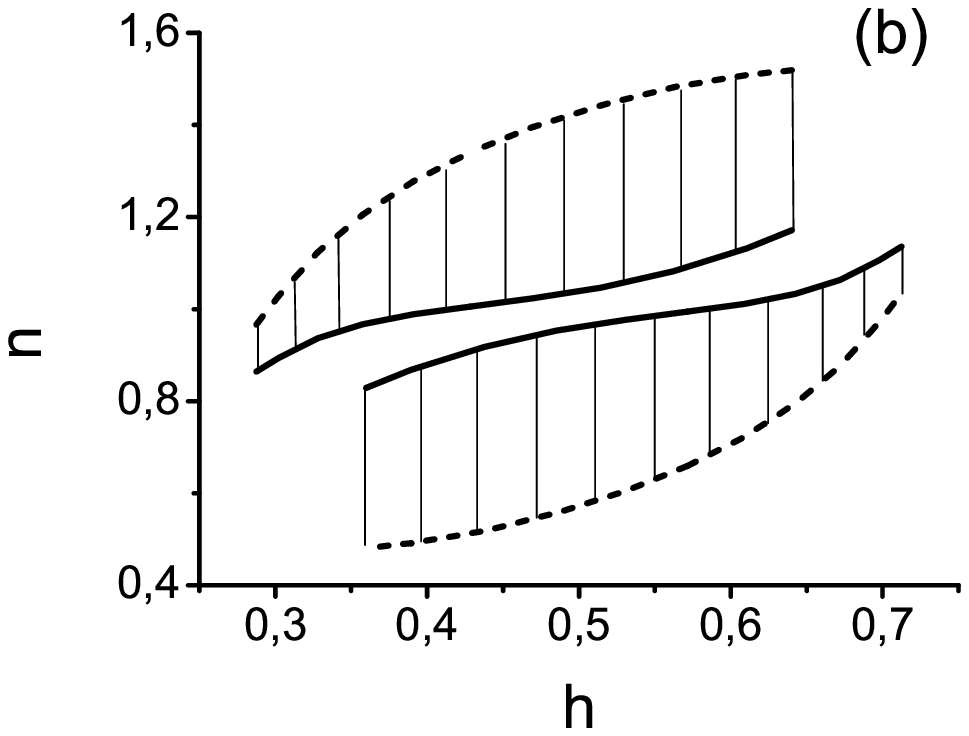,width=2.1in,angle=0} }
\vspace*{8pt}
 \caption{  $(n-h)$ phase diagram, $\Omega=0$. Phase separation
 \index{phase!diagram}
 regions
 are shown for different temperatures: a) $T=0.008$,
 b) $T=0.08$. Dashed lines denote
 the borders of the uniform phase,
 thick solid lines denote the borders of the phase with doubly
 modulated lattice period.}
 \label{ista:f28}
 \end{figure}

An existence  of the first order phase transitions between uniform
and   doubly modulated phases, shows  the possibility of a
separation into these two phases. This takes place at certain values
of the
 electron concentration. The corresponding $(n-h)$ diagrams are shown in
 Fig. \ref{ista:f28}.
 The borders of the separation regions were obtained from the
 convexity condition of the free energy defined as
 $F/N=\Phi/N+n\mu$.
 As temperature increases, the separation area
 narrows, but in the middle of it there appears a region of  the chess-board phase existence.
 This is an additional feature which
 supplements the picture  of separation shown in Fig. \ref{ista:f24}. At high
 enough temperatures, only the second order phase transition into the doubly
 modulated state remains   and the phase separation region
 disappears.

 \subsubsection{Pair Correlation Function and Susceptibilities}\label{ista:p3.2.3}

 The analysis of thermodynamically stable equilibrium states of
 the PEM in the case of weak coupling can be supplemented by an
  investigation of temperature and wave vector dependencies of the
 pseudospin, electron density and mixed pair correlation function.
 \index{function!pair correlation}
 The corresponding Green's functions were calculated in
 Ref.~[\refcite{ista:31}] within
 the GRPA scheme.  The cases of isothermal response \index{response!isothermal} were considered, where the
 isothermal susceptibility
 \begin{eqnarray}
\chi_T({\bf q},\omega_n)=\int^{\beta}_0
 \langle T_{\tau} M(0) M(\tau) \rangle_{{\bf q}} {\rm e}^{i\omega_n \tau} {\rm d}\tau
  -\beta \langle M  \rangle^2 \delta(\omega_n)\label{ista:eq1.60}
 \end{eqnarray}
is expressed in terms of the Matsubara Green's
function\index{function!Green's!Matsubara}  and the so-called
``isolated'' response, \index{response!``isolated''} which is
described by means of the two-time Zubarev
 Green's function\index{function!Green's!Zubarev}
 \begin{eqnarray}
\chi_I({\bf q},\omega) \sim\langle\langle M|M \rangle\rangle_{{\bf
q}\omega}.\label{ista:eq1.61}
 \end{eqnarray}
 The dipole moment $M_{i}$ of the unit cell was taken in the
form: $M_i=d_e n_i+ d_s S^{z}_i$; here the electron contribution due
to nonhomeopolarity of occupancy of the electron orbitals was taken
into account besides the pseudospin contribution.

Such an expression for dipole moment comes from the form of
transverse component of polarization in the case of the YBaCuO
structure. \index{YBaCuO-type superconducting crystals} The electron
component corresponds to the charge transfer in a perpendicular
direction from/to the Cu$_{2}$O$_{2}$ layers (with the participation
of the Cu-O chains), while the pseudospin component is connected
with the redistribution of the ionic charges when O$_{4}$ ion moves
from one equilibrium position to another.\cite{ista:26,ista:59}

Similarly to the above considered  strong coupling case, the simple
sequences of loop diagrams are taken  into account in the
diagrammatic representation for Matsubara's correlators. The
connections between loops are accomplished by semi-invariants or by
the boson (pseudospin) Green's functions. The contribution that
corresponds to the separate link is (see Ref.~[\refcite{ista:31}])
\begin{equation}
\Sigma_{\bf q}(\omega)=\frac12\sin^{2}\vartheta[K_{\bf
q}^{0}({\omega_{n}})+K_{\bf q}^{0}(-{\omega_{n}})]\langle
\sigma^{z}\rangle_{0}-M_{\bf q}(\omega_{n}).\label{ista:eq1.62}
\end{equation}
Here
\begin{eqnarray}
&&K_{\bf q}^{0}(\omega_{n})=\frac{1}{i\omega_{n}-\lambda};\quad
M_{\bf q}(\omega_{n})=\beta
b'\cos^{2}\vartheta\delta(\omega_{n}),\nonumber\\
&&\sin\vartheta=\Omega/\lambda;\quad
\langle\sigma^{z}\rangle_{0}=b=\frac12\tanh\frac{\beta\lambda}{2};\quad
b'=\frac{\partial b}{\partial(\beta\lambda)}.\label{ista:eq1.63}
\end{eqnarray}

The pseudospin Green's function $K_{\bf q}^{0}(\omega_{n})$ is
constructed of operators of the transverse pseudospin components
acting in the rotated reference system ($K_{\bf q}^{0}\sim\langle
T\sigma^{+}\sigma^{-}\rangle$;
$\sigma_{i}^{z}=S_{i}^{z}\cos\vartheta-S_{i}^{x}\sin\vartheta$;
$\sigma_{i}^{x}=S_{i}^{x}\cos\vartheta+S_{i}^{z}\sin\vartheta$) and
has a pole at $\lambda$ (the pseudospin reversal energy). The
semi-invariant $M_{\bf q}(\omega_{n})$, describing the correlation
of longitudinal pseudospin components ($\sim
T\langle\sigma^{z}\sigma^{z}\rangle^{c}$), is proportional to
$\delta(\omega)$. There is no such  contribution  in $\Sigma_{\bf
q}(\omega)$ when the Green's functions are calculated using the
equation of motion  and the decoupling procedure for Zubarev's
functions. Thus, the isothermal $(\chi_{T})$ and isolated
$(\chi_{I})$ susceptibilities do not coincide  for
PEM.\cite{ista:31}
It should be mentioned that a similar result is also obtained in the
cases $U\rightarrow\infty$\cite{ista:26} and
$t_{ij}\rightarrow0$.\cite{ista:25}

In the GRPA scheme, the summation of loop sequences leads to the
expression for pseudospin correlator
\begin{equation}
\langle TS^{z}S^{z}\rangle_{{\bf q},\omega}=-\frac{\Sigma_{\bf
q}(\omega)}{1-g^{2}\Sigma_{\bf q}(\omega) \Pi_{\bf
q}(\omega)},\label{ista:eq1.64}
\end{equation}
where
\begin{equation}
\Pi_{\bf q}(\omega)=\frac{2}{N}\sum_{\bf k}\frac{n(t_{\bf
k})-n(t_{{\bf k}-{\bf q}})}{\omega+t_{\bf k}-t_{{\bf k}-{\bf
q}}}\label{ista:eq1.65}
\end{equation}
is the standard electron loop contribution.

The condition  $\langle TS^{z}S^{z}\rangle_{{\bf
q},\omega=0}\rightarrow\infty$ (which corresponds to divergence of
isothermal susceptibility $\chi_{T}$) indicates an instability with
respect to transition into modulated (at $q\neq0$) or another
uniform (at $q=0$) phase. The thermodynamic parameter values, at
which $\chi_{T}\rightarrow \infty$, determine the spinodal
points\index{point!spinodal}.

The equation
        \begin{eqnarray}
&&\lambda^2+g^2\sin^2\theta\lambda\langle\sigma^z\rangle_0\Pi_{\bf
q} +\lambda^2 g^2 \beta b' \cos^2 \theta \Pi_{\bf
q}=0\label{ista:eq1.66}
       \end{eqnarray}
was solved together with equation (\ref{ista:eq1.45}) for the mean
values $\eta$ and $n$ written in MFA. The function $\Pi_{\bf q}$ was
calculated numerically by the direct momentum summation for square
lattice.

\begin{figure}[!h]
\centerline{\psfig{file=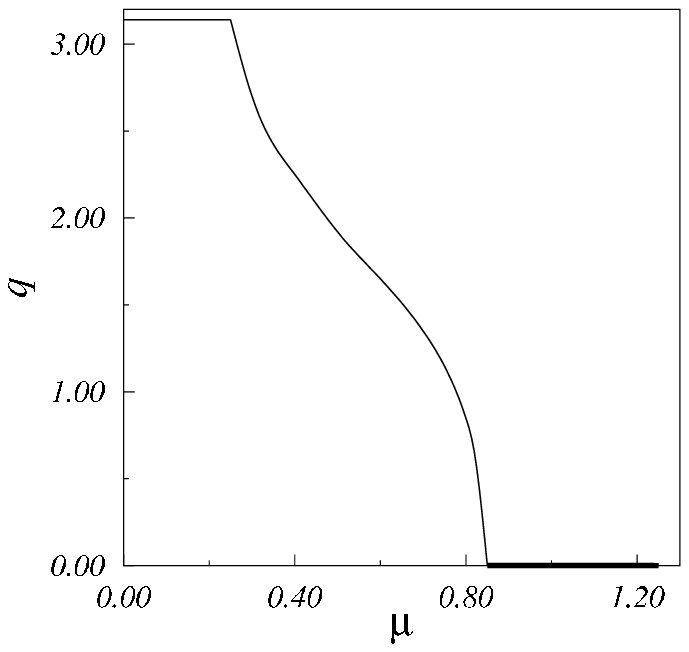,width=2.0in,angle=0} \quad
\psfig{file=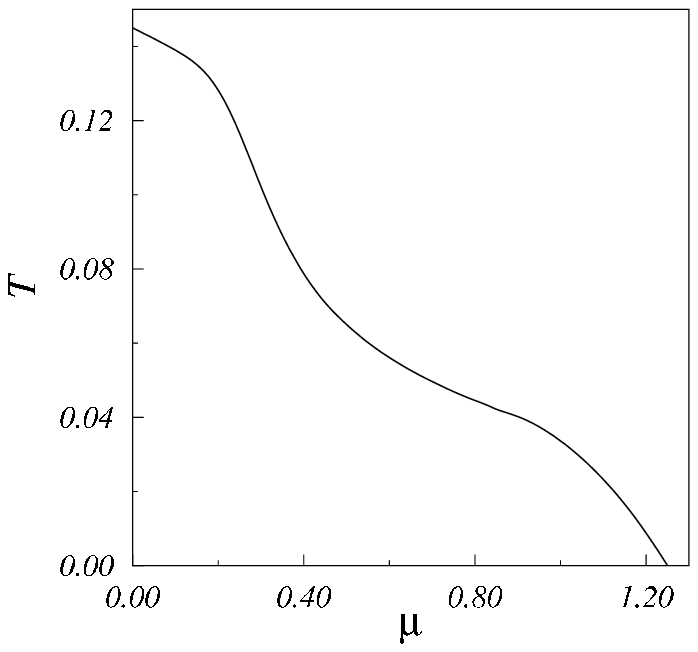,width=2.0in,angle=0} }
\vspace*{8pt}
 \caption{The dependence of the modulation wave vector {\bf q}=(q,q) and the
 temperature of absolute instability of high--temperature phase on the chemical potential,
$\Omega=0$, $g=0.5$.}
 \label{ista:f29}
 \end{figure}

At fixed values of  chemical potential, the critical point can be
 defined as an upper point of spinodal \index{spinodal} (on the $(T,h)$ plane) with the
 highest temperature depending on the wave vector ${\bf q}$ value.
 Fig.~\ref{ista:f29} shows the dependencies of
 the critical temperature
  and the corresponding
 wave vector on the chemical
 potential in the case $\Omega=0$ (only positive values of the chemical
 potential are  shown; at  $h=g$,  the picture is symmetrical
 with respect to the point $\mu=0$ which coincides
 with the centre of the energy band).
 We can see that the case
 ${\bf q}=(\pi,\pi)$
is realized when $|\mu|\lesssim 0.25$
 at chosen parameter values, which means that the
system can
 pass into the phase with doubly modulated lattice period. The case
${\bf q}=0$  (transition into the uniform phase)
 is realized when
 $ 0.85\lesssim|\mu|\le 1.25$.\footnote{$1.25=W+g/2$; this value corresponds to the upper edge
 of the band when $\langle S^z \rangle =\frac{1}{2}$.}
 The system undergoes  transition to the incommensurate phase \index{phase!incommensurate} at
 intermediate values of the chemical potential. The presence of
 tunneling splitting narrows the interval of values of $\mu$ at
 which the above mentioned transitions take place; at high enough
 values of $\Omega$, the transition into the chess-board phase
 occurs only.

 The given results at $\Omega=0$ generally
 correspond  to the
 picture of phase transitions in the FK model obtained
  in DMFT
in the case of weak coupling.\cite{ista:51,ista:57,ista:60}
It was shown for FK model
that at
 small values of $n$ the phase separation can be realized, while near
 half-filling of the band the chess-board phase is preferable and, finally,
 at intermediate values of $n$ the appearance of phase with an incommensurate
 modulation is possible. The transitions into one or another phase were observed
 by the divergences of corresponding susceptibilities.  It should be
 mentioned  that such a procedure did not enable the
 authors to reveal the thermodynamically stable states in
 the regions where the instabilities of the both types (at ${\bf q}=0$
 and ${\bf q}=(\pi/a,\pi/a,...)$) are superimposed; this problem can be
 solved based on the analysis of the behaviour of the grand canonical potential.

The GRPA scheme used here  to investigate the PEM is advantageous in
interpreting the dielectric susceptibility divergences due to the
explicit dependence of the $\chi_{T}({\bf q}, 0)$ function on the
wave vector.
 In the DMFT approach at $d\rightarrow\infty$ such a
dependence enters only through the function $ X({\bf
q})=\frac{1}{d}\sum\limits_{j=1}^{d}\cos q_j$ that
 leads to some difficulties in considering the  incommensurate ordering. Besides, in
 Refs.~[\refcite{ista:51,ista:57,ista:60}] there was used a regime of a fixed
concentration of localized particles; in the PEM this corresponds to
the regime $\langle S^z\rangle =\mathrm{const}$.
 In this case, the  authors came to a conclusion
 that the transition to the chess-board phase at $g<W$ is
always continuous and spinodals are the lines of phase
transitions.\cite{ista:55}
 In such a situation the separation into the uniform
and modulated phases would be impossible.
Contrary to that,  by analysing the behaviour of the grand canonical
potential we showed that the transition to the
 chess-board phase can be both of the  second order (the
spinodals
  are the phase transition lines) and of the first order (spinodals do
   not coincide with phase transition lines). Due to  the
    first order phase transition to the chess-board phase, the possibility
     of the phase separation into the uniform and the chess-board phases
  in the case of weak coupling
      was demonstrated. Such a possibility, as was shown in Ref.~[\refcite{ista:58}],
       see also Sec.~\ref{ista:p3.1.2}, exists in the case of large values of coupling
        constant.
         It should be mentioned that
        in Ref.~[\refcite{ista:55}] only the possibility
       of phase separation into
        different uniform phases was
         investigated.

Though the phase transitions in the PEM at weak coupling are similar
to the transitions revealed in this model
%
in the case of  strong interaction, $g\gg W$,\cite{ista:28,ista:29}
 the physical mechanisms of transitions are to a greater extent
 distinct. In the case of strong coupling, the electron spectrum is always
 split  due
 to the one-site interaction. The mechanism which ensures the
 advantage of the transition is connected with the different character of
 the electron spectrum reconstruction
in the subbands and with the corresponding
 redistribution of the electron density of states. At weak coupling, a new phase, which
 appears at the transition between
 uniform phases, is stabilized due to the shift of the electron band as a whole.
The phase with a double modulation appears due to energy gain at
 the splitting of the initial band at the Fermi level
 (the effect is similar to the Peierls instability at the interaction with
 phonons). Besides that, the dependencies of the critical temperatures on
 the coupling constant $g$ are different in the both cases:
   $T_c$ (or $T^{*}_c$) is proportional to $g^2$ at $g\ll W$, while at $g\gg
 W$ the critical temperatures decrease ($\sim \frac{1}{g}$) when $g$
 increases (such a type
  of  behaviour of $T^{*}_c$ for the FK model was obtained
 in Refs.~[\refcite{ista:45,ista:57}]).

\subsection{Superconductivity in the PEM}\label{ista:p3.3}

Due to the presence of intrinsic dynamics, the PEM at $\Omega\neq0$
also possesses an instability with respect to transition into
superconducting (SC) state. \index{state!superconducting (SC)}
%
%
Under certain conditions the transition to SC  state  will compete
with the transition to modulated phase (CDW). Among others, close
attention to this problem for an electron system interacting with
anharmonic structure units was paid in Ref.~[\refcite{ista:61}].
%
%
It was shown that, when there is no electron correlation, the
transition temperature to CDW, $T_{p}$, is higher than that to SC,
$T_{c}$. The authors, however, considered only the case where  the
local potential is symmetric and the electron filling  is close to a
half. The question concerning the appearance of superconductivity
for a wider range  of the model parameters and electron
concentrations has not been examined.
It was just the point of investigation performed in
Ref.~[\refcite{ista:32}] within the GRPA. To examine the possibility
of the appearance of SC phase we calculated the static
susceptibility $\chi^{SC}$ in the superconducting channel.
Taking into account the diagrams which correspond to ladder
approximation (with the parallel directions of lines of the fermion
Green's functions), we obtained the Bethe-Salpeter
equation\index{equation!Bethe-Salpeter} for the superconducting
   vertex part $\Gamma_{\omega_1,\omega_2}({\bf k}_1,{\bf k}_2)$\cite{ista:32}
   \begin{eqnarray}
   &&\Gamma=\Gamma^0+T\sum_{{\bf k}_3,\omega_3}\Gamma^0\chi^0\Gamma,\label{ista:eq1.67}
   \end{eqnarray}
    where $\chi^0_{\omega_1}({\bf k}_1)=\frac{1}{N}
     G^0_{{\bf k}_1}(\omega_1)G^0_{-{\bf k}_1}(-\omega_1)$;
     $\Gamma^0=g^2\langle TS^zS^z\rangle_{{\bf k}_2-{\bf k}_1,\omega_1-
     \omega_2}$ is given by the expression  (\ref{ista:eq1.64}).
The susceptibility
\begin{equation}
\chi^{SC}=\frac{1}{N}\sum_{{\bf k},{\bf
q}}\int\limits_{0}^{\beta}\langle T_{\tau}a_{\downarrow {\bf
k}}(\tau)a_{\uparrow-{\bf k}}(\tau)a_{\uparrow-{\bf
q}}^{+}a_{\downarrow {\bf q}}^{+}\rangle {\rm e}^{i
\omega_{n}\tau}{\rm d}\tau \label{ista:eq1.68}
\end{equation}
is connected with the vertex part $\Gamma$ in the following way
\begin{equation}
\frac{1}{T}\chi^{SC}=\sum_{\omega,{\bf k}}\chi_{\omega}^{0}({\bf
k})+T\sum_{{\bf k}_{1},{\bf k}_{2},\omega_{1},\omega_{2}}
\chi_{\omega_{1}}^{0}({\bf
k}_{1})\Gamma_{\omega_{1},\omega_{2}}({\bf k}_{1},{\bf
k}_{2})\chi_{\omega_{2}}^{0} ({\bf k}_{2})\, .\label{ista:eq1.69}
\end{equation}
Both  approximations, the ladder one for the
$\Gamma_{\omega_{1},\omega_{2}}({\bf k}_{1},{\bf k}_{2})$ vertex
construction and the chain (GRPA) one for the construction of
$\Gamma^{0}_{\omega_{1},\omega_{2}}({\bf k}_{1},{\bf k}_{2})$, were
employed by analogy with what was done in the $t-J$ model for the SC
description.\cite{ista:62} The  mentioned approximations correspond
to the known Migdal-Eliashberg
(ME)\index{approximation!Migdal-Eliashberg} one, which is usually
used in considering the electron-phonon systems, particularly in the
Holstein model.  Considering the phonon frequencies to be small in
comparison with the transfer integral,\cite{ista:63} one gets
results in qualitative agreement with the quantum Monte Carlo
simulations.\cite{ista:22} Since the PEM is similar to the Holstein
model, and may be considered as a double-level approximation of the
latter one, the ME approximation is expected to be satisfactory in
the limits of non-half  filling, low temperatures, and small
$\Omega$.

A way to find the SC transition temperature is to determine a
temperature, at which the susceptibility in the superconducting
channel diverges. It corresponds to the condition when the
scattering matrix
\begin{equation}
T_{\omega_{1}\omega_{2}}({\bf k}_{1},{\bf
k}_{2})=T\chi_{\omega_{1}}^{0}({\bf
k}_{1})\Gamma^{0}_{\omega_{1}\omega_{2}}({\bf k}_{1},{\bf k}_{2})
\label{ista:eq1.70}
\end{equation}
has an eigenvalue which is equal to unity.\cite{ista:63,ista:64} In
calculations performed in Ref.~[\refcite{ista:32}], the
approximation $\Gamma^{0}\approx-g^{2}\Sigma$, similar  to the
non-renormalized ME approximation in the Holstein
model,\cite{ista:63} was used. This can be done in the
high-temperature phase, when the system
 has not still passed to the CDW state. In such a case the
unperturbed vertex part does not depend on the wave vector and thus
the unit eigenvalue of the matrix
\begin{equation}
\widetilde{T}_{\omega_{1}\omega_{2}}=T\sum_{{\bf
k}_{1}}\chi_{\omega_{2}}^{0}({\bf k}_{1})
\Gamma_{\omega_{1}\omega_{2}}^{0}\label{ista:eq1.71}
\end{equation}
should be found at first. Of all the temperatures within the $(T,
h)$ plane which satisfy this condition, the highest one is chosen as
the critical SC transition temperature.

\begin{figure}[!h]
\centerline{\psfig{file=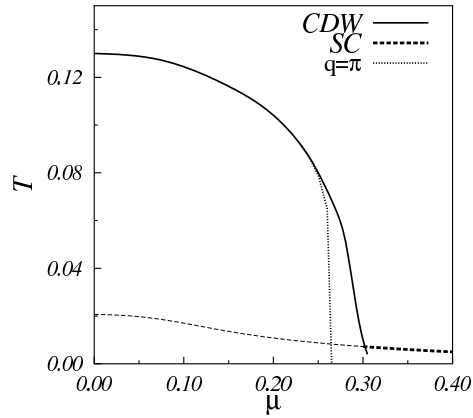,width=2.4in,angle=0}}
\vspace*{8pt}
 \caption{The temperature of the absolute instability of the high--temperature phase with respect
 to the transitions to CDW (solid line) and SC (bold dashed line) as functions of the chemical
 potential, for $\Omega=0.2$ and $g=0.5$. Dotted line shows the
 temperature of the transition to the phase with the modulation
 wave vector
  ${\bf q}=(\pi,\pi)$.}
 \label{ista:f30}
 \end{figure}

Numerical calculations performed in Ref.~[\refcite{ista:32}] show
that the SC transition in the simplified PEM is possible (in the
weak coupling case) at the electron concentrations away from
half-filling (when the chemical potential is placed near the
electron band edges) and outside  the region of the above described
transitions with the modulation of the electron and pseudospin
density (see Fig.~\ref{ista:f30}). Such a picture is similar to that
obtained within the DMFT for the Holstein model\cite{ista:63} (but,
in the latter case, the incommensurate phase does not appear at the
intermediate values of $\mu$). This is also consistent  with the
results obtained by quantum Monte-Carlo
simulations.\cite{ista:22,Su03} These papers just established that
SC could appear at low temperatures; the transition to
incommensurate CDW was not studied.


The estimates performed in Ref.~[\refcite{ista:32}] show that at
$W\approx0.5$ eV and $g\approx0.25$~eV, the maximum value of the SC
transition temperature is  $T_{max}^{SC}\approx10-40$ K. This agrees
with the conclusions obtained for PEM in Ref.~[\refcite{ista:61}],
where calculations were performed by analogy with the scheme used
for the Eliashberg equations in the limit of a weak electron-phonon
interaction, provided that the renormalization of the pseudospin
excitation energy is neglected. According to the estimations carried
out in Ref.~[\refcite{ista:65}], if there were no CDW, the SC
transition temperature would be $T^{SC}\approx40$ K at the band
filling close to a half.

%
%
The question of what  occurs in the systems as the tunneling
splitting \index{tunneling splitting} frequency  grows further is
also of great interest. It was shown  in Ref.~[\refcite{ista:30}]
that, at $\mu=0$, the critical temperature  of   transition to the
phase with double lattice period modulation, $T_{c}^{\rm CDW}$,
decreases with the increase in $\Omega$ according  to the
exponential law: $T_{c}^{\rm
CDW}\sim\exp\left(-\frac{\pi\sqrt{\Omega W}}{g}\right)$, see
(\ref{ista:eq1.59}). It follows from the analysis  of the behaviour
of the $\tilde{T}_{\omega_{1}\omega_{2}}$ matrix
elements\cite{ista:32} that the SC transition temperature,
$T_{c}^{\rm SC}$, changes with $\Omega$ in a similar way. In this
case, for $\mu=0,\,\, T_{c}^{\rm SC}$
$(\Omega\rightarrow\infty)\approx T_{c}^{\rm
CDW}\,\,(\Omega\rightarrow\infty)$. As  is seen  from
Fig.~\ref{ista:f30}, with the increase in $|\mu|$, the CDW
transition temperature falls more  rapidly  than  the SC one. Thus,
for the non-zero values of the chemical potential, provided that
$\Omega$ is sufficiently large, the SC transition temperature is
expected to be higher than the CDW one, and there  will only be the
transition to SC. However, as  was  noted  above, to make the
correct analysis  of the  competition  between these  transitions,
the renormalized $\Gamma^{0}$ vertex should be used  when  solving
Eq. (\ref{ista:eq1.70}) and  determining  $T_{\rm SC}$. Moreover,
when $\Omega$ values  are  sufficiently  large,  the applicability
of the approximation in deriving this  equation
 turns out to be
 unjustified.

The mechanism that  leads in the PEM to SC, which we have considered
here, as well as the traditional  phonon one, does not result in
high values of $T_{c}$, and apparently it does not explain the HTSC
phenomenon.
\section{Thermodynamics of PEM at Finite $U$ Values; the $U\rightarrow\infty$ Limit}\label{ista:p4}

The analysis similar to the one given above was performed in
Refs.~[\refcite{ista:24}--\refcite{ista:26},\refcite{ista:66}--\refcite{ista:69}]
for the PEM with $U\neq0$. The presence of the electron-electron
on-site interaction leads to some differences in the behaviour  of
the model with respect to the case of FK model (even if $\Omega=0$).
The consideration was based on expansions in terms of electron
transfer $t_{ij}$, as  in the strong coupling limit $g\gg W$.

The single-site Hamiltonian
\begin{equation}
H_{i}=Un_{i\uparrow}n_{i\downarrow}-\mu(n_{i\uparrow}+n_{i\downarrow})+
g(n_{i\uparrow}+n_{i\downarrow})S_{i}^{z}-hS_{i}^{z}-\Omega
S_{i}^{x}\label{ista:eq3.1}
\end{equation}
that includes the $U$-term can be reduced to the diagonal form using
the rotation transformation
$|R\rangle=\alpha_{Rr}^{(\varphi_{r})}|r\rangle$,\cite{ista:26,ista:66}
where $|R\rangle=|n_{i\uparrow},n_{i\downarrow}, S_{i}^{z}\rangle$
is the single-site basis of states
\begin{eqnarray}
|1\rangle=|0,0,1/2\rangle&\qquad\qquad\qquad\qquad&|\tilde{1}\rangle=|0,0,-1/2\rangle%
\nonumber\\%
|2\rangle=|1,1,1/2\rangle&\qquad\qquad\qquad\qquad&|\tilde{2}\rangle=|1,1,-1/2\rangle
\nonumber\\%
|3\rangle=|0,1,1/2\rangle&\qquad\qquad\qquad\qquad&|\tilde{3}\rangle=|0,1,-1/2\rangle
\nonumber\\%
|4\rangle=|1,0,1/2\rangle&\qquad\qquad\qquad\qquad&|\tilde{4}\rangle=|1,0,-1/2\rangle,%
\label{ista:eq3.2}
\end{eqnarray}
and
\begin{equation}
\cos\varphi_{r}=\frac{n_{r}g-h}{\sqrt{(n_{r}g-h)^{2}+\Omega^{2}}}\,
. \label{ista:eq3.3}
\end{equation}
In terms  of Hubbard operators $X^{rs}=|r\rangle \langle s|$, acting
on the new basis, the transformed Hamiltonian of the model is

\begin{equation}
H=\sum_{ir}\lambda_{r}X_{i}^{rr}+\sum_{i_{j}\sigma}t_{ij}a_{i\sigma}^{+}a_{j\sigma}
\label{ista:eq3.4}
\end{equation}
with
\[
\lambda_{r_{i}\tilde{r}}=U\delta_{r,r}+E_{0}n_{r}\pm\frac12\sqrt{(n_{r}g-h)^{2}+\Omega^{2}}
\]
and
\begin{equation}
a_{i\sigma}^{+}=\sum_{mn} A_{mn}^{\sigma}X_{i}^{mn},\quad
a_{i\sigma}= \sum_{mn}A_{mn}^{\sigma}X_{i}^{nm}\, .
\label{ista:eq3.5}
\end{equation}
Here $n_{1}=0$, $n_{2}=2$, $n_{3}=n_{4}=1$ $(n_{\tilde{r}}=n_{r})$;
expressions for $A_{mn}^{\sigma}$ are given in
Ref.~[\refcite{ista:26}].

\begin{figure}
\centerline{\psfig{file=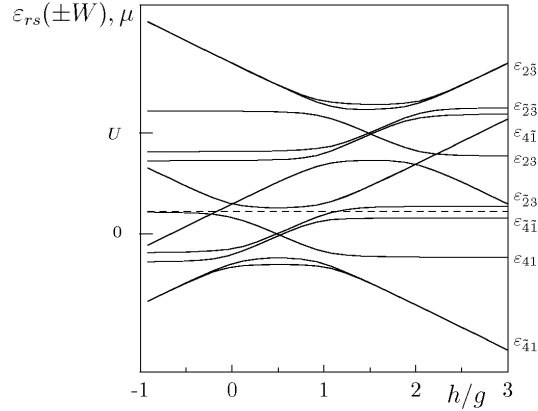,width=3.1in,angle=0}}
\vspace*{8pt}
\hfill%
\caption{Single-electron spectrum depending on $h/g$ in the
$\mu=$const regime; $U=2.2$, $g=1$, $h=0.7$, $\Omega=0.3$, $W=0.2$,
$\mu=0.5$.} \label{ista:f31}
\end{figure}

The detailed consideration of single--electron spectrum of model
(\ref{ista:eq0}) at $U\neq0$ was performed in
Refs.~[\refcite{ista:23,ista:67}] in the Hubbard--I approximation.
The interaction with  the anharmonic (pseudospin) mode  splits the
energy levels of ordinary Hubbard  model $0$, $E_0$ and  $2E_0 +U$
into sublevels $\lambda_{r,\tilde{r}}$. As a result, each Hubbard
single--electron band splits into four subbands. In the independent
subband approximation
\begin{equation}
\varepsilon_{rs}({\bf q})=\lambda_r-\lambda_s+
    t_{{\bf q}}\: (A_{rs}^\sigma)^2\langle X^{rr}+X^{ss}\rangle,
    \label{ista:eq3.6}
\end{equation}
where $(rs)=(41)$, $(\tilde{4}\tilde{1})$, $(\tilde{4}1)$,
$(4\tilde{1})$, $(23)$, $(\tilde{2}\tilde{3})$, $(\tilde{2}3)$,
$(2\tilde{3})$ for $\sigma=\uparrow$  and substitution
$4\leftrightarrow 3$ should be done for  $\sigma=\downarrow$. The
widths  and  statistical weights of the subbands are determined by
 parameters $A_{rs}^\sigma$.
The four lower and four higher bands correspond to the hole and
electron pair motion, respectively. Their positions depend on the
value of the asymmetry field $h$ (see Fig.~\ref{ista:f31}, where the
band edges as functions of $h$ are shown at the fixed value of
$\mu$).

Thermodynamics and dielectric properties of PEM with $U\neq0$ were
studied in Refs.~[\refcite{ista:26,ista:68}]. Being interested in
the problem of lattice instabilities in high-$T_{c}$
superconduc\-tors of YBaCuO - type we calculated the transverse
dielectric susceptibility (corresponding to the $\varepsilon_{zz}$
component for YBaCuO - structure), using, as above (see
Sec.~\ref{ista:p3.2.3}), the expression
$P_{i}=d_{s}S_{i}^{z}+d_{e}n_{i}$ for local dipole moment.  We can
respectively
separate  ion,  electron  and mixed components in the total
susceptibility
\begin{eqnarray}
\chi_\perp({\bf q},\omega_n) &=& d_S^2 \chi^{SS}({\bf q},\omega_n)
+ d_e^2 \chi^{nn}({\bf q},\omega_n)\nonumber\\
 &+& d_S d_e \left (
\chi^{Sn}({\bf q},\omega_n) + \chi^{nS}({\bf q},\omega_n)\right ),
    \label{ista:eq3.7}
\end{eqnarray}
where $\chi^{AA'}=K^{AA'}$ in the regime $\mu=\rm const$ (when we
fix the value of the chemical potential and   permit the charge
redistribution between conducting sheets Cu$_{2}$O$_2$ and other
structural elements, having in mind an application of the model to
the YBaCuO-type  crystals) and
\begin{equation}
\chi^{AA'}=K^{AA'}-\frac{K^{An}K^{nA'}}{K^{nn}}
   \label{ista:eq3.8}
\end{equation}
in the regime $n=\rm const$ (when we fix the electron concentration
$n$ in the conducting sheets Cu$_{2}$O$_2$).\cite{ista:26} Here
$K^{AA'}({\bf q},\omega_n)$ are Fourier--transforms of
semi--invariant Matsubara's Green's functions
\begin{equation}
K_{lm}^{AA'}(\tau-\tau') = \langle
T\tilde{A}(\tau)\tilde{A}^{\prime}_m(\tau')\rangle^c
\label{ista:eq3.9}
\end{equation}
constructed of the operators $S^z_i$, $n_i$.

In the case of zero hopping ($t_{ij}=0$) the exact expressions for
correlation functions $\chi^{ss}(\omega_{n})$,
$\chi^{nn}(\omega_{n})$, $\chi^{ns}(\omega_{n})$ can be  easily
obtained.\cite{ista:25}

\begin{figure}
\centerline{\psfig{file=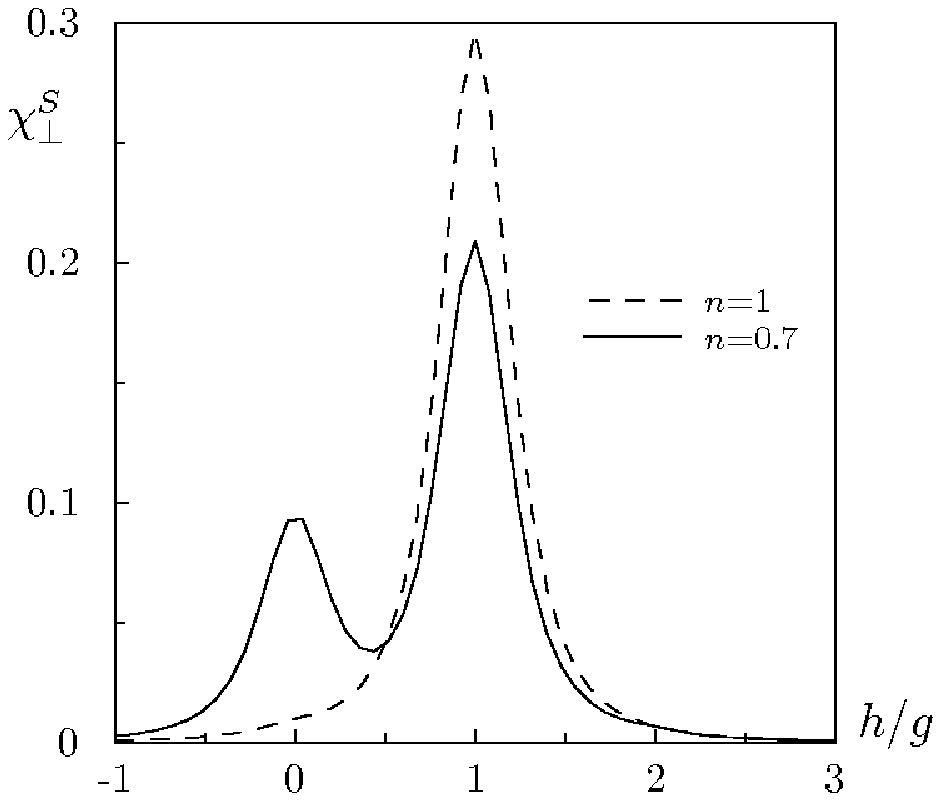,width=2.4in,angle=0}
\psfig{file=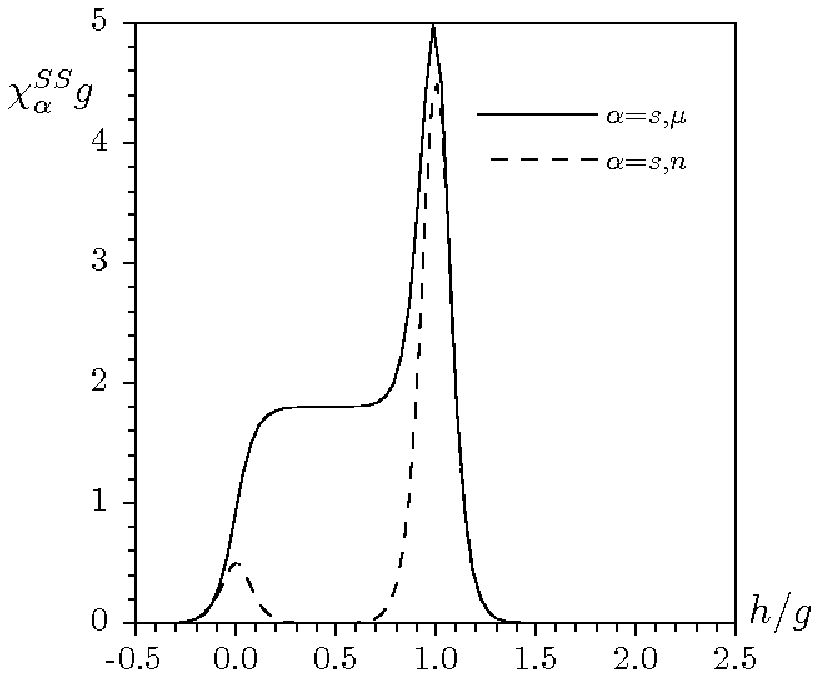,width=2.1in,angle=0}
  }
  \centerline{\quad(a) \hspace{5.2cm}(b)}
 \vspace*{8pt}
\caption{(a): Dielectric susceptibility as a function of $h/g$ at
different electron concentrations;
 $U=2.2$, $g=1$, $h=0.7$, $\Omega=0.3$, $W=0.2$, $d_{S}=0.6$.
(b): Pseudospin component of susceptibility $\chi_{\alpha}^{ss}g$
$vs$ $h/g$ for $\Omega=0$ ($\omega_{n}=0$, $n=0.9$, $T/g=0.05$).
  }
  \label{ista:f32}
  \end{figure}

For $n$=const regime  the  main contribution into susceptibility is
produced by the pseudospin subsystem. The susceptibility
\begin{equation}
X^{ss}\cong
\frac{ds^{r}}{vc}\sum_{r=1}^{4}\frac{\Omega^{2}}{[(n_{r}g-h)^{2}+\Omega^{2}]^{3/2}}\langle
X^{\tilde{r}\tilde{r}}-X^{rr}\rangle \label{ista:eq3.10}
\end{equation}
as a function of $h$ generally possesses three peaks. The maxima at
$h=0$, $g$ and $2g$  correspond to the  points of possible
dielectric instabilities (that can appear due to the proximity of
corresponding energy subbands). The intensities of peaks are
redistributed with the change of $n$; this is illustrated in
Fig.~\ref{ista:f32}a for the case $U/W\gg1$ ($\Omega=0.3 g$,
$0<n<1$), when only two peaks are present;  the first peak
disappears at $n\rightarrow1$ and the second one disappears at
$n\rightarrow0$. The effect remains the same at
$t_{ij}\rightarrow0$.

In the $\mu=$const regime there are three regions of $h$ values
($h<0$, $0<h<g$ and $h>g$) with different field and temperature
behaviour of susceptibility (see Fig.~\ref{ista:f32}b). This is a
result of the difference in the direction and value of the total
effective field $h-n_i g$ acting on  the pseudospin at site $i$ when
electron occupation on this site changes ($n_i=0$ or $1$), which
testifies to a strong correlation between electron and pseudospin
subsystem.
%

The behaviour of  transverse dielectric susceptibility
\index{dielectric!susceptibility} $\chi^{ss}$ as a function of the
field $h$ and temperature becomes more complicated when we take into
account the electron transfer.
The calculations of components of susceptibility (\ref{ista:eq3.7})
were performed\cite{ista:25,ista:26} at $t_{ij}\neq0$ in the
particular case of no tunneling splitting in anharmonic potential
well ($\Omega=0$).

Green's  functions (\ref{ista:eq3.9}), constructed of the operators
\begin{equation}
n_{i\sigma}=\sum_{r=1}^{4}n_{r}(X_{i}^{rr}+X_{i}^{\tilde{r}\tilde{r}}),
\qquad S_{i}^{z}=\frac12
\sum_{r=1}^{4}n_{r}(X_{i}^{rr}-X_{i}^{\tilde{r}\tilde{r}}),\label{ista:eq3.11}
\end{equation}
can be expressed in terms of functions
\begin{equation}
K_{lm}^{(pq)}(\tau-\tau')=\langle
TX_{l}^{pp}(\tau)X_{m}^{qq}(\tau')\hat{\sigma}(\beta)\rangle_{oc}.\label{ista:eq3.12}
\end{equation}
The perturbation theory with respect to electron hopping term
$t_{ij}$ and the corresponding diagrammatic technique for Hubbard
operators\cite{ista:52} were used to calculate these functions. The
diagrammatic series were summed up in the GRPA, where the sequences
of electron ``loops'' connected by semiinvariants (similarly to the
case of the simplified PEM, see Sec.~\ref{ista:p3.1.1}) or by
vertices with the three fermion lines (such vertices appear only at
$U\neq0$) are taken into account.

The  Fourier  transforms $K^{(pr)}({\bf q},{\omega}_n)$ of functions
(\ref{ista:eq3.9}) can be presented in the
form\cite{ista:25,ista:26}
\begin{eqnarray}
 & K^{pp'}\left ( {\omega}_n, {\bf q}\right ) = \beta\left \{
      \bar{\bar{b}}_{pp'}({\bf q}) +
      \left [ {}^\prime\Pi(0,{\bf q}) \bar{\bar{b}}({\bf q}) \right ]_{pp'}\! +
      \left [ \bar{\bar{b}}({\bf q}) \Pi^\prime(0,{\bf q}) \right ]_{pp'}\!
      \right . &\nonumber
 \label{ista:eq3.13}\\
 & \left .
    +  \left [ {}^\prime\Pi(0,{\bf q}) \bar{\bar{b}}({\bf q})
\Pi^\prime(0,{\bf q})\right ]_{pp'}\right \}\delta\left
({\omega}_n\right ) +
      \Pi_{pp'}^{\prime\prime}\left ( {\omega}_n,{\bf q} \right ) .&
\end{eqnarray}
The first term in the right hand side of (\ref{ista:eq3.13}) is the
``full'' semi--invariant of the  second order that satisfies the
Dyson--type equation
\begin{equation}
        \bar{\bar{b}}_{pq} = \tilde{b}_{pq} +
                \left ( \tilde{b}\Pi\bar{\bar{b}}\right )_{pq}.
  \label{ista:eq3.14}
\end{equation}
Here $\tilde{b}_{pq} = \langle X^{pp}X^{qq}\rangle_{0c}$ is the
second--order semi--invariant calculated in mean--field
approximation and the full ``loop'' contributions $\Pi$,
$\Pi^\prime$, ${}^\prime\Pi$, $\Pi^{\prime\prime}$ are determined
from the Bethe--Salpeter type equations
\begin{eqnarray} \Pi^{\prime\prime}
& = & \Pi^{\prime\prime}_0 + \Pi^{\prime\prime}_0\, \Pi^\prime
        + {}^\prime\Pi_0\, \Pi^{\prime\prime},
\nonumber \\
\Pi^\prime & = & \Pi^\prime_0 + \Pi^\prime_0\, \Pi^\prime
        + \Pi_0\, \Pi^{\prime\prime},
\nonumber \\
{}^\prime\Pi & = & {}^\prime\Pi_0 + {}^\prime\Pi_0\, {}^\prime\Pi
        + \Pi_0\, \Pi^{\prime\prime},\nonumber
\label{ista:eq3.15} \\ \Pi & = & \Pi_0 + \Pi_0\, {}^\prime\Pi +
\Pi^\prime_0\, \Pi \, .
 \end{eqnarray}

Zero--order polarization loops $\Pi_0$, $\Pi^\prime_0$,
${}^\prime\Pi_0$, $\Pi^{\prime\prime}_0$ are constructed of the
single--electron Green's functions which are calculated in the
Hubbard--I approximation (corresponding to the summation of chain
fragments of diagrams).\cite{ista:25,ista:26} Loops $\Pi^\prime_0$,
${}^\prime\Pi_0$, $\Pi^{\prime\prime}_0$ are determined only by
intraband transitions whereas loop $\Pi_0$ is determined by the
interband transitions as well
\begin{equation}
\Pi_0(mr,np) = \frac{1}{N} \sum\limits_{\bf k} t_{\bf k} t_{{\bf
k}+{\bf q}} \frac{n_+\left [ \varepsilon_{mr}\left ( {\bf k}\right )
\right ] -
             n_+\left [ \varepsilon_{np}\left ( {\bf k} + {\bf q}\right ) \right ] }
        {i{\omega}_n + \varepsilon_{mr}\left ( {\bf k}\right ) -
                        \varepsilon_{np}\left ( {\bf k} + {\bf q}\right ) }.
 \label{ista:eq3.16}
\end{equation}
where $\varepsilon_{mr}({\bf k})$ are determined in
(\ref{ista:eq3.6}) and $n_+(\lambda)$ is Fermi distribution.
%

\begin{figure}
\centerline{\psfig{file=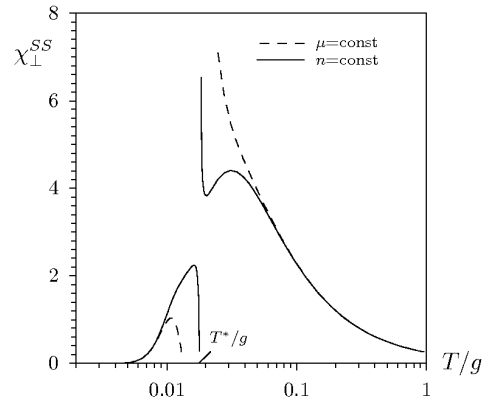,width=2.8in,angle=0}}
\vspace*{8pt}
  \caption{Temperature dependence of $\chi^{ss}$ at $n=0.95$ and ${\bf q}=0$
  ($U\rightarrow\infty$, $h/g=1.05$, $W/g=0.2$ ).}
  \label{ista:f34}
  \end{figure}

\begin{figure}
\begin{picture}(150,140)
\put(-20,00){ \psfig{file=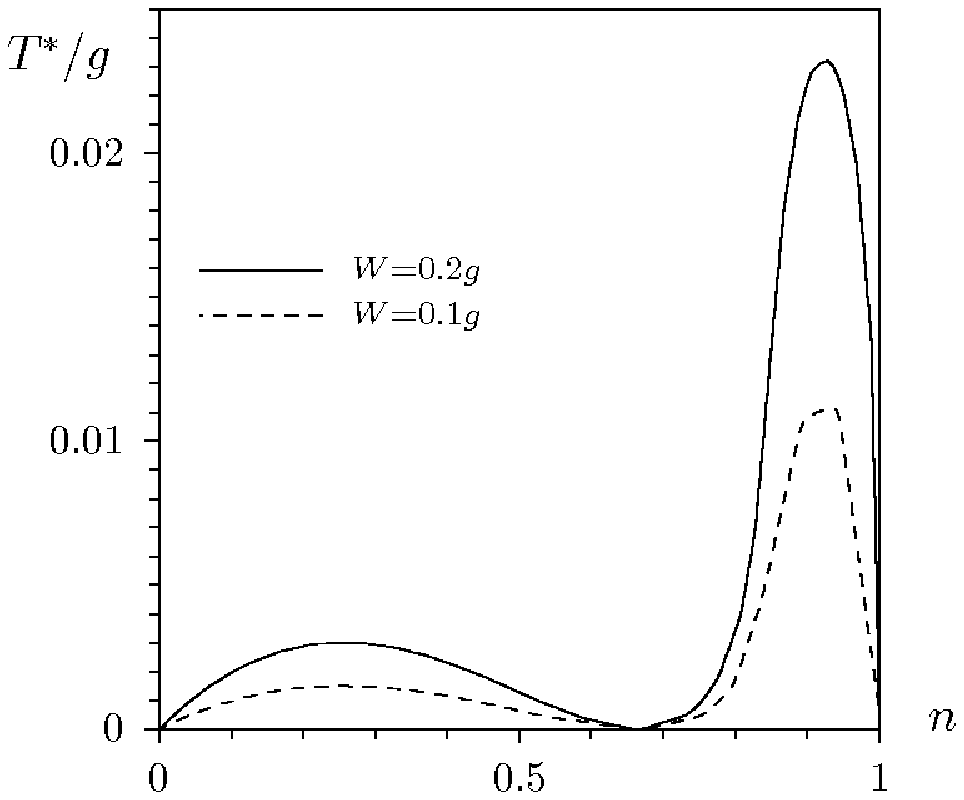,width=2.4in,angle=0}}
\put(150,10){ \psfig{file=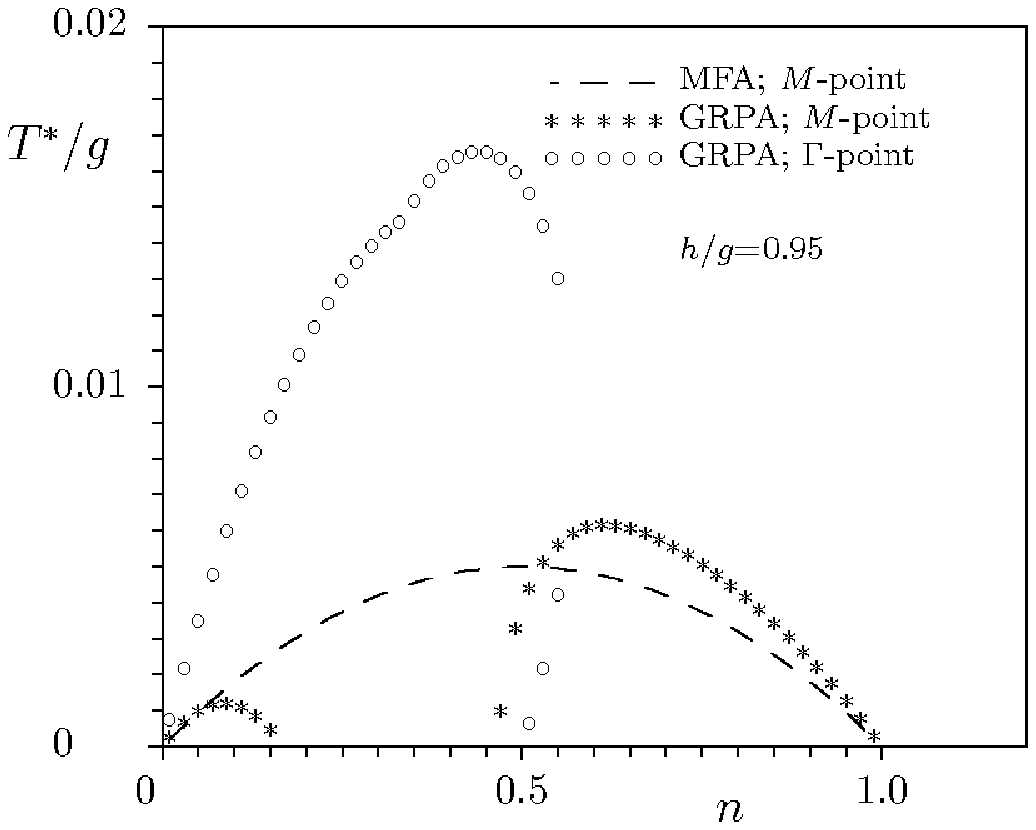,width=2.5in,angle=0}}
\end{picture}
  \centerline{\quad(a) \hspace{5.2cm}(b)}
  \caption{Dependence of the instability temperature $T^{*}$ on the electron concentration $n$
  in the cases $h>g$ and $0<h<g$.
  (a)\,\, $U\rightarrow\infty$, $\Omega=0$, $n/g=1.05$ (dielectric instability, ${\bf
  q}=0$).
  {(b)}\,\,$U\rightarrow\infty$, $\Omega=0$, $n/g=0.95$ ($\circ\circ\circ$ -- dielectric instability, ${\bf q}=0$;
  $***$-- instability with respect to CDW ordering, ${\bf q}=(\pi,\pi)$; dashed line
  shows the MFA results).}
  \label{ista:f35}
  \end{figure}

Numerical calculations of the static dielectric susceptibility
$\chi_{\bot}({\bf q},0)$ performed in
Refs.~[\refcite{ista:25,ista:68}] along the $(0,0)\div (\pi, \pi)$
line in the $2D$ Brilloine zone, revealed that, similarly to the
above considered  case $U=0$, the essential feature  of the model is
the presence of divergences on the temperature dependencies  of
functions $\chi_{n}^{ss}$ and $\chi_{\mu}^{ss}$
(Fig.~\ref{ista:f34}). They appear in a certain range of the model
parameter values. For $h>0$
and $h<g$ such divergencies  exist  only  at ${\bf q}=0$
($\Gamma$--point) and can  be  treated as the manifestation of the
dielectric type instabilities which appear in the pseudospin
subsystem (i.e., the system of anharmonic oscillators) under the
influence of the effective interactions. As in the case of a
simplified model, it means that the  system can  transform to
another uniform phase.
The corresponding phase diagram $T^*$  vs $n$ is  shown  in
Fig.~\ref{ista:f35}(a), where the lines limiting the stability
region are plotted. For $0<h<g$, besides the dielectric instability
\index{dielectric!instability} at $\Gamma$-point at $n\gtrsim0$, the
instability at ${\bf q}=(\frac{\pi}{a},\frac{\pi}{a})$ ($M$--point)
with respect to the charge ordering (double modulation) occurs
(Fig.~\ref{ista:f35}(b)).
Such  an instability is realized when the electron band (that
corresponds to the lower Hubbard subband at $U\rightarrow\infty$) is
nearly fully occupied.

The above mentioned divergences on the temperature  dependencies and
an increase of the dielectric susceptibility which  takes place in
the vicinity of $h=0$  and  $h=g$  values  (for $U\to\infty$, $n<1$)
are  connected,  first  of  all,  with  the polarization
contributions $\Pi_0$ from the electron transitions between  hole
subbands ($\tilde{4}\tilde{1}$) and ($\tilde{4} 1$) (or
($\tilde{4}\tilde{1}$) and ($4\tilde{1}$)). Just for $h\sim 0$ and
$h\sim g$ these subbands come nearer to  each other.
%

%
It should  be noted that the  results presented here  for dielectric
response \index{dielectric!response} of PEM in the $U\to\infty$
limit were used in Ref.~[\refcite{ista:25}]  in describing the
dielectric anomalies which have been observed in the high-$T_{c}$
superconductors of the YBaCuO type  at an early stage of
investigations. In particular, the temperature dependence of
$\chi_{n}^{ss}$ given in Fig.~\ref{ista:f34} was related to  the one
obtained experimentally for
YBa$_{2}$Cu$_{3}$O$_{7-\delta}$\cite{ista:9} where the similar
behaviour of dielectric permittivity $\varepsilon_{c}$ was found at
the  temperature near (and above) the point of the superconducting
phase  transition.

\begin{figure}
\begin{picture}(150,80)
\put(-20,00){ \psfig{file=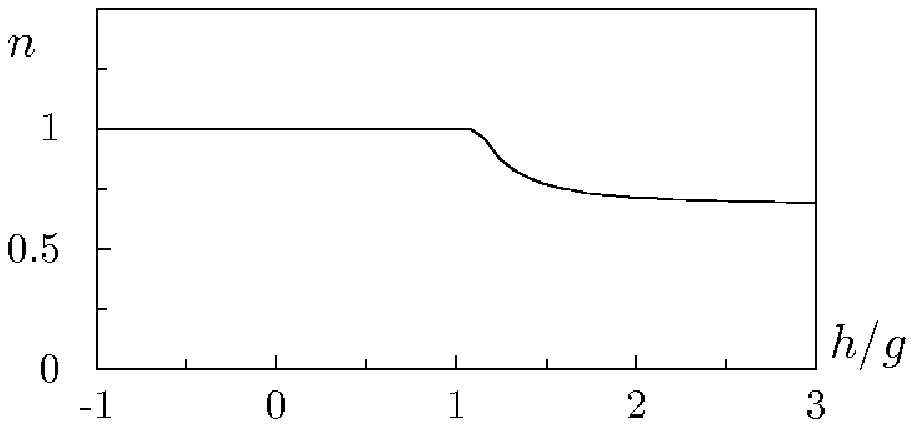,width=2.4in,angle=0}}
\put(150,00){ \psfig{file=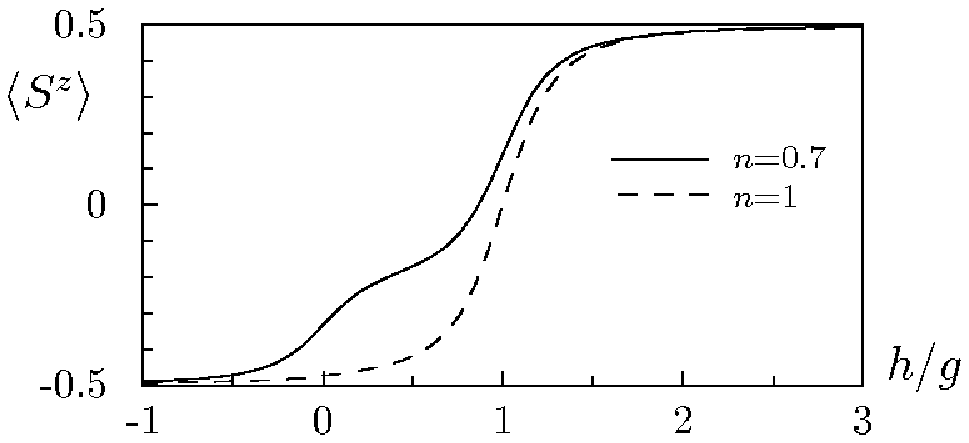,width=2.5in,angle=0}}
\end{picture}
  \centerline{\quad(a) \hspace{5.2cm}(b)}
\caption{Mean electron concentration $n$ (a) and pseudospin mean
value (b) as a function of  $h/g$ in the $\mu$=const regime.}
  \label{ista:f36}
  \end{figure}

An interesting feature of the model at large values of $U$ is the
possibility of change of the electron concentration $n$ under the
influence of $h$. Such an effect can take place in the $\mu$=const
regime; this becomes possible when the chemical potential leaves (or
enters) the energy subband\cite{ista:24} (Figs. \ref{ista:f31} and
\ref{ista:f36}(a)). The change  of $n$ is accompanied by the
corresponding change of the pseudospin mean value
(Fig.~\ref{ista:f36}(b)). In the case of YBaCuO structure, this
means that if $\mu$ moves inside the subband with the change of e.g.
external electric field, then the electron concentration (average
occupancy of states) in Cu$_{2}$O$_{2}$ layers changes. This causes
a redistribution of O$_{4}$ ions in their equilibrium positions. All
this may correspond to the so-called electric-field effect observed
in HTSC compounds (see, for example, Ref.~[\refcite{ista:70}]).

The obtained results show that in the $U\rightarrow\infty$ limit the
PEM also exhibits a tendency to transform into  another uniform or
doubly modulated  phase at low temperatures. It resembles the
behaviour  of the FK model in the strong coupling
case.\cite{ista:51} The plots  in Fig.~\ref{ista:f35} are given
without the effect of phase separation. When the segregation on the
regions with different $n$ values takes place, the corresponding
uniform phases or both the uniform and the modulated ones will
coexist.

\section{Two-Sublattice Pseudospin-Electron Model}\label{ista:p5}

Two-sublattice PEM\index{model!pseudospin-electron
(PEM)!two-sublattice}\cite{ista:33,ista:34,ista:35,ista:36,ista:69}
appeared as a generalization of the usual PEM with the aim of more
realistic description of
the anharmonic  subsystem of the apex oxygen ions in the YBaCaO type
structures. In this case the pseudospin energy in the internal field
is of the form $h\sum_{i}(S_{i1}^{z}-S_{i2}^{z})$ that is the
reflection of the mirror  symmetry of  the problem. Hamiltonian of
the model is of the form\cite{ista:33}
%
%
\begin{eqnarray}
\label{ista:eq3.17}
&&\hat{H}=H_{e}+H_{s}+h_{l-s}+H_{s-s},\nonumber\\
&&H=H_e+H_s+H_{e-s}+H_{s-s}, \\
&&H_e=-\mu\sum_{n,s}(n_{n1}^s+n_{n2}^s) + U\sum_{n}(n_{n1}^\uparrow
n_{n1}^\downarrow + n_{n2}^\uparrow
n_{n2}^\downarrow) \nonumber \\
&&+\sum_{ij}\sum_{s\alpha}t_{ij}a_{is\alpha}^{+}a_{js\alpha},
\nonumber\\ &&H_s= -h\sum_n (S_{n1}^z-S_{n2}^z) - \Omega\sum_n
(S_{n1}^x+S_{n2}^x), \nonumber\\ &&H_{e-s}= g \sum_{n,s}
(n_{n1}^sS_{n1}^z-n_{n2}^sS_{n2}^z), \nonumber\\ &&H_{s-s}= -J
\sum_n S_{n1}^z S_{n2}^z - \frac{1}{2}\sum_{n,n'}\sum_{\alpha,\beta}
J_{nn'}^{\alpha\beta} S_{n\alpha}^z S_{n\beta}^z . \nonumber
\end{eqnarray}
%
Here,  $n_{n\alpha}^{s}$ and $S_{n\alpha}^{z}$ are  operators of the
electron occupation  number \mbox{$(S=\uparrow,\downarrow)$} and
pseudospin, respectively, in the $n$-th unit cell ($\alpha$=1,2
corresponds to the two apex oxygen $\rm O_{4}$ in the cell).
Additionally to the term $H_{e-s}$ describing the interaction
between electrons and pseudospins, the direct interaction $H_{s-s}$
between pseudospins is included; the $J S_{n1}^{z}S_{n2}^{z}$
interaction  within the one-cell clusters is separated. Hamiltonian
(\ref{ista:eq3.17}) is invariant with respect to the particle-hole
transformation: $n_{n\alpha}^{s}\rightarrow1-n_{n\alpha}^{s}$,
$h\rightarrow2g-h$, $\mu\rightarrow-\mu-U$.

Thermodynamics of the system described by Hamiltonian
(\ref{ista:eq3.17}) without a term related to the electron transfer
was studied in Refs.~[\refcite{ista:33,ista:34}]. The mean-field
approximation combined with the exact treatment of interactions
within the one-cell clusters was used. For this purpose the
single-clusters basis of states
$|R_{i}\rangle\equiv|n_{i1}^{\uparrow},n_{i1}^{\downarrow},
n_{i2}^{\uparrow}, n_{i2}^{\downarrow}\rangle\oplus|S_{i1}^{z},
S_{i2}^{z}\rangle$, which consist of sixty-four state vectors, was
introduced. In such a way the grand canonical potential $\Phi_{MF}$
and the mean values $\langle S_{\alpha}^{z}\rangle$ and
$\sum\limits_{s}\sum\limits_{\alpha=1}^{2}\langle
n_{n\alpha}^{s}\rangle=n$ were calculated.\cite{ista:33}

Instead of the $\langle S_{\alpha}^{z}\rangle$ parameters, we can
use their linear combinations: $\eta=\langle
S_{1}^{z}+S_{2}^{z}\rangle$ (the order parameter for the ordered
ferroelectric-like phase) and $\xi=\langle
S_{1}^{z}-S_{2}^{z}\rangle$ (the parameter which is responsible for
the in-phase reorientation of both pseudospins in the unit cell). In
the regime $\mu=const$ we obtain the equations for $\eta$ and $\xi$
from the condition of the minimum of the grand canonical potential
$\Phi_{MF}$
\begin{eqnarray}
\left\{
\begin{array}{c}
(\frac{\partial \Phi_{MF}}{\partial \eta})_\mu = 0,\\
(\frac{\partial \Phi_{MF}}{\partial \xi})_\mu = 0.
\end{array}
\right. \label{ista:eq3.18}
\end{eqnarray}
%
\begin{eqnarray}
\textstyle \lefteqn{\textstyle\Phi_{MF}= \frac{1}{4}\{
(J_{11}+J_{12})\eta^2 + (J_{11}-J_{12})\xi^2 \}  } \\ &&\textstyle
-T{\rm ln}\left[
  2\left\{ e^{\beta \frac{J}{4}}\cosh\beta \frac{(J_{11}+J_{12})\eta}{2} +
  e^{-\beta \frac{J}{4}} \cosh\beta \left( h+ \frac{(J_{11}-J_{12})\xi}{2}\right)\right\}  \right.
  \nonumber\\
&&\textstyle\left. +8e^{\beta \mu}\left\{ e^{\beta \frac{J}{4}}
\cosh\beta \frac{(J_{11}+J_{12})\eta}{2} \cosh\beta \frac{g}{2}
  + e^{-\beta \frac{J}{4}} \cosh\beta \left( h \right. \right. \right. \nonumber \\
&&\textstyle + \left. \left.
  \frac{(J_{11}-J_{12})\xi}{2} - \frac{g}{2} \right) \right\}
\left.+  8e^{2\beta \mu}\left\{ e^{\beta \frac{J}{4}} \cosh\beta
\frac{(J_{11}+J_{12})\eta}{2}   \right. \right. \nonumber
\\ &&\textstyle  \left. \left.  +
  e^{-\beta \frac{J}{4}} \cosh\beta \left( h+ \frac{(J_{11}-J_{12})\xi}{2} -g \right)\right\} \right]
  \, . \nonumber
 \label{ista:eq3.19}
\end{eqnarray}
In the case $n=$const the condition of the minimum value of the free
energy $F_{MF}=\Phi_{MF}+\mu n$, supplemented by the equation for
chemical potential
\[
-\frac{\partial\Phi_{MF}}{\partial\mu}=n,
\]
was used to determine the thermodynamically stable equilibrium
states.

\begin{figure}[!h]
\centerline{\psfig{file=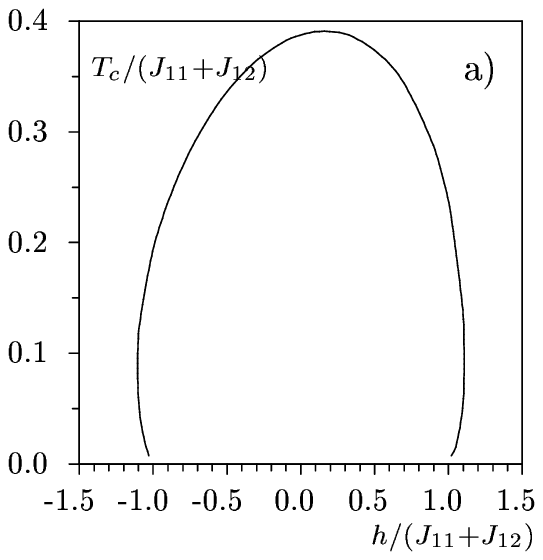,width=1.85in,angle=0} \quad
\psfig{file=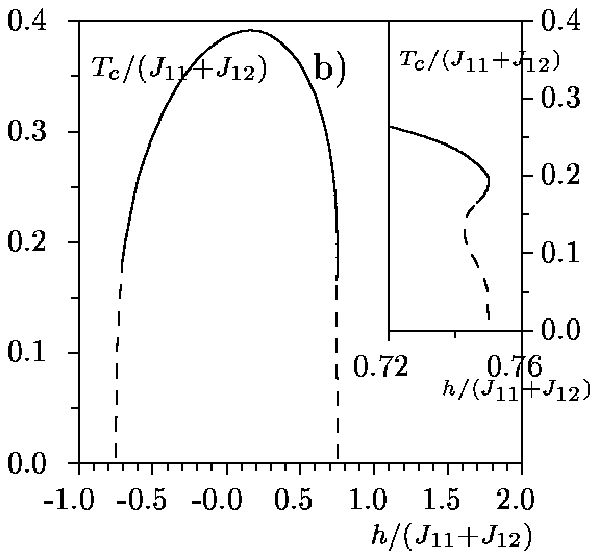,width=2.05in,angle=0}} \vspace*{8pt}
%
\caption{ $(T_{c}-h)$ phase diagrams
         at different values of parameter $J_{11}-J_{12}$ in the
         regime $\mu=const$:
         a) $\frac{J_{11}-J_{12}}{J_{11}+J_{12}}=-1$,
         b) $\frac{J_{11}-J_{12}}{J_{11}+J_{12}}= 0$.
         Other parameters: $J/(J_{11}+J_{12})=-1$, $g/(J_{11}+J_{12})=1$,
         $\mu /(J_{11}+J_{12})=-1$.
   The phase transitions can be of the second (solid lines) or of
    the first order (dotted lines).
} \label{ista:f37}
\end{figure}

When the set of equations (\ref{ista:eq3.18}) has a non-zero
solution for $\eta$ and the corresponding thermodynamic potential
has a minimum, then our system is in the polar (ferroelectric)
phase.  The obtained phase diagrams  in the $\mu=$const
case\cite{ista:33,ista:34} are shown in Fig.~\ref{ista:f37} (all
parameters are normalized here  by $J_{11}+J_{12}>0$). One can see
that at $\frac{J_{11}-J_{12}}{J_{11}+J_{12}}=-1$ (the case of
pseudospin-pseudospin interaction  between different sublattices
only) the phase transition into ferroelectric phase
\index{phase!ferroelectric} is of the second order. The presence of
the intra-sublattice interaction leads to the possibility of change
of the phase transition order and the appearance of tricritical
points. \index{point!tricritical} An increase of the
$\frac{J_{11}-J_{12}}{J_{11}+J_{12}}$ parameter causes the narrowing
of the ferroelectric region; its width is also proportional to $J$.
At $J=0$, $J_{12}=0$ the model transforms into the one-sublattice
PEM; the ferroelectric phase disappears and we have the first-order
phase transition with zero value of the order parameter $\eta$ and a
sharp change of parameter $\xi=\langle S_{1}^{z}-S_{2}^{z}\rangle$.

%
%

\begin{figure}
\centerline{\psfig{file=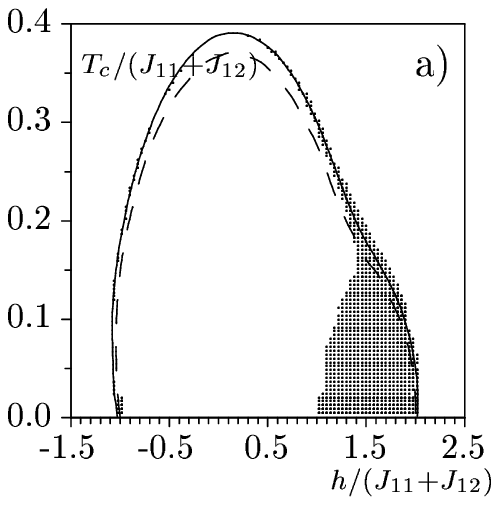,width=2.0in,angle=0} \quad
\psfig{file=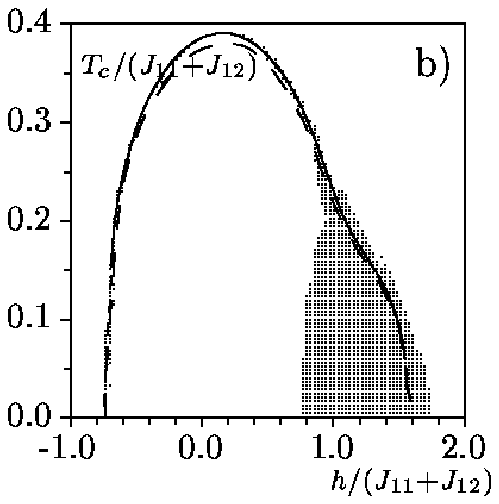,width=2.0in,angle=0}} \vspace*{8pt}
\caption{The $h$ dependence of the temperature of ferroelectric
phase transition $T_{\mathrm{c}}$ at different values of parameter
$J_{11}-J_{12}$ in the regime $n$=const: a)
$(J_{11}-J_{12})/(J_{11}+J_{12})=-1$, b)
$(J_{11}-J_{12})/(J_{11}+J_{12})= 0$.
Other parameters: $J/(J_{11}+J_{12})=1$, $g/(J_{11}+J_{12})=1$,
$n=0.4$. Solid lines and dashed lines represent the second order and
the first order phase transitions, respectively. The widely spaced
dashed line corresponds to a one-loop approximation. Dots represent
a separation area. } \label{ista:f39}
\end{figure}

In the $n$=const regime, the region of $h$ values at which the
ferroelectric phase exists becomes broader. The corresponding phase
diagrams  are shown in Fig.~\ref{ista:f39}
at different values of long range interaction $J_{\alpha\beta}=
\sum_{n'} J_{nn'}^{\alpha\beta}$ and fixed concentration $n=0.4$.
%

\begin{figure}[!h]
\centerline{\psfig{file=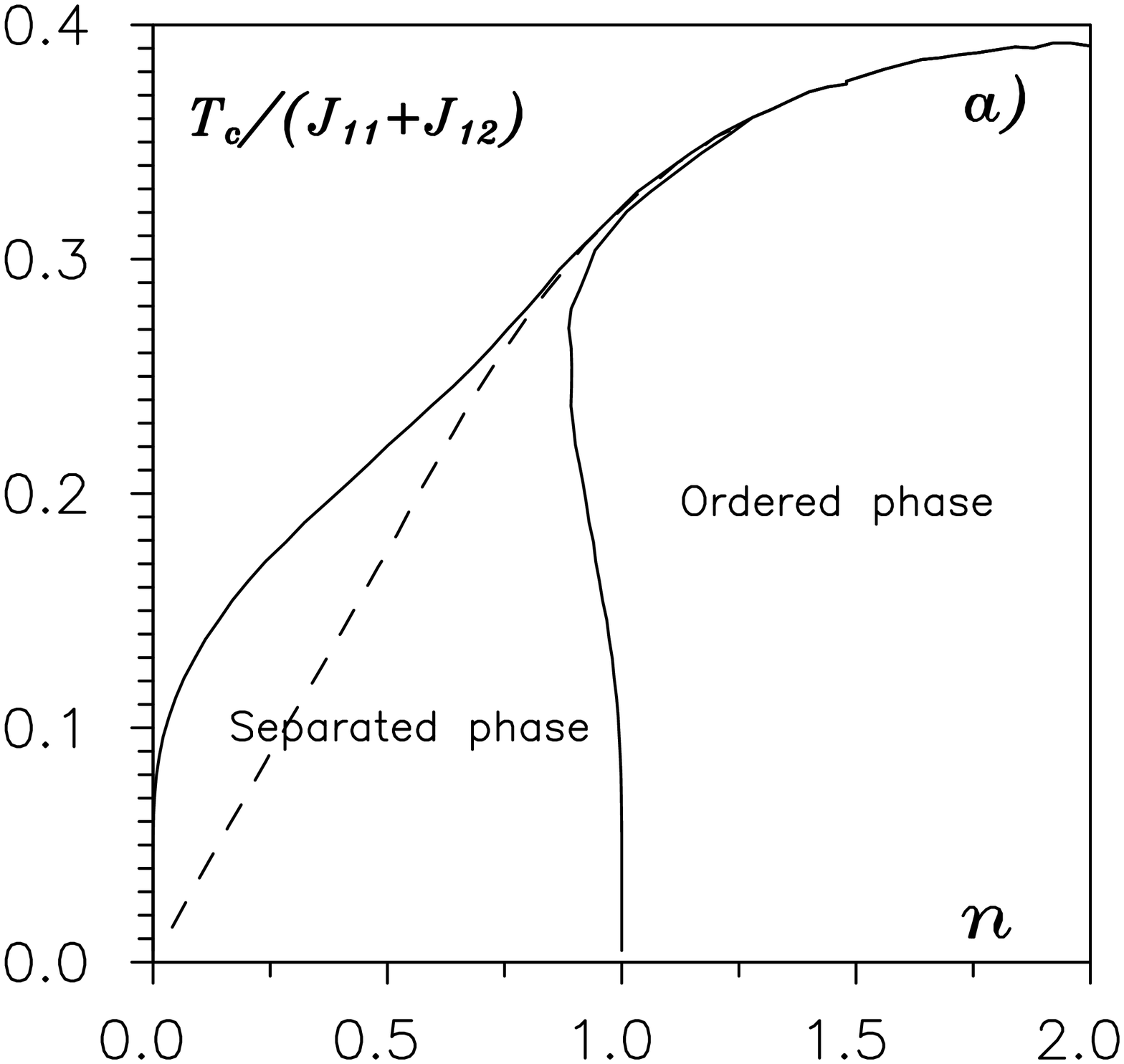,width=2.0in,angle=0} \quad
\psfig{file=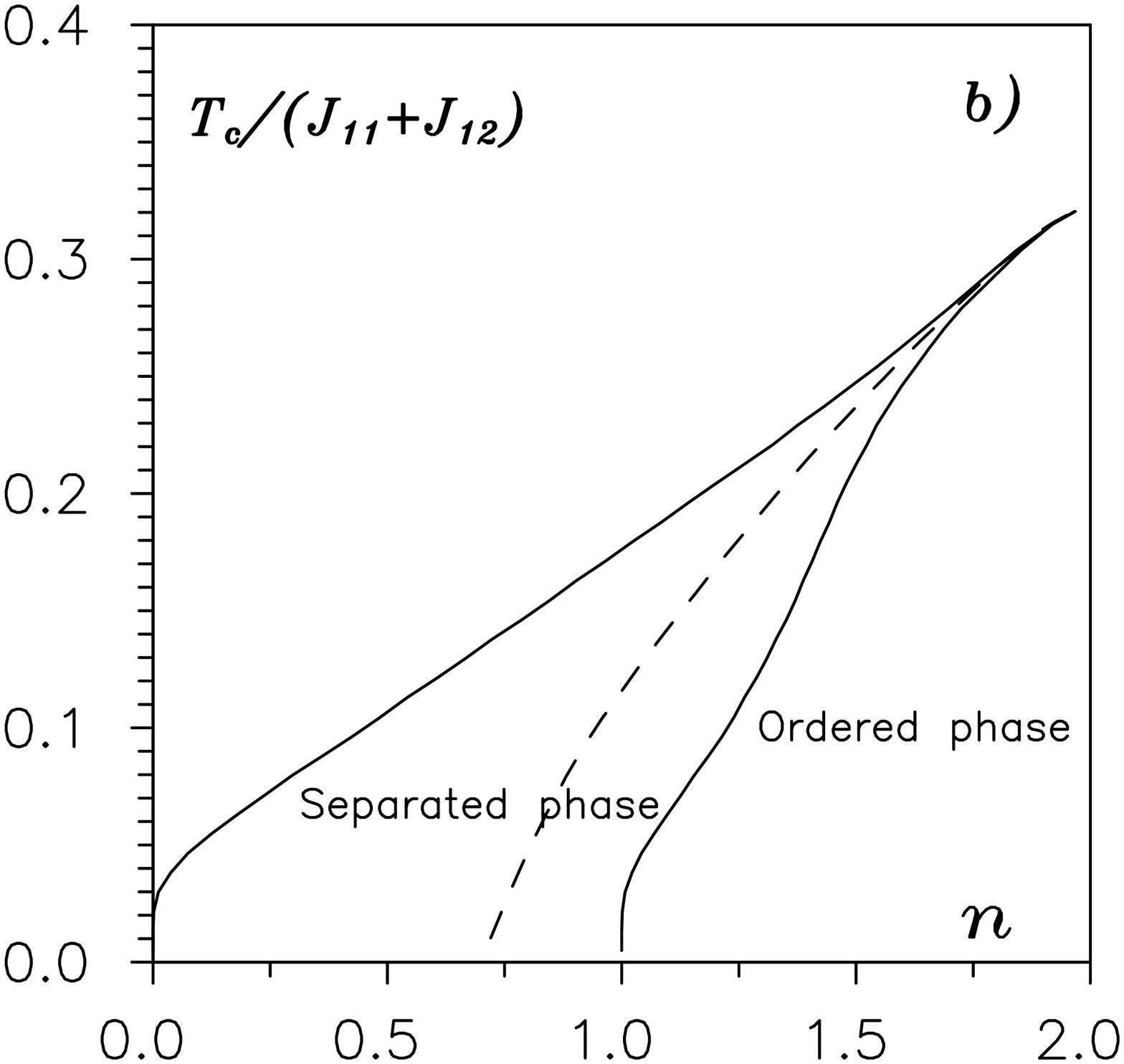,width=2.0in,angle=0}} \vspace*{8pt}
\caption{
Phase ($T-n$) diagram in a mean field approximation. The phase
separation region is limited by solid lines. The dashed lines point
to the region of ferroelectric instabilities. Parameters:
$(J_{11}-J_{12})/(J_{11}+J_{12})= 1$,
$J/(J_{11}+J_{12})=g/(J_{11}+J_{12})=1$; a) $h/(J_{11}+J_{12})=1$,
b) $h/(J_{11}+J_{12})=1.35$\,. } \label{ista:f40}
\end{figure}

 For comparison, there are also shown   the phase transition lines
obtained in the so-called one loop-approximation (in which the
higher order corrections to the MFA described by two-tailed
diagrams\cite{ista:34} are taken into account). In the areas marked
by points, the system is separated into two regions with
concentrations $n_1$ and $n_2$ ($n_1<n<n_2$). Also, one can notice
that phase separation  takes place near the border of stability
region of two phases (the ordered phase with nonzero polarization
and the disordered  one). That is why the ordered phase spreads
wider and  extends up to the edge of the separated area.
Fig.~\ref{ista:f40} illustrates such a behaviour. Here the dashed
lines
 represent the region of ferroelectric type instabilities.
These lines would separate the ferroelectric phase if there were no
phase separation. Figure~\ref{ista:f40}(b) also shows that at a
fixed value of asymmetry parameter $h$ at concentration $n<0.75$ (at
low temperature), the ordered phase is possible only due to the
phase separation.
%
%

Now, let us focus briefly on the effect of electron transfer. As
above, we restrict ourselves to the limit $U\rightarrow\infty$. The
case is considered, when the chemical potential is placed in the
region of energy subbands $(\alpha=1,2)$
\begin{eqnarray}
\varepsilon_{\alpha}^{31}({\bf k})=\varepsilon_{\alpha}^{41}({\bf
k})=-(-1)^{\alpha}\frac{g}{2}+t_{{\bf k}}\langle
X_{\alpha}^{44}+X_{\alpha}^{11}\rangle_{0}, \nonumber\\
\varepsilon_{\alpha}^{\tilde{3}\tilde{1}}({\bf
k})=\varepsilon_{\alpha}^{\tilde{4}\tilde{1}}({\bf k})=
(-1)^{\alpha}\frac{g}{2}+t_{{\bf k}}\langle
X_{\alpha}^{\tilde{4}\tilde{4}}+X_{\alpha}^{\tilde{1}\tilde{1}}\rangle_{0}
\label{ista:eq3.20}
\end{eqnarray}
separated by the gap equal to $g$. The Hubbard-I approximation in
which  the expressions (\ref{ista:eq3.20}) are given here, was
improved in Ref.~[\refcite{ista:36}] by the mean-field corrections
(being of the form of the loop-like inclusions) to the electron
Green's functions. Thus,  renormalization  of spectrum due to the
 shift of subbands dependent on electron concentration was taken
into account. Besides, the mean values of Hubbard operators in
(\ref{ista:eq3.20}) were determined self-consistently; the free
energy of the pseudospin-electron systems was calculated in the
above described  GRPA approach. The pseudospin part of the
two-sublattice Hamiltonian was taken in the MFA.

\begin{figure}[b]
\centerline{\psfig{file=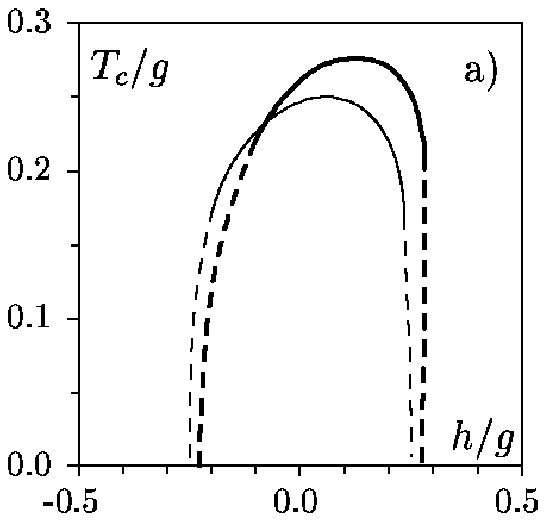,width=1.9in,angle=0} \quad
\psfig{file=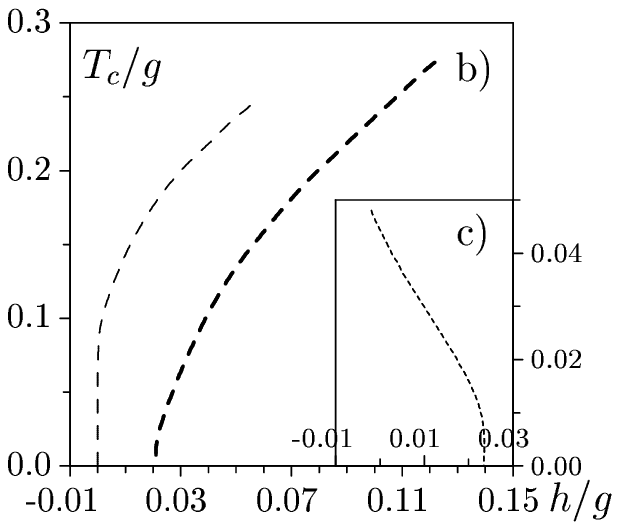,width=2.1in,angle=0}} \vspace*{8pt}
\caption{Dependence of temperature of the phase transition $T_c$ on
the parameter $h$ at different values of interaction parameters
$J_{11}$ and $J_{12}$ in the regime
 $\mu=const=-g$.
Thick lines correspond to the case  $t_{ij}/g=0.2$; thin lines
correspond to the case $t_{ij}=0$. a) $J_{11}=J_{12}=g/2$, b)
$J_{11}=g, J_{12}=0$. c) $J_{11}=J_{12}=0, t_{ij}/g=0.2$. Phase
transitions are either
 of the second order (solid lines) or of the first order (dashed lines).}
\label{ista:f41}
\end{figure}
Figure~\ref{ista:f41} illustrates the effect of electron subsystems
on the shape of the phase diagram. With respect to the standard
Mitsui model\index{model!Mitsui} (which corresponds to the $g=0$ and
$t_{ij}=0$ limit), the phase boundary  becomes asymmetric and the
region of the ferroelectric phase existence is shifted to the higher
values of $h$. As it has already been noted, at $J_{12}=0$,
ferroelectric phase does not exist  and only a discontinuous change
of $\xi$ on the transition line takes place. Such a transition also
remains in the case when $J_{11}=0$ and only the indirect
interaction via electron subsystem is present.

\begin{figure}[!h]
\begin{picture}(150,280)
\put(00,140){ \psfig{file=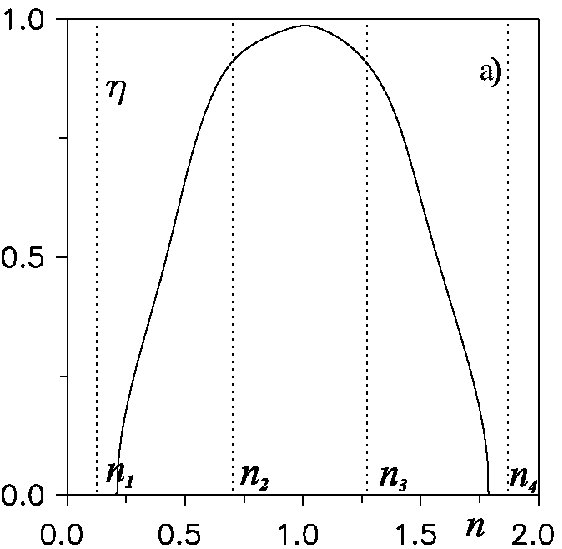,width=1.9in,angle=0}}
\put(150,140){ \psfig{file=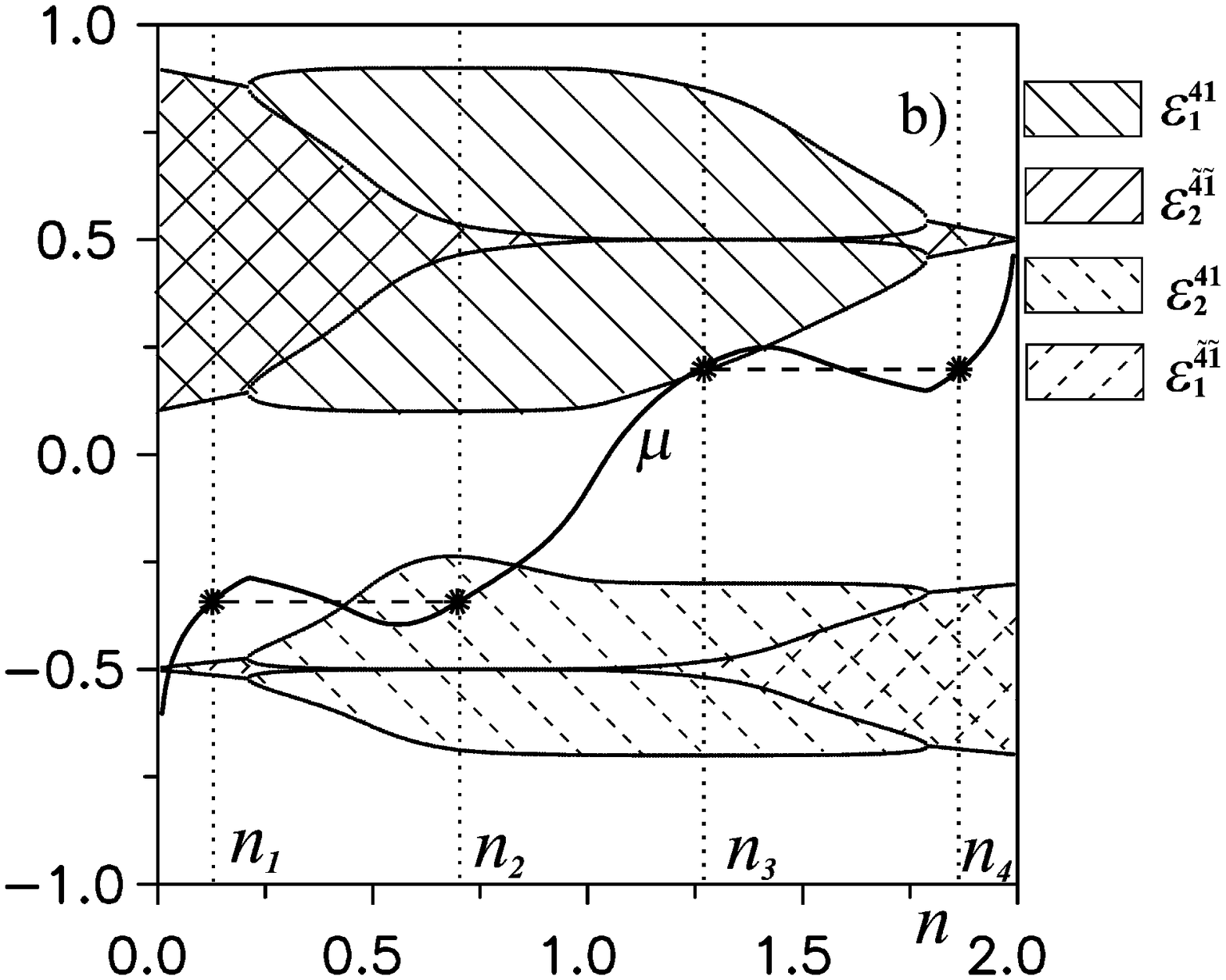,width=2.4in,angle=0}}
\put(80,00){ \psfig{file=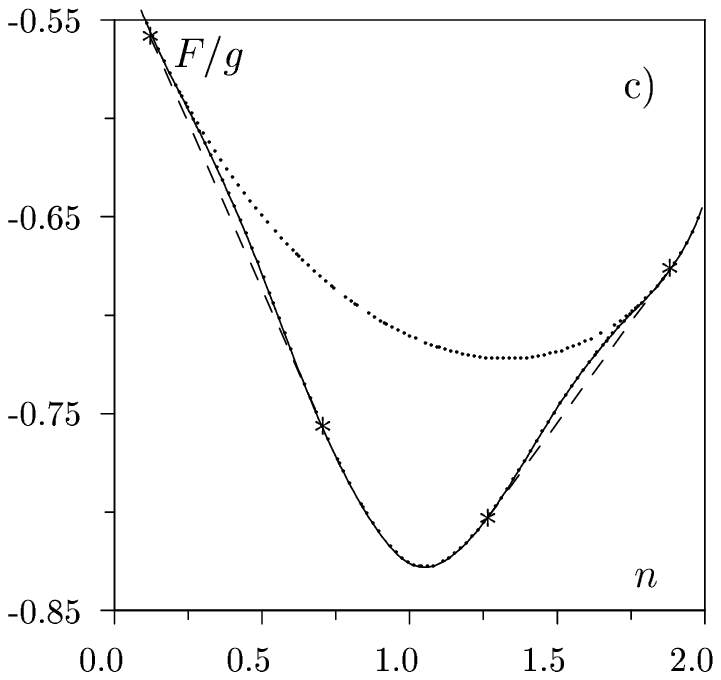,width=1.9in,angle=0}}
\end{picture}
\caption{ Dependence of the order parameter $\eta$ (a), band
spectrum (b), and free energy (c) on electron concentration. The
parameter values are: $J_{11}=J_{12}=g/2$, $T/g=0.1$,
$t_{ij}/g=0.1$, $h/g=0.5$. }  \label{ista:f42}
\end{figure}

 The changes in electron spectrum (see Refs.~[\refcite{ista:35,ista:36}])
are demonstrated in Fig.~\ref{ista:f42}(b). In the case shown in
Fig.~\ref{ista:f42}, when at $n=$const the separation into
ferroelectric and nonpolar phases takes place, the $\eta(n)$ and
$\mu(n)$ dependencies indicate that the ferroelectric phase appeared
before the separation.  This is also supported by the presence of
concavity in the free energy (dashed tangent lines in
Fig.~\ref{ista:f42}(c) link the points with concentration values
$n_{1}$, $n_{2}$ and $n_{3}$, $n_{4}$ on which the separation takes
place). Hence, there is a separation into paraelectric and
ferroelectric phases at concentrations $n_{1}<n<n_{2}$ and
$n_{3}<n<n_{4}$. The pure ferroelectric phase exists in the
concentration range $n_{2}<n<n_{3}$. The separation area in the
$(h,n)$ plane changes its shape depending on the $t_{ij}$ value. In
general, the electron transfer narrows the separation
region.\cite{ista:35} The reverse effect also takes place:  the
electron spectrum is modified by a phase separation being sensitive
to concentration value.

One can relate the phase transitions and ferroelectric  type
instabilities described  by the herein considered two-sublattice PEM
to the observed  dielectric and thermodynamic properties of the
YBaCuO-type superconducting  crystals.  In some experiments,
YBa$_{2}$Cu$_{3}$O$_{7-\delta}$ was found to be both pyroelectric
and piezoelectric, implying the existence of macroscopic
polarization directed along the c-axis.\cite{ista:71} The
possibility of the existence of the ferroelectric-like phase was
not, nevertheless,  unambiguously confirmed. In the two- sublattice
PEM the  ordered  polar phase is present in a rather restricted
region  of the model parameter values, where  the ground state is
degenerated; for example, its width along the $h$-axis is determined
by the interaction constant  between the pseudospins (describing
apical O$_{4}$ ions) within the unit cell cluster. In this
connection, it should be mentioned that presence of the oxygen
vacancies  in the chain element of structure  at $\delta>1$
effectively influences   the value of field $h$.\cite{ista:72} In
its turn, instability with respect to the polar phase appearance can
show up only at a certain nonstoichiometry (at certain values of
parameter $\delta$).

The phase separation in the two-sublattice case is also  related to
the real structure of the YBa$_{2}$Cu$_{3}$O$_{7-\delta}$ crystal.
Besides the question of microscopic nature of the so-called ``stripe
phases'' there is a problem of the genesis of structural
inhomogeneities  in a single crystal of the YBaCuO type observed in
the experiments using Raman spectroscopy\cite{ista:6} and mesoscopic
structural investigations.\cite{ista:73} The results obtained within
the model approach show the possibility of  the Mitsui - type (due
to both direct and indirect, via conducting electrons, interactions
between anharmonic structure units) mechanism of  the mentioned
instabilities.

\section{Conclusions}

The present investigations of thermodynamics and energy spectrum of
the PEM show  a variety of phases and phase transitions. Depending
on the thermodynamic equilibrium  regimes, they manifest themselves
as $(i)$ transitions between different uniform phases, or between
uniform and modulated  phases, with the commensurate or
incommensurate period of modulation (in the $\mu$=const regime) or
as $(ii)$ transitions into phase  separated states (in the $n=$const
regime). The latter takes place when at $\mu=$const, the
corresponding phase transition is of the first order. Such phase
transitions can be realized at the change of temperature $T$, field
$h$, chemical potential $\mu$ (in the case ($i$)) and other
parameters of the model. The corresponding phase diagrams are built
in  case of  strong or weak coupling ($g\gg W$ and $g<W$,
respectively).

A microscopic  reason for phase transitions in  standard PEM
(without a direct pseudospin-pseudospin interaction) is related to
the indirect effective coupling between pseudospins arising due to
electron transfer and possessing a dynamic character. The form of
such coupling depends on the electron concentration, temperature and
the model parameter values. Consequently, the modulated phase
appears  at intermediate values of $\mu$ ( this corresponds to the
electron occupation near half-filling) both at strong  and at weak
coupling. However, the formation mechanisms   of the effective
interaction are different in these cases.

The two-sublattice PEM is a special case, close to the real HTSC
systems of the YBaCuO type. The phase transitions described here,
connected with the appearance of  the ferroelectric state  or with
the jump-like change  of the mutual orientation of pseudospins in
sublattices, can have a relation to the segregation or bistability
phenomena as well as to the  development of ferroelectric-type
instabilities in the mentioned systems.

 The analysis carried out in the weak coupling case elucidates  the
role of the transverse field $\Omega$ (having a meaning of the
tunneling-like splitting parameter) in the obtained picture of phase
transitions. As a whole, a topology of the phase diagrams does not
change at $\Omega\neq0$ in  comparison with the case $\Omega=0$. The
critical temperatures (including $T_{SC}$) decrease with $\Omega$.
In the case of DOS with logarithmic singularity (at dimensionality
$d=2$) a peculiar effect is revealed: the critical temperature
remains finite at  any large values of $\Omega$ (tending to zero at
$\Omega\rightarrow\infty$, exclusively).  This is the difference
with respect to the behaviour of the systems with direct interaction
(e.g., Ising model with transverse field). The important property is
that at $\Omega\neq0$ the superconducting phase  can appear in the
PEM. Such a phase competes with the modulated one and is stable at
the electron occupancy near the upper (or lower) edge of the
electron band.

As is seen from the results obtained in DMFT for the simplified
model, the structure of electron spectrum of the PEM is  different
in cases $g\gg W$ and $g< W$: the split subbands due to interaction
 (even at $U=0$) or a single  band, respectively.  The
similar spectrum is obtained in the approximations, based on the
Hubbard-I scheme (e.g. GRPA) in the first case or the Hartree-Fock
approach in the second one. As was shown by Zeyher, Kuli\'c, and
Gehlhoff,\cite{Kulic94,Gehlhoff95,Zeyher96}  GRPA keeps in a
systematic way all terms of the leading order of a so-called $1/N$
expansion, where $N$ is the  local spin component number on a
lattice site. Though such approximations do not adequately reproduce
all the features of spectrum, the obtained phase diagrams are  in a
good agreement with the ones constructed in DMFT. An interesting
feature of spectrum is that there exists a critical value of $g$: at
$g>g_{c}$ a gap appears (in the case of simplified PEM) and a
metal-insulator type transition takes place.


Pseudospin-electron model (PEM) can be considered as a
  generalization of the Falicov-Kimball (FK) model
  to the case of
  different thermodynamic equilibrium regimes as well as an
   extension of the latter
    model due to the inclusion of the pseudospin
   dynamics and the Hubbard type correlations. The PEM possesses a
   similar variety of phase transitions but there are differences
    in the conditions of their realization and in  the criteria of
    the appearance of different phases.
In the above considered   cases such  differences are  discussed and
a comparison with the results for the FK model is made.

From the point of view of theoretical studies, the investigations of
the PEM are  far from complete. Another interesting  problem is
connected with the thermodynamics of the PEM with electron transfer
at $U\neq0$ and $\Omega\neq 0$. The investigations performed
revealed only the existence of instabilities connected with certain
values of the wave vector ${\bf q}$ but the phase diagrams
determining the regions of existence  of different phases have not
been built so far. The intermediate coupling case ($g\sim W$), that
was not adequately investigated  even for the simplified PEM, calls
for more detailed  consideration. An important point is to complete
the study of the  collective excitation spectrum (connected with
pseudospin reorientation, electron transitions and polaron effect)
and the dynamic susceptibility. Among the possible generalizations
of the model, one can note an extension to the cases with the
asymmetric electron transfer (in the spirit of the asymmetric
Hubbard model) and with pseudospin $S>1/2$. They appear to be quite
promising in connection  with investigations of the ionic transport
based on the  lattice models as well as in the study of the ion
intercalation processes (see e.g. Ref.~\refcite{Zhou06} for a recent
development in this field).

\section*{Acknowledgements}
\addcontentsline{toc}{section}{Acknowledgements}

The author is grateful to A. Shvaika and T. Mysakovych for reading
the manuscript and technical help.

\printindex
\end{document}